\documentclass[manuscript]{aastex}
\newcommand{\lsim}{\, \lower2truept\hbox{${< \atop\hbox{\raise4truept\hbox{$\sim$}}}$}\,}
\newcommand{\gsim}{\, \lower2truept\hbox{${> \atop\hbox{\raise4truept\hbox{$\sim$}}}$}\,}

\newcommand{\oneskip}{\vskip\baselineskip}

\newcommand{\puncspace}{\ifmmode\,\else{\ifcat.\C{\if.\C\else\if,\C\else\if?\C\else%
\if:\C\else\if;\C\else\if-\C\else\if)\C\else\if/\C\else\if]\C\else\if'\C%
\else\space\fi\fi\fi\fi\fi\fi\fi\fi\fi\fi}%
\else\if\empty\C\else\if\space\C\else\space\fi\fi\fi}\fi}
\newcommand{\SP}{\let\\=\empty\futurelet\C\puncspace}

\begin{document}

\title{Models for Evolution of Dusty Galaxies and E/S0s
Seen in Multiband Surveys}
\author{C. K. Xu,  C. J. Lonsdale, D. L. Shupe}
\affil{Infrared Processing and Analysis Center, Jet Propulsion Laboratory, 
Caltech 100-22, Pasadena, CA 91125, USA}
\author{A. Franceschini}
\affil{Dipartimento di Astronomia, Universita' di Padova, 
Vicolo dell'Osservatorio 5, I-35122, Padova, Italy}
\author{C. Martin, D. Schiminovich}
\affil{California Institute of Technology, Pasadena, CA 91125, USA}

\received{7.12.2002}
\accepted{12.13.2002}

\begin{abstract}
Phenomenological models for evolution of dusty galaxies
and E/S0 galaxies, respectively, are developed to
address two major questions concenring
galaxy populations in deep infrared (IR) surveys:
(1) Do normal late-type galaxies
or starburst galaxies (including galaxies with obscured AGNs)
dominate among sources in deep IR surveys?
(2) How much do E/S0 galaxies contribute to the
counts in deep mid-infrared (MIR: 3 -- 20$\mu m$) surveys?
Among three new models for evolution of dusty galaxies,
it is assumed in Model S1  
that starburst galaxies are the dominant population,
and in Model S2 that normal galaxies dominate. Model S3
is an intermediate model. Comparing the model predictions
with a wide range of observational data collected from
the literature, we find that none of these models
can be ruled out, given the uncertainties of the data.
We show that the most direct method to distinguish
these models is to compare the predicted color distributions 
of IR galaxies with observations, which will soon be available from the
SWIRE survey. The models for E/S0 galaxies 
follow a simple passive evolution approach. Among the three
E/S0 models (E1, E2 and E3) investigated in this paper,
Model E2 which is specified by a peak formation redshift 
$z_{peak}=2$, and an e-folding formation time scale 
$\omega=2$ Gyr, fits the data best. This suggests a synchronization
between the evolution of E/S0 galaxies and of starburst galaxies, 
in the sense that the peak of the
formation function of E/S0s ($z_{peak}=2$)
is close to the peak of the evolution functions of starburst
galaxies ($z_{peak} = 1.4$). We find that
E/S0s contribute about 10 -- 30\% of the
counts in the MIR bands of $<10\mu m$, and 
up to 30 -- 50\% of the optical/NIR
counts in the bright end. Their contributions
to counts in the UV (2000{\AA}) and in the 
longer wavelength IR ($\geq 12\mu m$) bands are negligible.
Taking into account this contribution, new predictions 
for counts and confusion limits in the SIRTF bands are 
presented.

\end{abstract}


\section{Introduction}

Many new windows have recently
been opened in various wavebands for observations of 
high redshift galaxies. Particularly, deep ISO surveys
(e.g. Rowan-Robinson et al. 1997; Franceschini et al. 1997;
Kawara et al. 1998;
Elbaz et al. 1999; Aussel et al. 1999; Flores et al. 1999b;
Puget et al. 1999; Dole et al. 2001;
Clements et al. 1999; Serjeant et al. 2000) in the infrared
and SCUBA surveys 
(Hughes et al. 1998; Barger et al. 1998; Blain et al. 1999) in the
submm wavebands have shed lights on the dark side of galaxy 
formation and evolution. This stimulated a surging wave of 
empirical models (Xu et al. 1998; Blain et al. 1999; 
Roche \& Eales 1999; Dole et al. 2001; Xu 2000; Rowan-Robinson 
2001; Xu et al. 2001; Franceschini et al. 2001; Pearson 2001; Malkan \&
Stecker 2001; Chary \& Elbaz 2001) to interpret the new ISO and SCUBA
sources. The main aim of these empirical models is to better constrain some
important characteristics of the galaxy evolution, including the
peak epoch of the star formation rate in the universe and
the roles played by different star formation modes (e.g. the quiescent mode
against the interaction stimulated mode),
for which the earlier optical deep surveys
such as the Hubble Deep Field Survey (Williams et al. 1996)
may have provided biased information (Madau et al. 1996) by neglecting
the dust extinction (Rowan-Robinson  et al. 1997; Lonsdale 1999). 
This will certainly have impact on the theoretical simulations
of galaxy formation and evolution, which 
(e.g. Somerville et al. 2001) have just started 
facing the fact that most of the early star formation may be
hidden behind a thick veil of dust,
making the incorporation of effects of dust extinction and 
emission in the framework essential (Granato et al. 2000).

Some of the most recent empirical models (Rowan-Robinson 2001;
Xu et al. 2001; Franceschini et al. 2001; Pearson 2001;
Chary \& Elbaz 2001) have the feature that they can
be constrained by counts and the cosmic background radiation
in various IR and submm bands simultaneously. This is achieved
by using SED templates to link sources in different bands.
The SED library of Xu et al. (2001) contains realistic SEDs of a 
complete sample of 837 IRAS 25$\mu m$ selected galaxies,
enabling prediction of counts as well as color distributions,
which provide additional constraints to the model.
Templates in Rowan-Robinson (2001), Xu et al. (2001) and Chary
\& Elbaz (2001) extend from the IR-submm to optical and UV wavebands,
hence these models can predict also the contributions from
IR sources to optical and UV counts, relating the evolution of
IR sources to that of galaxies seen in earlier optical and UV surveys.
All of these works found strong evolution among IR sources in
the redshift range of $0<z<1.5$. Assuming that the narrow 
sub-mJy bump on the Euclidian normalized
differential counts of ISOCAM 15$\mu m$ surveys (Elbaz et al. 1999)
is due to the K-corrections caused by the strong unidentified 
broad band emission features (UIB) at 6 -- 8$\mu m$,
which are shifted into the ISOCAM 15$\mu m$ band filter (LW3)
when $z\sim 1$, Xu (2000) concluded that typical IR galaxies 
at $z\sim 1 $ have $L_{15\mu m} \sim 10^{11}\; L_\sun$, namely
about 20 times more luminous than their local counterparts,
while their comoving density is about the same as their local
counterparts. As argued by Xu et al. (2001), 
the location of the 15$\mu m$ bump 
($f_{15\mu m} \sim 0.4 mJy$) provides a strong constraint
on the luminosities of IR sources at $z\sim 1$.
This has been confirmed by the detailed study of 
ISOCAM sources in the HDF-north by Elbaz et al. (2002),
who obtained redshifts for nearly all of these sources
(40 out of total 41).
The strong evolution has also been supported 
by surveys in other ISOCAM bands. For example,
Clements et al. (2001) found that the luminosity functions of their ISO
12$\mu m$ sources are consistent with a pure
luminosity evolution of rate $\sim (1+z)^{4.5}$, as derived by
Xu (2000) from the ISOCAM 15$\mu m$ counts. 

On the other hand, different authors 
identified different populations of galaxies 
as the major carriers of the evolution of IR sources.
It was found in early IRAS studies
(Franceschini et al. 1988; Rowan-Robinson \& Crawford 1989)
that IR sources can be divided into 3 different populations:
normal late-type galaxies ('cirrus galaxies'), starburst
galaxies, and galaxies with AGNs. Assuming that all IR sources
undergo pure luminosity evolution with the 
same evolution rate, Rowan-Robinson (2001) found that
'cirrus galaxies' dominate the ISOCAM $15\mu m$ counts, the
SCUBA 850$\mu m$ counts, and the cosmic IR background radiation.
Xu et al. (2001), motivated by results of optical identifications
of ISO galaxies by Flores et al. (1999b) and Aussel et al. (1999)
which show a larger percentage of these sources are in
interacting systems, assumed that most of the evolution 
of the IR sources is due to starburst galaxies
(defined by warm FIR colors: $f_{60\mu m}/f_{100\mu m} \geq 0.5$).
Obscured AGNs with relatively low
$f_{25\mu m}/f_{60\mu m}$ ratios (due to the
very high extinction affecting even the MIR fluxes) may be
misclassified as starbursts in Xu et al. (2001), who
classified galaxies with AGNs using the criterion
$f_{25\mu m}/f_{60\mu m} \geq 0.2$.
Franceschini et al. (2001) also assumed that the strong evolution
is confined to the starburst population. Their starburst 
galaxies include all 'active' galaxies not
classified as Sey~1, i.e. including the Sey~2s
and the LINERs, in Rush et al. (1993). 
Pearson (2001) found that a separate population
of ultra-luminous galaxies (ULIRGs) with L$\sim 10^{12} L_\sun$,
confined in a very narrow range of redshift centered at $z\sim 1$,
is mostly responsible for the strong evolution seen in the ISOCAM
15$\mu m$ counts. Since all these models can fit the IR-submm
counts and the CIB within uncertainty limits, there is 
apparently a degeneracy concerning the population of
galaxies that carry most of the evolution of IR sources.
The degeneracy still remains even when the new, 
detailed information of the 15 micron universe
(source counts, redshift distributions, luminosity functions
at different redshifts, etc.; Elbaz et al. 2002;
Franceschini et al. 2001) is considered (see Section 2). 

One of the central issues in the current agenda of hierarchical
galaxy formation simulation studies concerns the roles played
by two different star formation modes (eg. Somerville et al. 2001;
Kauffmann et al. 2001), namely the quiescent star formation mode
and the interaction-induced star formation mode. There are apparent
links between the normal quiescent galaxies 
and the quiescent star formation
(as in the Milky Way and M31) and between starburst
galaxies and the interaction-induced star formation (as in 
the Antennae Galaxies and in M82). Therefore, the answer to the
question whether the normal galaxies or starburst galaxies dominate
the faint IR counts will have important impact on galaxy
evolution theories.

Another deficiency of the current models for IR sources
is the neglection of early type galaxies. Although these 
galaxies have little dust emission, therefore 
do not contribute significantly to IR counts at 
wavelengths longer than $\sim 10\mu m$, they
are found as an important population in
ISO 6.7$\mu m$ counts (Flores et al. 1999b; Aussel et al.
1999; Serjeant et al. 2000).
Future mid-IR surveys such as those planned for SIRTF
IRAC cameras (4 bands centered at 3.6$\mu m$, 4.5$\mu m$, 
5.8$\mu m$ and 8$\mu m$, respectively) will certainly detect
many of these galaxies.

In this paper, we shall address the above two issues,
namely (1) developing new models to investigate how to
break the degeneracy concerning the major
evolution population in IR counts (using SIRTF data 
in particular), and (2) modeling
the evolution of E/S0 galaxies and predicting their
contributions to the MIR counts. The plan of the paper is as follows:
in Section 2 a set of new models for evolution of dusty galaxies
are presented,
while in Section 3 models for evolution of E/S0 galaxies are 
developed and compared to the observations; predictions
for the UV, optical and NIR counts by a set of composite models,
each consisting of a dusty galaxy evolution model and an E/S0 evolution 
model, are presented and compared with data in Section 4.
The predictions for counts and confusion limits in SIRTF bands by the same
models are presented in Section 5. In Section 6 we show how to
use color distributions of ISO sources and future SIRTF
sources to answer the question whether starburst galaxies or
normal late-type galaxies are dominant in the IR sources. 
Section 7 is devoted to discussion. A summary is given
in Section 8.\footnote{More detailed results on predicitons of
models presented in this paper are available on request
to C.K.X..}
The $\Lambda$-Cosmology ($\Omega_\Lambda =0.7$, $\Omega_m = 0.3$, H$_0$=75
km sec$^{-1}$ Mpc$^{-1}$) is assumed throughout the paper.

\section{New Models for Evolution of Dusty Galaxies}
\subsection{Previous Models}
In all the multi-band models for the evolution of dusty galaxies
in the literature, the different populations of galaxies are
separated by their characteristic spectral energy distributions
(SEDs): the normal galaxies have relatively low $f_{60\mu m}/f_{100\mu m}$
and $f_{25\mu m}/f_{12\mu m}$ ratios and very prominent UIB features,
the starburst galaxies have relatively high $f_{60\mu m}/f_{100\mu m}$
and $f_{25\mu m}/f_{12\mu m}$ ratios and less prominent UIB features
over a steep rising MIR continuum, and galaxies with
AGNs have relatively high $f_{25\mu m}/f_{60\mu m}$ ratios and 
very week UIB (contributed by ISM dust not associated with the AGN).
Models by Xu et al. (2001) have also considered the luminosity
dependence of the SED in each population, while some other models
(Rowan-Robinson 20001;
Franceschini et al. 2001; Pearson 2001; Chary \& Elbaz 2001)
separate high luminosity starbursts (Arp220 type) from moderate
luminosity starbursts (M82 type). Because of their different SEDs,
these different populations of galaxies give different relative
contributions to counts at different wavelengths. However,
it appears that the counts are rather loose constraints to
these relative contributions. At least 2 factors are responsible
for this degeneracy: 
(1) there are many parameters in the evolution
functions (both the luminosity evolution and the density evolution)
when different populations are allowed to evolve
differently, and (2) there are still significant uncertainties
in the current ISO counts (e.g. counts of the same ELAIS 15$\mu m$
sources reported by Serjeant et al. 2000 and by Gruppioni et al. 2002
differ with each other by 
as much as a factor of 3), due both to calibration errors and
the field-to-field variation.
In principle, the degeneracy can be broken by comparing model
predictions on redshift distributions of deep
FIR ($\lambda \gsim 60\mu m$) selected samples with
observations: Models assuming warmer IR SEDs (starbursts dominant)
usually predict larger mean redshifts than models assuming 
cooler IR SEDs (normal-galaxies dominant) do. However, similar to
the SCUBA galaxies, the large beams of FIR detectors (such as the
ISOPHOT cameras) make it rather difficult to pin-down the
optical counterparts (usually quite faint) of faint FIR sources.

A more direct method to constrain the relative contributions of
different populations to the IR counts is to compare the predicted
and observed color distributions of IR sources. 
Among the multi-band models for evolution of dusty galaxies
in the literature, those of Xu et al. (2001) have the most
sophisticated algorithm in dealing with the SEDs, and are the
only ones that can predict both the mean colors and 
their dispersions. Therefore, we choose to
build our new models using the same algorithm as Xu et al. (2001).

\subsection{New Models}
In the new models, the simulation code of Xu et al. (2001)
is modified in the following aspects:
\begin{description}
\item{(1)} The luminosity and density evolution functions 
for the starburst galaxies and normal late-tpye galaxies, respectively,
have the following new form:
\begin{eqnarray}
F_i(z) = & (1+z)^{u_i}\times 
         ((1+(1+z)/(1+z_1))/(1+1/(1+z_1)))^{v_i-u_i} \times \nonumber \\ 
         & ((1+1/(1+z_2))/(1+(1+z)/(1+z_2))^{v_i+w_i} \;\; (z\leq 7);
\end{eqnarray}
and
\begin{eqnarray}
G_i(z) = & (1+z)^{p_i}\times 
          ((1+(1+z)/(1+z_1))/(1+1/(1+z_1)))^{q_i-p_i} \times \nonumber \\ 
         & ((1+1/(1+z_2))/(1+(1+z)/(1+z_2))^{q_i+r_i} \;\; (z\leq 7).
\end{eqnarray}
These smoothly-joined 3-piece power laws, in contrast with the
sharply-joined 2-piece power laws used in Xu et al. (2001),
will allow softening in the low redshift end ($z<0.5$) of the evolution 
functions to improve the fit to the redshift distribution
of IRAS 60$\mu m$ sources. 
\item{(2)} The 25$\mu m$ local luminosity functions (LLF) of 3 populations
used in Xu et al. (2001) did not themselves 
take into account of the evolutionary
effects. In the new models, new 25$\mu m$ LLFs corrected for
these effects are used (see Appendix A for the details of the
new LLFs). It is found that these 
new LLFs are only marginally different from the old ones.
\item{(3)} The UV portion ($\lambda < 3000${\AA}) of the SEDs
in the SED lib, which is an important part of the code,
is better constrained in this work (Appendix B). 
\end{description}

Given the large number of parameters, it is out the scope of this
paper to explore the entire parameter space for models allowed by
available data. Instead, we concentrate on three models which
predict very different relative contributions 
by starburst galaxies and normal late-type galaxies
to the counts in the IR surveys. For each of them, parameters
have been fine-tuned so that the model can fit
all available data as well as possible.
These models for dusty galaxies are presented in
Table 1 and Fig.1. 

Model S1 is similar to the 
models in Xu et al. (2001) and in Franceschini et al. (2001) in
the sense that the starburst galaxies are assumed to dominate
the evolution of the IR sources. In Model S2,
similar to Rowan-Robinson (2001), 
all three populations of dusty galaxies
are assumed to have pure luminosity evolution with
the same evolution rate. This is actually the 
same evolutionary scenario adopted by Xu (2000),
who assumed that the entire body
of IR sources evolve as a single population. 
Note Rowan-Robinson (2001) assumes that
there is no SED evolution, while in Model S2
we assume SEDs evolving with luminosity. 
In Model S3, it is assumed that the normal galaxies
and starburst galaxies give about equal contributions to
the ISOCAM 15$\mu m$ counts, as hinted at by the optical identifications
of ISO sources (Flores et al. 1999b; Aussel et al. 1999).
For Model S1 and Model S3, the luminosity evolution function of galaxies with
AGNs is adopted from the optical QSO luminosity evolution function
of Boyle et al. (2000)
\begin{equation}
F_{AGN} (z) = 10^{1.36z-0.27z^2} \;\; (z\leq 7).
\end{equation}
This is slightly different from the power law function used
in Xu et al. (2001).


\noindent{\bf Table1. New models for dusty galaxies}\\
\hskip-1.5truecm\begin{tabular}{lcccccccccccccccc}
\hline
    &     &     & \multicolumn{3}{c}{Normals}  & & \multicolumn{6}{c}{Starbursts} & & \multicolumn{3}{c}{AGNs} \\
\cline{4-6}\cline{8-13}\cline{15-17}
model & $z_1$ & $z_2$ & $u_1$ & $v_1$ & $w_1$ & & 
$u_2$ & $v_2$ &$w_2$ & $p_2$ & $q_2$ &$r_2$& & $u_3$ & $v_3$ & $w_3$ \\
\hline
&&&&&&&&&&&&&&&\\
S1 & 0.5 & 0.85 & 1.5 & 1.5 & 2 & & 2 & 11.5 & 1.2 & 1 & 9.8 & 1.2 & &\multicolumn{3}{c}{$1.36z-0.27z^2$} \\
&&&&&&&&&&&&&&&\\
S2  & 0.5 & 1.2 & 3.2 & 5.7 & 2.5 & & 3.2 & 5.7 & 2.5 & 0 & 0 & 0 & &3.2 & 5.7 & 2.5 \\
&&&&&&&&&&&&&&&\\
S3  & 0.5 & 1 & 2 & 8 & 2 & & 2 & 10.2 & 1.5 & 2 & 3 & 1.5 & &\multicolumn{3}{c}{$1.36z-0.27z^2$} \\
&&&&&&&&&&&&&&&\\
\hline
\end{tabular}
\oneskip

\vskip1truecm

\subsection{Predictions for Counts and Redshift Distributions}
In Fig.2, Fig.3 and Fig.4, predictions by these
3 models for the 15$\mu m$, 60$\mu m$, 90$\mu m$,
170$\mu m$, 850$\mu m$ counts and for the CIB 
are compared with the data
\footnote{The units used in the plots throughout
this paper are rather heterogeneous. This is becuase
many data points collected from the literature
are measured from figures that used different
units. Often it is difficult to convert them to the
same units, in particular for errors bars.}, respectively.
The data points are taken from a large pool of measurements
found in the literature, which is still evolving rapidly.
This is particularly true for the measurements involving
ISO data, because more and more new data reduction tools 
are becoming available for these rather complex data.

Model S1, which is otherwise same as the 'peak model'
in Xu et al. (2001) except for the modifications
listed in Section 2.2, predicts that
the contribution from the starburst population dominates
almost everywhere in these plots. 
The model predictions are in reasonably good agreements
with data in all plots, given the large dispersions
in the data sets. Particularly,
the new results of Gruppioni et al.
(2002) on the 15$\mu m$ counts of ELAIS
sources (filled 4-point stars in plot)
are about a factor of 3 lower than the counts from the same 
survey reported by
Serjeant et al. (2000) (open 4-point stars),
and are also significantly lower than other ISOCAM
measurements in the flux range of 1~mJy $< f_{15\mu m} <$ 10~mJy.
Model S1 predicts a less prominent sub-mJy peak 
in the 15$\mu m$ counts
than the 'peak model' of Xu et al. (2001),
because the sharp peak at z=1.5 in the evolution functions of
the old model is replaced
by a smooth peak in the new model (at $z\simeq 1.3$, see Fig.1).
It is worth to note that, in the bright end of the SCUBA counts
($f_{850\mu m} \gsim 5 mJy$), the model predicts that
the contribution from galaxies with AGN exceeds that from starbursts.
This result seems to be consistent with the limited knowledge 
we have about the bright SCUBA sources: out of the 
7 SCUBA sources of $f_{850\mu m} \ge 8 mJy$ 
in the sample of Smail et al. (2002), 4 show signs of AGN. 
On the other hand,
the model predictions on the SCUBA 850$\mu m$ counts 
are close to the lower boundary of the data.

In contrast to Model S1, Model S2 predicts that for 15$\mu m$ counts,
850$\mu m$ counts and for the CIB, the population of normal galaxies
dominate (Fig.3). This is in agreement
with the results of Rowan-Robinson (2001). 
The same model predicts that normal galaxies and
starburst galaxies give about equal contributions to
the 170$\mu m$ counts, and starbursts dominate
the 90$\mu m$ and 60$\mu m$ counts. This model can also
fit the data in all these plots well (Fig.3).

By design, the contributions from
normal galaxies and from starburst galaxies
are indeed nearly equal in the predictions of
Model S3 for 15$\mu m$ counts (Fig.4). 
The same model predicts that the 60$\mu m$,
90$\mu m$ and 170$\mu m$ counts are dominated by 
starbursts. For the SCUBA 850$\mu m$ band, the model
predicts comparable contributions from
the normal galaxies and from the starbursts to
the counts in a wide flux range, and the
contribution from AGNs dominant the counts
at a few 10s of mJy level while being negligible
at sub-mJy level. The CIB predicted
by this model is dominated by the contribution from
the starbursts in the
wavelength range $20\mu m \lsim \lambda \lsim 300\mu m$,
and by that from normal galaxies at other wavelengths.

In Fig.5, Fig.6 and Fig.7, 
predictions by
the 3 models for the redshift distributions
for the IRAS 60$\mu m$, ISOCAM 15$\mu m$,
ELAIS 90$\mu m$, FIRBACK 170$\mu m$, and SCUBA
850$\mu m$, and future SWIRE 24$\mu m$ surveys 
are plotted. The model predictions
for the IRAS 60$\mu m$ and  ISOCAM 15$\mu m$ surveys
are compared to the data. All 3 models give
more satisfactory fits to redshift distribution of
the IRAS 60$\mu m$ sources than the `peak model' (favorite model)
of Xu et al. (2001) does, though
Model S1 still over-predicts the number of 
IRAS 60$\mu m$ sources with $z>0.2$ by about a factor of
3. For the redshift distribution of 
ISOCAM 15$\mu m$ sources in the HDF-north, the predictions
of the Model S1 gives the best fit among the 3 models
although, given the large error bars of the data, the difference among
model predictions is subtle.

All the 3 models predict
bi-modal redshift distributions for ELAIS 90$\mu m$ 
and FIRBACK 170$\mu m$ sources, with normal galaxies
dominating the first peak and starburst galaxies 
dominating the second peak. Model S1 predicts that
most of ELAIS 90$\mu m$ 
and FIRBACK 170$\mu m$ sources have $z\geq 0.5$.
 In contrast,
Model S2 predicts that
most of these sources have $z\leq 0.5$, being
in the first peak of the redshift distributions.
This difference is due to the so-called
'temperature-redshift' degeneracy for infrared
galaxies (Blain 1999): In the 
long wavelength bands such as the 90$\mu m$ and the 170$\mu m$,
the sources in Model S1 tend to have larger redshifts
than those in Model S2 in order to compensate their
warmer IR SEDs. Interestingly, this indicates that
we can break the degeneracy over these models
by comparing the predicted and observed redshift
distributions of the ISOPHOT sources in these bands.
Serjeant et al. (2001) obtained
high confidence redshifts for 16 (out of 37)
sources of $f_{90\mu m} \geq 0.1 Jy$
in the ELAIS S1 field (3.96 deg$^2$), none of them
has redshift larger than 0.5. This sets a lower limit
of 43$\pm 11$\% for the percentage of
the 90$\mu m$ sources in this field to have $z < 0.5$.
Taken at the face value, only the prediction of S2
(53\%) is comfortably above this lower limit,
while the prediction of S3 (37\%) is marginally
consistent with it, and the prediction of S1 
(30\%) is marginally below the limit. 
The above comparisons seem to favor Model S2.
However, given the fact that this data set was taken
from a small region and can be seriously affected
by any clustering effect, more observations are
needed for any definitive conclusions.

Model S1 predicts a broad redshift distribution,
peaking between $2<z<3$, for bright SCUBA sources
($f_{850\mu m} > 8$ mJy), broadly consistent
with observational constraints (Smail et al. 2002;
Ivison et al. 2002; Fox et al. 2002). For the same sources,
Model S2 predicts a prominent peak around z=1.5
and less than 25\% of sources at redshifts $>2$,
not favored by observations which in
general suggest a mean redshift larger than 2 
(see, e.g. Ivison et al.
2002). Model S3 also predicts a peak around z=1.5 for
the $f_{850\mu m} > 8$ mJy sources, though with
a much wider high-redshift wing (44\% of sources
have $z \geq 2$). 

\subsection{Predictions for ISO LFs and SFH}
The ISOCAM $15\mu m$ sources in HDF-North have been thoroughly
studied in the literature (Rowan-Robinson et al. 1997;
Aussel et al. 1999; Elbaz et al. 1999; Cohen et al. 2000;
Franceschini et al. 2001; Elbaz et al. 2002). All but 1 of a total 
of 41
sources with $f_{15\mu m} \geq 0.1$ mJy have spectroscopic redshifts
(Franceschini et al. 2001;
Elbaz et al. 2002). Exploiting these redshifts, we derived the 
15$\mu m$ luminosity functions (LFs) in three redshift intervals:
$0.4 < z \leq 0.7$, $0.7 < z \leq 1$, and 
$1 < z \leq 1.3$ using the classical V$_{max}$ method.
These results are compared to the predictions of the three
models in Fig.8. 

Given the small size of the data set (40 galaxies),
it is understandable that the derived luminosity functions
have substantial uncertainties. Particularly,
significant
fluctuations can be found in the redshift distributions
(Fig.5--7), presumably due to clustering effects. 
Compared to these uncertainties,
the difference in the predictions by the three different
models is again subtle. They all lie near the lower ends 
of the error-bars (Poisson error) of the two data points
in the panel of $0.4 < z < 0.7$. This is likely to be
due to a bias in the data caused by
the over-density in the $0.4 < z <0.6$ bin (Fig.5--7).  
In the panels of $0.7 < z < 1$ and $1 < z < 1.3$, the 
predictions of Model S1 are slightly higher and fit
the data slightly better than those of other two models.

In Fig.9a, the star formation history (SFH) predicted by
the three models are compared to observations. 
The red lower limits are derived from the ISOCAM
15$\mu m$ LFs (Fig.8). The lower limit of 15$\mu m$ 
luminosity density is derived by summing up the 
contributions in the bins where the LF is actually measured 
(i.e. no extrapolations are applied
to {\it measured} LFs), then converted to total
IR luminosity using the formula $L_{ir} = 11.1\times L_{15\mu m}$
(Elbaz et al. 2002). The star formation rate is then estimated
from the IR luminosity density using the formula 
of Kennicutt (1998):
SFR ($M_\sun$ yr$^{-1}$)=$L_{ir}\times 1.7\; 10^{-10}$ L$_\sun$.
The predictions of the three models are very close to each other
(within 50\%).

In Fig.9b, Fig.9c and Fig.9d, these predictions 
are broken into contributions by different populations.
Indeed Model S1 predicts that, except for
in the low z end ($z\lsim 0.2$), the starburst
galaxies overwhelmingly dominate the star formation
in the universe. Model S2 predicts just the opposite:
the star formation in the universe has been always dominated by
normal late-type galaxies, while in Model S3 the contributions
of these two populations are more comparable. All three models
predict minor contributions from galaxies with AGNs. Note that
since some of the IR emission from galaxies with AGNs is powered
by the gravitational energy released in the AGN, not by star
formation, the model predictions plotted here should be
treated as upper limits.

In the literature, the steep decline of the star formation rate
since $z \sim 1$ has been well established since early works
of Lilly et al. (1996) and Madau et al. (1996). However,
it is still controversial on what happened to the star formation
rate in earlier universe, particularly before $z = 2$.
Most information on this issue is obtained from
observations of Lyman-Break galaxies (LBGs, Steidel et al. 1999).
However, the extinction correction for these rest-frame
UV selected galaxies is very uncertain. In addition, many high
redshift star-forming galaxies may be completely missed
in the surveys of LBGs because of heavy extinction, 
causing systematic underestimation when using LBGs to
determine the star formation rate in high z universe
(see, e.g., Rowan-Robinson et al. 1997).
In this respect, the predictions of our models 
follow closely the data points determined from LBG surveys
after the extinction correction (Steidel et al. 1999). 
As discussed in Xu et al. (2001),
the evolution of IR galaxies at redshifts $>2$ is mainly constrained
by the sub-mm data in the SCUBA surveys and in the 
CIB observations, due to the negative
K-correction. Too much star formation in those large
redshifts will result in too many sub-mm counts and 
too high sub-mm background radiation.
In the literature, other models (e.g. Gispert et al.
2000; Rowan-Robinson 2000;
Chary \& Elbaz 2001) which also invoked 
sub-mm data to constrain the evolution of $z > 2$
IR galaxies found the similar trend 
(a peak at $z\sim 1$ and a shallow/flat slope in
$z\gsim 2$) in the SFR evolution. The claim that
the universal star formation rate increases monotonically
with redshift to very high z
($\gsim 7$) by Lanzetta et al. (2002) is
not supported by our and other studies on evolution of
IR galaxies.

\section{Passive Evolution Models for E/S0 Galaxies}
The models follow a simple, passive evolution approach
(Pozzetti et al. 1996).
The basic assumption is that there has been no
star formation in an E/S0 galaxy since its initial formation.
Consequently, its radiation in different bands (i.e.
the SED and the L/M ratio) evolves
passively with the ever-aging stellar population.
Instead of assuming that all E/S0's formed at once together
(as in the classical monolithic galaxy formation scenario),
the E/S0 galaxies are assumed to form in a broad
redshift range (Franceschini et al. 1998), specified
by a truncated Gaussian function. The SEDs of different ages are
calculated using the GRASIL code of Silva et al. (1998).
No dust emission is considered in these models.

\subsection{Evolution Function}
The evolution function $\Psi(M,t)$,
specifying how many E/S0s are formed in a unit volume, in a unit
time interval, and in a unit mass interval,
is assumed to have the following form:
\begin{equation}
\Psi (M,t) =  \psi (M)\; T(t)
\end{equation}
where the time dependence function $T(t)$ is a truncated Gaussian:
\begin{eqnarray}
T(t) & = & exp(-(t-t(z_{peak}))^2/\omega^2) \;\;\;\; \;\;
(t \geq t(z_0)) \nonumber \\
     & = & 0 \;\;\;\;\;\;\; (t < t(z_0)).
\end{eqnarray}

In this prescription, for the sake of simplicity,
the time dependence is
assumed to be independent of the mass. This implies that 
E/S0s of different mass have the same formation history
(i.e. no differential evolution).
The time dependence function $T(t)$ is
fully defined by 3 free parameters: the peak formation
time $t(z_{peak})$, the time scale $\omega$ (in Gyr), and the 
time when the formation of E/S0s started $t(z_0)$.

Accordingly, for a given redshift z, the number
of E/S0's in a unit mass interval and a unit volume
can be found from the following integration:
\begin{equation}
\Phi(M,z) =  \psi (M)\; \int_{t(z_0)}^{t(z)}T(t')\; dt'.
\end{equation}

If the local mass function $\Phi(M,z=0)$ is known, then
the mass dependence function $\psi (M)$ can be derived:
\begin{equation}
\psi (M) = {\Phi(M,z=0)\over \int_{t(z_0)}^{t(z=0)} T(t')\; dt'}.
\end{equation}
where $t(z=0)$ is the current age of the universe.

In this work, the local mass function is constrained
using the local K-band luminosity function of Kochanek
et al. (2001) and the $L_K/M$ ratio predicted by
GRASIL code (Silva et al. 1999).

\subsection{Galaxy Age Distribution and SED Assignment}
Eq(6) predicts galaxy number for a given redshift in
a unit mass interval and a unit volume. These galaxies 
have different ages spanning between $\tau=0$ and
$\tau=t(z) - t(z_0)$. The age distribution function,
$G(z,\tau)$, is also a truncated Gaussian:
\begin{eqnarray}
G(z,\tau) & = & G_0\; exp(-(t(z)-\tau-t(z_{peak}))^2/\omega^2)
 \;\;\;\;\;\;\;\;  0\leq \tau \leq (t(z)-t(z_0)) \nonumber \\
     & = & 0 \;\;\;\;\;\;\; (otherwise).
\end{eqnarray}

For every galaxy simulated,
an age is assigned to it according to the above
distribution (i.e. probability) function. According to the passive
evolution model, galaxies of different ages have different
SEDs and different L/M ratios. 
The SEDs of different ages are calculated using GRASIL (Silva et al. 
1998). Again, for the sake of simplicity, we assume a simple
`single-burst' scenario, in analogy to the merger
scenario of E/S0 formation as hinted at in the studies of
ULIRGs (Kormendy \& Sanders 1992), 
to model the formation of E/S0s. Namely, we assume that
all the stars in an E/S0 galaxy were formed in a short burst (lasting
$10^8$ yrs), after which all the ISM was blown out. Accordingly,
the following parameters are adopted for
the input of GRASIL: $t_{win}=0.1$ (Gyr), $k_{sch}=1.0$,
$\nu_{sch}= 20.0$, and $\tau_{inf}=0.01$.
Then, in the simulation of E/S0s, we include only sources older
than 1 Gyrs. Here we implicitly assume that merger remnants
younger than 1 Gyr do not look like E/S0s because these systems
may have not fully relaxed. Namely they
should be classified as post-starbursts or E+As,
not E/S0s. Optical follow-up observations of 
ISOCAM sources (Flores et al. 1999b; Aussel et al. 1999;
Cohen et al. 2000; Elbaz et al. 2002) have shown that E+A
galaxies are IR bright and have similar
IR properties as active starburst galaxies.

\subsection{Three Models for E/S0 Evolution} 
We consider 3 different E/S0 models in this paper, 
specified by different
$z_{peak}$ and $\omega$ (Table 2, Fig.10). 
Model E1
($z_{peak}=5$, $\omega=0.5$ Gyr) mimics the classical
monolithic scenario (Eggen et al. 1962). Model E2 assumes a later
and broader E/S0 formation epoch 
($z_{peak}=2$, $\omega=2$ Gyr). Model E3 
($z_{peak}=1$, $\omega=3$ Gyr) assumes that
most of E/S0s are formed at $z>1$, as hierarchical
galaxy formation workers have advocated (Kauffmann et al. 1993;
Kauffmann \& Charlot 1998).
\vskip1truecm

\noindent{\bf Table 2. Evolution models for E/S0s}
\nopagebreak

\hskip-0.5truecm\begin{tabular}{lccc}\hline
Models & $z_{peak}$ & $\omega$ (Gyr) & $z_0$ \\ 
\hline
&&& \\
E1 & 5 & 0.5 & 7 \\
&&& \\
E2 & 2 & 2 & 7 \\
&&& \\
E3 & 1 & 3 & 7 \\
&&& \\
\hline
\end{tabular}

\vskip1truecm

In Fig.11 we compare the model predictions
on the redshift dependence of optical/NIR colors with the
data from Franceschini et al. (1998). Predictions
of all three models can fit the overall trend of the
data well. On the other hand, Model E1 predicts
very little scattering, which is not confirmed by
the data. The other two models predict significantly 
larger dispersion, in better agreement with the data. 

In Fig.12 the redshift distribution of early type
galaxies of $K_{ab} \leq 20.15$, taken from Franceschini
et al. (1998), is compared to the model predictions.
Note that only 15 out of 35 redshifts in Franceschini et al. (1998)
are spectroscopically measured, the rest are photometric
redshifts. Among the three models, Model E2 
fits the data best.
It should be noted that, as discussed in Benitez et al. (1999) and
Rodighiero et al. (2001), some high z E/S0s
may have been missed in the Franceschini et al. (1998)
investigation.

\section{Counts in UV, Optical, NIR and MIR Bands}
Both the E/S0 galaxies and the dusty galaxies contribute
significantly in these bands. Since
the three E/S0 models predict almost identical counts
(difference $\lsim 0.1$ dex), in the rest of the paper we shall present
only the results from Model E2 ($z_{peak}=2$, $\omega=2$ Gyr).

In Fig.13, Fig.14 and Fig.15 the contributions
of E/S0 galaxies to the counts in these bands (as predicted by
Model E2) are added to the contributions
of dusty galaxies as predicted by
Model S1, Model S2 and Model S3, respectively.
These model predictions are compared to
observations in the vacuum UV (2000{\AA}),
the B and R, the NIR K, and the
MIR 6.7$\mu m$ band (ISOCAM) and
12$\mu m$ (IRAS and ISO CAM) bands. 
Between the three figures,
the differences in the total counts in the
bands plotted are generally within
the uncertainties of data.
The model predictions can
account for all counts in the R, K and MIR bands.
On the other hand, the model predictions are 
significantly lower than the counts observed in
the 2000{\AA} band (Milliard et al. 1994).
For the B band, the model predictions are slightly
lower ($\sim 0.3$ dex) than the observed counts
for $B>20$ mags. These results strongly
hint at a population of star forming galaxies which
are infrared quiet (very low dust 
attenuation/emission), therefore seen only in 
the UV and the blue bands (see Section 7.3 for
more discussion). 

In the bright end of the K-band counts,
the prediction by S1+E2
(Fig.13) is in good agreement with the morphologically
segregated counts reported by Huang et al. (1998),
namely at K$\lsim 17$ the E/S0 galaxies contribute
about 50\% of the counts, and at fainter magnitudes
the contribution from late-types becomes more
and more dominant. The predictions by S2+E2
and S3+E2 on the E/S0 contribution to K-band
counts at K$=16$ are $\sim 30\%$ and $\sim 40\%$,
slightly less than the observational result of 
Huang et al. (1998).

The three models fit the faint, sub-mJy
ISOCAM 6.7$\mu m$ counts very well.
However, predictions of S1+E2 are about
a factor of 3 lower than the ELAIS
counts (Serjeant et al. 2000) at few mJy level. 
For these counts, predictions by S3+E3
have the best agreement with, 
though being still about 50\% less than,
the data. It should be noted that,
as indicated by the large discrepancy 
between the ELAIS 15$\mu m$ counts of Serjeant at al. (2000)
and of Gruppioni et al. (2002), the uncertainties
of the ELAIS 6.7$\mu m$ counts of Serjeant et al.
(2000) may be signicantly larger than reported.

The scatters of the ISOCAM 12$\mu m$ data
(Clements et al. 1999) are large, indicating
significantly larger uncertainties than
the Poisson noise (error bars in the plot).
The three brightest ISO points may suffer serious
biases caused by errors in galaxy/star separation
(Clements et al. 1999), so are likely to be
less reliable than other data points.
If this is indeed the case then, among the
three models, the predictions of
S1+E2 have the best agreement with the data.
The effect of the Local Supercluster (Lonsdale et al.
1990) can be seen in the bright IRAS 12$\mu m$ 
counts. 

\section{Counts and Confusion Limits in SIRTF Bands}
 
The E/S0 galaxies contribute significantly only
to the four IRAC bands (3.6$\mu m$, 
4.5$\mu m$, 5.8$\mu m$, and 8$\mu m$), while
the 3 MIPS bands (24$\mu m$, 70$\mu m$ and
160$\mu m$) will see only dusty galaxies.

Model predictions of contributions by E/S0 galaxies
and by different populations of dusty galaxies
to the counts in three SIRTF bands 
($3.6\mu m$, $24\mu m$ and $70\mu m$ bands)
are plotted in Fig.16. The total counts
by the three different models are very
close to each other, though the relative
contributions from normal galaxies and from
starbursts are very different. 
This can be understood by the fact
that, tuned to fit the same counts in
the ISO bands which covers a wide wavelength
range from 6.7$\mu m$ to 170$\mu m$, 
these models are forced to be similar with each other.

The 3-$\sigma$ confusion limits are given in Table 3
for all 7 SIRTF bands.
These are calculated by the method described in Xu et al. (2001),
and assuming that SIRTF (85 cm dish)
is diffraction limited
in all bands (so the beams can be approximated by
the Airy function). Since this idealized assumption
may not be true, particularly for the short wavelength
IRAC bands (e.g. the 3.6$\mu m$ and 4.5$\mu m$ bands), 
the results for those bands should be
treated as lower limits. The confusion limits predicted
by the 3 different models differ by up to 80\%.
The largest difference occurs for the 70$\mu m$ band.
These differences reflect the real uncertainties,
mostly due to the uncertainties in the ISO data
which are the major constraints to the models.  

\vskip1truecm


\noindent{\bf Table 3. Confusion limits (3$\sigma$) of SIRTF bands} 
\nopagebreak

\hskip-0.5truecm\begin{tabular}{lccccccc}\hline
Models & 3.6$\mu m$ & 4.5$\mu m$ & 5.8$\mu m$ & 8$\mu m$ & 24$\mu m$ & 70$\mu m$ & 160$\mu m$ \\ 
\hline
S1+E2 & 0.12 $\mu$Jy & 0.21 $\mu$Jy & 0.43 $\mu$Jy & 1.02 $\mu$Jy & 
65 $\mu$Jy & 6.15 mJy & 81.5 mJy \\  
S2+E2 & 0.18 $\mu$Jy & 0.35 $\mu$Jy & 0.65 $\mu$Jy & 1.32 $\mu$Jy & 
57 $\mu$Jy & 3.35 mJy & 57.5 mJy \\  
S3+E2 & 0.14 $\mu$Jy & 0.27 $\mu$Jy & 0.53 $\mu$Jy & 1.17 $\mu$Jy & 
57 $\mu$Jy & 4.45 mJy & 68.5 mJy \\  
\hline
\end{tabular}
\vskip1truecm

\section{Colors as Model Discriminators for Dusty Galaxies}

In empirical evolution models, different populations of
dusty galaxies are distinguished by different SEDs. Therefore,
the most direct method to distinguish models 
predicting different dominant populations for IR sources
is to compare the predicted color distributions 
with the observations.

In Fig.17, we compare predictions of the three models
for evolution of dusty galaxies (Table 1) for colors of ISO
galaxies. The detection limits are set to be $f_{6.7\mu m} \geq 30\; 
\mu Jy$, $f_{15\mu m} \geq 100\; \mu Jy$, $f_{90\mu m} \geq 100\; mJy$,
$f_{170\mu m} \geq 180\; mJy$, $R \leq 24\; mag$, and $K \leq 20\; mag$.
The predictions for $f_{15\mu m}/f_R$ and $f_{15\mu m}/f_K$ 
are compared to the data of ISOCAM $15 \mu m$ sources in HDF-N
(Aussel et al. 1999; Cohen et al. 2000; Hogg et al. 2000).
It appears that, for ISO galaxies, only the $f_{15\mu m}/f_K$ color
is a good model discriminator in the sense that the peaks of
the color distributions predicted by the three models
are separated from each other. The data seem to favor  
Model S1 which predicts a peak in the $f_{15\mu m}/f_K$
color distribution right at the place where the
data peaks. However, this data set is again too small
(30 galaxies) and error bars too large to
distinguish the three models.

SWIRE\footnote{http://www.ipac.caltech.edu/SWIRE.}, a SIRTF
Legacy Science program, 
will survey 65 deg$^2$ of sky in all 7 SIRTF bands
(3.6, 4.5, 5.8, 8, 24, 70, and 160$\mu m$) 
down to a depth comparable to, 
or even deeper than, that of ISOCAM 15$\mu m$ surveys.
The documented 5$\sigma$ sensitivity limits of SWIRE are:
$f_{3.6\mu m} = 7.3\; \mu Jy$,
$f_{4.5\mu m} = 8.7\; \mu Jy$,
$f_{5.8\mu m} = 27.5\; \mu Jy$,
$f_{8\mu m} = 32.5\; \mu Jy$, 
$f_{24\mu m} = 0.45\; mJy$, and 
$f_{70\mu m} = 2.75\; mJy$,
$f_{160\mu m} = 17.5\; mJy$.
Compared to predicted confusion limits in Table 3, it appears
that SWIRE surveys will be confusion limited in the
the 70$\mu m$ and 160$\mu m$ bands.
Extensive ground based follow-up observations in the optical,
NIR and radio bands will be carried out.
These survey areas will also be observed in the far-UV
(1500{\AA} and 2300{\AA}) by 
GALEX\footnote{http://www.srl.caltech.edu/galex.} in the deep
survey mode. In Fig.18, model predictions for
the distributions of 6 colors of SWIRE galaxies
are plotted. These colors are selected among many
possible combinations to illustrate how the colors
of SWIRE galaxies can discriminate the models.
Model simulations of sky coverage of 5 deg$^2$ are carried out.
For each plot, the samples of simulated sources are 
selected according to SWIRE's sensitivity limits or,
for the 70$\mu m$ and 160$\mu m$ bands,
the confusion limits. Also it is required that
$R \leq 24$ mag and $ K\leq 20$ mag
when the R and K band data are invovled.
In four ($f_{70\mu m}/f_{8\mu m}$,
$f_{24\mu m}/f_{3.6\mu m}$, $f_{24\mu m}/f_{8\mu m}$,
and $f_{24\mu m}/f_{K}$) of the six colors plotted,
the peaks of Model S1 (starburst dominant) and Model S2
(normals dominant) are clearly separate.
Among these, the peak of the distribution predicted
by Model 3 (intermediate model) is close to that of
S1 in the $f_{70\mu m}/f_{8\mu m}$ color plot,
close to that of
S2 in the $f_{24\mu m}/f_{8\mu m}$ color plot,
and more ambiguous in the other two plots.
Given the large sky coverage of SWIRE, which
is 13 times of what is simulated here, it is very
hopeful that these color distributions will indeed
provide clues to the question of which population is 
dominant among IR sources.

\section{Discussion}

\subsection{Evolution of SEDs of Starburst Galaxies}
As pointed out in Xu et al. (2001), the most
important assumption in our models for the
evolution of dusty galaxies
is that high redshift (i.e. $z \gsim 1$) star-forming galaxies have the
same SEDs as their local counterparts when the luminosity
is the same. Since our SEDs cover from
the UV all the way through to the radio waveband,
the validity of this assumption demands similarity between 
high redshift and local galaxies in many physical conditions,
a requirement that seems too strong to be fulfilled in the strict sense.
Particularly, the long term star formation
history plays an important role in the optical
and NIR emission of galaxies, and it is obvious that
high redshift galaxies have very different long term star formation
history (i.e. much younger) than that of local galaxies.
Does this mean that our predictions for
the optical/NIR flux densities of high redshift
dusty galaxies are flawed and therefore unreliable?
Our argument to dispute this suspicion is based on
the fact that high redshift
galaxies (particularly IR selected galaxies),
already detected or to be detected in the future surveys,   
are almost exclusively high luminousity galaxies
(less luminous galaxies with high redshift are
too faint to be detected). As their local counterparts,
these high redshift ULIRGs must host
very powerful starbursts or AGNs as energy sources which
contribute much of the emission
even in the NIR bands ($\sim 40\%$ for the local ULIRGs,
Surace et al. 2000; Scoville et al. 2000).
Therefore, as far as the detectable IR sources
are concerned, the difference in
the underlying old population between the high redshift
dusty starforming galaxies and their local counterparts 
does not affect very seriously our model predictions.

Recent multi-wavebands observations of SCUBA galaxies
(Smail et al. 2002), Lyman-Break Galaxies 
(Adelberger \& Steidel 2000), and ISOCAM
galaxies (Flores et al. 1999a; Aussel 
et al. 1999; Cohen et al. 2000; Elbaz et al. 2002)
are consistent with our assumption that
high redshift star-forming galaxies have similar 
SEDs as their local counterparts, particularly
when binned according to the luminosity
(as stressed by Adelberger \& Steidel 2000).
On the other hand, some new observations
hint at possible systematic differences between 
high redshift and local ULIRGs.
The HST image of SCUBA
galaxy SMM J14011+0252 (Ivison et al. 2001)
shows that the star formation
activity in that source is widely
spread (up to a few kpc), a situation
remarkably different from typical ULIRGs
found in the local universe (Sanders \& Mirabel 1998).
Compared to centrally concentrated starbursts
found in most of the local ULIRGs, galaxies with such
widely distributed starbursts are expected to have 
less steep MIR slopes
(i.e. smaller $f_{25\mu m}/f_{12\mu m}$ ratio)
and cooler FIR color (i.e. larger $f_{100\mu m}/f_{60\mu m}$ ratios).
This is due to the less intense radiation field
(less warm dust emission in the 25$\mu m$ and 60$\mu m$ bands)
and less dust opacity (less extinction for MIR fluxes).
Indeed, Chapman et al. (2002) found that two 
sources detected both 
by the FIRBACK 170$\mu m$ band survey (Puget et al. 1999;
Dole et al. 2001) and by SCUBA (Scott et al. 2000), 
one at z=0.91 and the other at z=0.46,
have significantly cooler dust temperatures ($T_{dust}\sim 30$K) compared 
to $T_{dust}\sim 50$K found for typical ULIRGs such as Arp220.
They argue that this may indicate the starbursts
in these systems are also extended. Given the small amount
of information and the possible bias for cooler
galaxies due to the sub-mm selection (Chapman et al. 2002), 
it is still too early to tell whether
high-redshift ULIRGa are systematically more extended,
and therefore have SEDs closer to less luminous
local interacting galaxies (such as the Antennae galaxies)
which are in earlier stages along the merger sequence.
Future surveys like SWIRE will address these questions.

\subsection{E/S0 evolution and starburst evolution}
Comparisons between predictions E/S0 evolution models and  
observational data (Fig.11--12) indicate that
the intermediate model (Model E2), which assumes
a peak formation redshift of $z_{peak}=2$, is the favorite
among the three models. This is in agreement with
some previous works (Franceschini et al. 1998; 
Rodighiero et al. 2001; Im et al. 2002). This model
can reproduce well the trend in the color-z plots
and also fit the redshift distribution very well.
However, particular in the B-K v.s. z plot, the data
show significantly larger dispersion than the model predictions.
There are two possible causes for this: 
\begin{description}
\item{(1)} In our simple
models, we have assumed that all of the E/S0s are formed
through the same starburst procedure, which means their
stellar populations have the same metallicity. It is known
(e.g. Worthy 1994) that local E/S0s have different metallicity
which is a major cause of the different M/L ratios and 
colors among these galaxies. Therefore, by neglecting
these effects, our models leave some of the scatters
in the color distributions unaccounted for.
\item{(2)} Some of the blue E/S0s in the data (Franceschini et
al. 1998) may not be true E/S0s. According to Im et al. (2002),
many of these 'blue interlopers' have strong,
narrow emission lines, suggesting that they are low-mass 
starbursts rather than  massive star-forming E/S0s. 
\end{description}

In fact, there is still a lack of consensus in the definition of
E/S0s in deep surveys. Using a strict algorithm selecting
the most symmetric and smooth galaxies, Im et al. (2002)
found far less blue sources among their E/S0 sample than,
e.g., Schade et al. (1999) and Menanteau (1999) whose samples
were selected with less strict algorithms. A related uncertainty
in our models is the choice on the exclusion of galaxies younger
than 1 Gyr (Section 3.2). This choice is not entirely arbitrary:
pushing the cut-off toward younger ages means
more galaxies with high L/M ratios, which in turn
results in too high model predictions for optical and NIR counts.
This shows that the question of where to put the boundary 
between E/S0s and post-starbursts deserves more investigation.

If indeed E/S0s are formed through mergers (Toomre 1978; Kormendy \& 
Sanders 1992), then their evolution is linked to the evolution of
starburst galaxies which are closely related to mergers (Sanders
\& Mirabel 1998). Our results indeed indicate some synchronization
between the two populations, in the sense that the peak of the
formation function of E/S0s in the best fit model ($z_{peak}=2$)
is close to the peak of the evolution functions of starburst
galaxies ($z_{peak} = 1.4$). Future works exploring this possible
link will provide important constraints to the evolution of
both populations.

\subsection{IR-quiet star-forming galaxies}
Our results (Fig. 13 -- 15) show that only a small fraction
(10 -- 20 \%) of UV selected galaxies (as in the sample 
of FOCA survey, Milliard et al. 1992) are IR bright,
as indicated by the small contribution of simulated dusty galaxies
to the UV counts (the E/S0 galaxies contribute even less).
This suggests a separate population of
IR quiet, UV bright galaxies which dominate the UV selected
samples. Such a population is also
needed for the B-band counts, because our simulations 
under-predict 30 -- 50\% of observed counts in that band,
while fully account for the R and K band counts
(Fig. 13 -- 15). The best candidates for such galaxies are
the low-metallicity (therefore low dust content) blue dwarf  
galaxies such as I Zw 18 (Searle \& Sargent 1972). 
There is an apparent link between this
IR quiet star-forming galaxy population and the
'faint blue galaxies' found in deep optical surveys
(see Koo \& Kron 1992 and Ellis 1997 for reviews).
What is the relation between this population and the 
dusty (IR bright) star-forming galaxies?
How does this population evolve (backwardly) with the redshift,
and how does this evolution correlate with the evolution of
IR bright galaxies? Are Lyman Break Galaxies, being
selected by the UV flux in the rest frame, more
closely related to the IR quiet star forming galaxies,
or to the dusty star forming galaxies (as argued by 
Adelberger \& Steidel 2000)? Answers to these 
questions will help to unify the pictures of 
galaxy formation/evolution seen in different wavebands.
We plan to address these questions in our future work,
particularly in connections with the forthcoming
GALEX and SWIRE missions.

\section{Summary}
New models for the evolution of extragalactic
IR sources are presented in this paper. 
The models for
dusty galaxies and for E/S0 galaxies,
the latter contributing significantly to 
counts at wavelengths $\lambda < 10\mu m$, have been
developed separately. 

Compared to previously
published models in Xu et al. (2001), 
the new models for evolution of dusty galaxies 
in this work have the following improvements:
\begin{description}
\item{(1)} The evolution functions have the form
of smoothly-joined 3-piece power-law (Fig.1), instead
of the sharply-joined 2-piece power-law.
\item{(2)} New local luminosity functions at 25$\mu m$,
which take into account the evolution effects, are used
for the three dust galaxy populations (i.e. normal late-type 
galaxies, starburst galaxies, and galaxies with AGNs).
\item{(3)} The UV portion ($\lambda < 3000${\AA}) of the SEDs
in the SED library is constrained by an empirical correlation
between $f_{UV}-f_{B}$ and $B-K$, instead of mere an
extrapolation of the optical SED ($\lambda > 4000${\AA}).
\end{description}
In order to address the question whether normal late-type galaxies
or starburst galaxies dominate among IR sources of
$z>0.5$, three new models are developed: 
\begin{itemize}
\item Model S1 --- starburst galaxies dominant ;
\item Model S2 --- normal galaxies dominant;
\item Model S3 --- intermediate between S1 and S2.
\end{itemize}
Predictions of these three models for counts in various
bands are fairly close to each other, therefore they
can hardly be distinguished using the counts. They can
also fit very well the luminosity functions of ISO $15\mu m$ sources in
three redshift intervals ($0.4< z \leq 0.7$, 
$0.7< z \leq 1$, $1< z \leq 1.3$). In principle, they
can be distinguished by redshift distributions of
FIR sources, but 
the large beams of FIR detectors (such as the
ISOPHOT cameras) make it rather difficult to pin-down the
optical counterparts (usually very faint) of faint FIR sources.
At the same time, the peaks in the distributions of 
several IR and optical colors predicted by
these models have separated locations.
We argue that these color distributions are the best tools
to distinguish these models. There are only very limited
amount of multi-waveband data available for high redshift
dusty galaxies in the literature
(mostly for ISOCAM 15$\mu m$ sources), a situation that will
be drastically improved when SIRTF is launched and 
SWIRE surveys are available.

The models for E/S0s follow a simple, passive evolution approach.
The basic assumption is that there has been no
star formation in an E/S0 galaxy since its initial formation.
Instead of assuming that all E/S0's formed at once together
(as in the classical monolithic galaxy formation scenario),
the E/S0 galaxies are assumed to form in a broad
redshift range (Franceschini et al. 1998), specified
by a truncated Gaussian function. The SEDs of different ages are
calculated using the GRASIL code of Silva et al. (1998).
No dust emission is considered in these models.
Three such models with different E/S0 formation histories
are calculated:
Model E1 ($z_{peak}=5$, $\omega=0.5$ Gyr) is close to the classical
monolithic scenario. Model E2 assumes a later
and broader E/S0 formation epoch 
($z_{peak}=2$, $\omega=2$ Gyr). Model E3 
($z_{peak}=1$, $\omega=3$ Gyr) assumes that
most of E/S0s are formed at $z>1$, as the hierarchical
galaxy formation works have advocated (Kauffmann et al. 1993;
Kauffmann \& Charlot 1998). Comparisons with limited data
(e.g. colors and redshift distribution)
available for morphologically identified E/S0 galaxies
at z$\sim 1$ indicate that model E2 can fit the data
best among the three models. 
This suggests a synchronization
between the evolution of E/S0 galaxies and of starburst galaxies, 
in the sense that the peak of the
formation function of E/S0s ($z_{peak}=2$)
is close to the peak of the evolution functions of starburst
galaxies ($z_{peak} = 1.4$). Combining model
predictions by E2 with those by S1, S2 and S3 
(dusty galaxy evolution models), comparisons with
number counts in different wavebands (Fig.13--15) indicate that
E/S0s contribute upto 30 -- 50\% of the optical/NIR
counts in the bright end, and about 20 -- 30\% of the
ISOCAM 6.7$\mu m$ band counts. Their contributions
to counts in the UV (2000{\AA}) and in the 
longer wavelength IR ($\geq 12\mu m$) bands are negligible.

Using these new models for extragalactic
IR sources, particularly including E/S0 galaxies, 
we made new predictions for the
counts and confusion limits in the SIRTF bands.
The results indicate that SWIRE surveys will
be confusion limited in the 70$\mu m$ and
160$\mu m$ bands. The confusion limits predicted
by different models differ by up to 80\%.
These differences reflect the uncertainties of
these predictions.

\clearpage

\vskip5truecm


\centerline{\bf Appendix}
\oneskip

\appendix
\section{New 25$\mu m$ Local Luminosity Functions for Three Populations}

In Xu et al. (2001), luminosity functions were presented for 
three population subsamples--AGNs, normal late-type galaxies
and starbursts. The subsamples were determined by IRAS colors.  We have
extended this work to take into account the effects of
luminosity evolution on the LLF estimates, and to include
a more comprehensive uncertainty analysis.

We recomputed the population LLFs including a luminosity
evolution term of the form $(1+z)^q$, for values of q =0,
3.0, and 4.5.  This factor is applied to both the source
luminosity, and the minimum luminosity detectable at the
source's redshift.  The shape parameters are tabulated
in Table A.1.


\noindent{\bf Table A.1. Paremeters of 25$\mu m$ Luminosity Functions}
\nopagebreak

\hskip-0.5truecm\begin{tabular}{cccccc}\hline
Population &   q  &   $\alpha$ & $\beta$ & $L_*/L_\sun$ & $\alpha+\beta$ \\
\hline
AGNs      & 0.0  &    0.336  &  1.691 &   6.9$\times 10^9$ &      2.027\\
AGNs      & 3.0  &    0.329  &  1.713 &   6.6$\times 10^9$ &      2.042\\
AGNs      & 4.5  &    0.326  &  1.724  &  6.5$\times 10^9$ &      2.040\\
&&&&& \\
Starbursts& 0.0  &    0.265  &  2.283  &  7.9$\times 10^9$ &      2.548\\
Starbursts& 3.0  &    0.265  &  2.275  &  7.7$\times 10^9$ &      2.540\\
Starbursts& 4.5   &   0.264  &  2.300  &  7.9$\times 10^9$ &      2.564\\
&&&&& \\
Normals   & 0.0   &   0.482  &  3.875  &  5.7$\times 10^9$ &      4.357\\
Normals   & 3.0   &   0.480   & 3.992  &  5.8$\times 10^9$ &      4.472\\
Normals   & 4.5   &   0.479   & 4.055  &  5.8$\times 10^9$ &      4.534\\
&&&&& \\
\hline
\end{tabular}

\vskip1truecm

The resulting LLFs (with same normalization constant applied)
are plotted as visibilities
in panel {\it a} of Fig.19 
(solid line is no evolution;
dashed is q=3.0; dashed-dot is q=4.5).  There is a small
difference between the LLFs at large luminosities. The
model calculations in this paper use the LLFs with q=3.0.

We performed an analysis of the covariance of the fitted
parameters, using the information matrix (Efstathiou et
al. 1988).  In panels {\it b}, {\it c}, and {\it d} of
Fig.19, we have plotted the 68\% confidence intervals
in pairs of the parameters, as ellipses under the assumption
of normally distributed uncertainties (Avni 1976;
Press et al. 1992).  
In general, the parameters are not strongly correlated,
except for $\alpha$ and $\beta$ for AGNs.  This correlation
is explained by the distribution of this population at
higher relative redshifts; the high-luminosity slope of
the luminosity function depends on the sum of $\alpha$
and $\beta$, so the parameters are degenerate when $\alpha$
is not well determined at low luminosities.  For the
same reason, the confidence intervals for AGNs for $\alpha$
and $L_*$ are larger for the other populations.  The
variations in the parameters with evolution exponent are for the
most part small compared to the uncertainties in the
parameters.
%


\section{Extrapolation of SEDs to the UV Bands}
In Xu et al. (2001), the UV (1000 --- 4000{\AA}) SEDs 
are extrapolations from data points in the B, J, H and K bands,
and therefore are not well constrained. In this work, 
this is improved by introducing the following constraints:
\begin{description}
\item{(1)} The UV-B v.s. B-K correlation. The correlation is established
using a sample of galaxies detected both in the vacuum UV bands
(1500{\AA} -- 2500{\AA}) and in the FIR bands (IRAS). The sample
is taken from Xu \& Buat (1995). The K-band magnitudes are found  
in the 2MASS database. In Fig.20, the UV-B v.s. B-K color-color plot
for this sample is presented. The sample is divided into galaxies with
AGNs ($f_{60\mu m}/f_{25\mu m} < 0.4$), galaxies with
IR excess ($f_{60\mu m}/f_{25\mu m} \geq 0.4$ and 
$L_{fir}/L_B > 0.5$), and
normal late-types ($f_{60\mu m}/f_{25\mu m} \geq 0.4$ and 
$L_{fir}/L_B \leq 0.5$). Data of the three ULIRGs observed 
by Trentham et al. (1999) in UV using  HST are also plotted.
They follow the same trend of IR excess galaxies,
though with larger scatters. We found that the trend for
normal late-types can be well fitted by the following function
\begin{eqnarray}
 \log (\nu f_\nu (2000{\AA})/\nu f_\nu(4400{\AA})& = -0.2-0.3\times((B-K)-2)^2
\;\;\;\;\;\; ((B-K) >2) \nonumber \\ 
   & =  -0.2-0.1\times((B-K)-2)\;\;\;\;\;\; ((B-K) \leq 2),
\end{eqnarray}
and for galaxies with AGNs and galaxies with IR excess, it 
can be well fitted by a two-step linear function
\begin{eqnarray}
 \log (\nu f_\nu (2000{\AA})/\nu f_\nu(4400{\AA})& = -0.4\times ((B-K)-2.) 
\;\;\;\;\;\;\; ((B-K) >2) \nonumber \\ 
   & =  0 \;\;\;\;\;\;\;  ((B-K) \leq 2).
\end{eqnarray}
\item{(2)} The UV slope. We use the following relation between the
UV slope $\beta$ ($F_\lambda \propto \lambda^{\beta}$, $1200{\AA} <
\lambda < 2600{\AA}$) and the $L_{fir}/L_B$ ratio, found by 
Calzetti et al. (1995), to constrain the slope of the UV SEDs
between 1200{\AA} and 2600{\AA}:
\begin{equation}
 \beta = 1.12\times \log(L_{fir}/L_B) -0.94.
\end{equation}
\item{(3)} Lyman-break. A sharp drop-off is imposed to the
SEDs shortward of 912{\AA}: The flux density decreases
a factor of 30 from 912{\AA} to 700{\AA}. There is a less
steep drop-off, about a factor of 5, from 1200{\AA} to 
912{\AA}, which is to mimic the effect of the Ly$\alpha$ 
absorption.
\end{description}

As an example, the model SED of M82 is compared
to data in Fig.21.

%


\vskip1cm 
This research has made use of the NASA/IPAC Extragalactic Database
(NED) which is operated by the Jet Propulsion Laboratory, California
Institute of Technology, under contract with the National Aeronautics
and Space Administration.  This work has made use of data
products from the Two Micron All Sky Survey (2MASS), which is a joint
project of the University of Massachusetts and the Infrared Processing
and Analysis Center/California Institute of Technology, funded by the
National Aeronautics and Space Administration and the National Science
Foundation. C. K. Xu,  C. J. Lonsdale and D. L. Shupe were supported 
by the SIRTF Legacy Science Program
provided by NASA through a contract with the Jet Propulsion Laboratory,
California Institute of Technology.


\clearpage

{\bf Figure Captions:}

\begin{figure}
\caption{Schematic plots of evolution functions of the three new models.} 
\end{figure}

\begin{figure}
\caption{
{\scriptsize Comparisons of predictions of Model S1 with observed IR
counts and CIB.  Data points in the plot of 15$\mu m$ counts: the
ELAIS counts reported by Serjeant et al. (2000) are 
plotted with open four-point stars, and the ELAIS counts 
reported by Gruppioni et al. (2002) are 
plotted with filled four-point stars.  
Other data points have the same symbols as in Elbaz
et al. (1999): A2390 (six-point stars); ISOHDF-North (open circles),
ISOHDF-South (filled circles), Marano FIRBACK Ultra-Deep (open
squares), Marano Ultra-Deep (exes), Marano FIRBACK Deep (asterisks),
Lockman Deep (open triangles), Lockman Shallow (filled triangles).
Filled squares with error bars are counts taken from Xu (2000). 
Note that the high points in the bright end ($f_{15\mu m} \geq 0.5$ Jy)
are due to Local Supercluster (Lonsdale et al. 1990). The
shaded area marks the range of counts estimated by Mazzei et al
(2001).  Data points in the IRAS 60$\mu m$ plot: Large filled
circles: Mazzei et al. (2001); Xs: Hacking \& Houck (1987); open
stars: Gregorich et al. (1995); open circles: Bertin et al. (1997);
small filled squares: counts in the South Galactic cap ($b^{II} <
-50^\circ $) by Lonsdale et al. (1990); open triangles: Saunders et
al. (1991); open squares: Rowan-Robinson et al. (1990).  Data
points in the 90$\mu m$ plot: filled circles: Efstathiou et al. (2000),
crosses: Linden-Voernle et al. (2000), open square: 
Matsuhara et al. (2000), open diamonds: total counts of
Juvela et al. (2000), open triangles: counts of 
multiple detections of Juvela et al. (2000). 
Data points in the 170$\mu m$ plot: filled circles: 
Dole et al. (2001), open square: 
Matsuhara et al. (2000), open diamonds: total counts of
Juvela et al. (2000), open triangles: counts of 
multiple detections of Juvela et al. (2000).  
The SCUBA 850$\mu m$ counts: crosses: Blain
et al. (1999); open circles: Hughes et al. (1998); open diamonds:
Eales et al. 2000; open squares: Barger et al. (1999); filled
diamonds: Scott et al. (2002).  The cosmic IR background: Filled
circles: Lagache et al. (1998); open squares: Finkbeiner et
al. (2000); open stars: Gorjian et al. (2000); filled star: Dwek and
Arendt (1998); large crosses: SCUBA source count results (Blain et
al. 1999); shadowed area: the range of COBE/FIRAS results (Fixsen et
al. 1998); diamonds and Xs with upper-limits: upper-limits from TeV
gamma-ray radiation of Mrk403 and Mrk501 (Dwek \& Slavin 1994; Stanev
\& Franceschini 1998).}  
}
\label{fig2}
\end{figure}

\begin{figure}
\caption{Comparisons of predictions of Model S2 with
observed IR counts and CIB. The sources of data points
are the same as in Fig.2.} 
\end{figure}

\begin{figure}
\caption{Comparisons of predictions of Model S3 with
observed IR counts and CIB. The sources of data points
are the same as in Fig.2.} 
\end{figure}

\begin{figure}
\caption{Predictions of Model S1 for redshift distributions
of various IR surveys. The observed redshift distributions
(histograms in corresponding plots)
of IRAS 60$\mu m$ sources and ISOCAM 15$\mu m$ sources
are taken from Rowan-Robinson (2001) and Franceschini et al.
(2001), respectively.}
\end{figure}
\begin{figure}
\caption{Predictions of Model S2 for redshift distributions
of various IR surveys. Otherwise same as in Fig.5.}
\end{figure}

\begin{figure}
\caption{Predictions of Model S3 for redshift distributions
of various IR surveys. Otherwise same as in Fig.5.}
\end{figure}

\begin{figure}
\caption{Comparisons of model predictions for
the 15$\mu m$ luminosity functions (LFs) with observations.
The data points for z=0 LF are taken from Xu (2000).
The solid line is the Schechter function fit of the data points.
Points in other panels are derived using data taken from
Elbaz et al. (2002).}
\end{figure}

\begin{figure}
\caption{Star formation history (SFH) predicted by new models.
Data points: 
Red arrows with error bars: lower limits derived
from the 15$\mu m$ LFs in Fig.8 (see text). 
Filled squares: SFR from ISOCAM surveys (Flores et al. 1999b).
Xs: SFR from Balmer line surveys (Gallego et al. 1995;
Tresse \& Maddox 1998; Yan et al. 1999;
Glazebrook et al. 1999). Open squares:
SFR from UV surveys (`extinction
corrected' data taken from Fig.9 of Steidel et al. 1999). 
Open triangle: SFR from 
SCUBA (Barger et al. 1999). 
The shaded area: predictions by Pei \& Fall (1995).
All data have been converted to 
the cosmology model specified by H$_0$=75 km/sec/Mpc,
$\Omega_m = 0.3$, and $\Omega_\Lambda = 0.7$.
} 
\label{fig9}
\end{figure}

\begin{figure}
\caption{Schematic plots of formation functions of three E/S0 models.} 
\end{figure}

\begin{figure}
\caption{Color v.s. redshift diagrams of E/S0 galaxies.
Model predictions 
compared with observations (open diamonds, Franceschini et al.
1998).
}
\end{figure}

\begin{figure}
\caption{Model predictions for the redshift distribution
of E/S0s in HDF-N compared with the data (histogram,
Franceschini et al. 1998).
}
\end{figure}

\setcounter{figure}{12}

\begin{figure}
\caption{Counts in the UV (2000{\AA}), B, R and K bands.
Predictions of Model S1 (for dusty galaxies)
plus predictions of Model E2 (for E/S0s) are compared to
data. Data points in the UV (2000{\AA}) plot:
Milliard et al. (1994).  Data points in the B-band plot:
Open squares: Williams et al. (1996);
filled squares: Metcalfe et al. (1995); 
Xs: Metcalfe et al. (1991);
open diamonds: Gardner et al. (1996).
Data points in the R-band plot: open squares: Lin et al. (1999);
open diamonds: Cohen (2002).
Data points in the K-band plot:
open triangles: Bershady et al. (1998);
open squares: Soifer et al. (1994); Xs:  Minezaki et al. (1998);
open diamonds: Gardner et al. (1996). 
Data points in the 6.7$\mu m$ band plot:
open squares: Taniguchi et al. 1997;
Xs: Altieri et al. 1999;
diamond: Flores et al. 1999a; shaded area:
Oliver et al. (2002) and Serjeant et al. (2000).
Data points in the 12$\mu m$ band plot:
open squares: Fan et al. 1998;
Xs: Clements et al. 1999.
}
\label{fig13}
\end{figure}

\begin{figure}
\caption{Counts in the UV (2000{\AA}), B, R and K bands.
Predictions of Model S2 (for dusty galaxies)
plus predictions of Model E2 (for E/S0s) are compared to
data. Data points are the same as in Fig.13.
}
\end{figure}

\begin{figure}
\caption{Counts in the UV (2000{\AA}), B, R and K bands.
Predictions of Model S3 (for dusty galaxies)
plus predictions of Model E2 (for E/S0s) are compared to
data. Data points are the same as in Fig.13.
}
\end{figure}

\begin{figure}
\caption{
Model predictions for counts in three SIRTF bands
($3.6\mu m$, $24\mu m$ and $70\mu m$). 
}
\end{figure}

\pagebreak

\begin{figure}
\caption{Predictions of Model S1, Model S2 and
Model S3 for color distributions of ISO sources.
The detection limits in these bands are set to be $f_{6.7\mu m} \geq 30\; 
\mu Jy$, $f_{15\mu m} \geq 100\; \mu Jy$, $f_{90\mu m} \geq 100\; mJy$,
$f_{170\mu m} \geq 180\; mJy$, $R \leq 24\; mag$, and $K \leq 20\; mag$.
The observed color distributions
(histograms in corresponding plots)
of ISOCAM 15$\mu m$ sources in the HDF-N field
are derived from data taken from Cohen et al. 
(2000) and Hogg et al. (2000).}
\end{figure}

\begin{figure}
\caption{Predictions of Model S1, Model S2 and
Model S3 for color distributions of SWIRE sources.
The detection limits in these bands are set to be:
$f_{70\mu m} \geq 6.15\; mJy$,
$f_{24\mu m} \geq 0.45\; mJy$, $f_{3.6\mu m} \geq 7.3\; \mu Jy$,
$f_{8\mu m} \geq 32.5\; \mu Jy$, $R \leq 25\; mag$, and
$K \leq 20\; mag$.
}
\end{figure}

\begin{figure}
\caption{
Panel {\it a}:  
25$\mu m$ local luminosity functions (LLFs) of
three populations of dusty galaxies.
Different lines denote different assumptions on
the evolution of the sources:
solid line --- no evolution;
dashed line --- q=3.0; dashed-dot line --- q=4.5. 
Panels {\it b}, {\it c}, and {\it d}:
Plots of pairwise covariance of the LLF parameters.
The ellipses show
the 68\% confidence intervals.
The lines are the same as in panel {\it a}.
}
\end{figure}

\begin{figure}
\caption{The UV-B v.s. B-K correlation of galaxies. 
Data for ULIRGs (large crosses) are taken from 
Trentham et al. (1999). Other data are taken from
the UV and IRAS selected sample of Xu and Buat (1996).
}
\end{figure}

\begin{figure}
\caption{The model SED of M82 compared to data.
Data points: filled squares: collections of Silva et al. (1998);
open squares: 2MASS and NED data.
} 
\label{fig21}
\end{figure}


\begin{references}

\reference{Adelberger00} Adelberger, K.L., Steidel, C.C.
2000, ApJ, 544, 218.

\reference{Altieri99} Altieri, B., Metcalfe, L., Kneib, J.P.,
McBreen, B., et al. 1999, A\&A, 343, L65.

\reference{Aussel99} Aussel, H., Cesarsky, C.J., Elbaz, D.,
Starck, J.L. 1999, \aap, 342, 313.

\reference{Avni76} Avni, Y. 1976, ApJ, 210, 642.

\reference{Barger98} Barger, A.J., Cowie, L.L., Sanders, D.B., et al.
1998, Nature, 394, 248.

\reference{Barger99} Barger, A.J., Cowie, L.L., Sanders, D.B. 1999,
ApJ, 518, L5.

\reference{Bertin97} Bertin, E., Dennefeld, M., Moshir, M.
1997, A\&A, 323, 685.

\reference{blain99} Blain, A.W. 1999, MNRAS, 309, 955.

\reference{blainetal99} Blain, A.W., Smail, I., Ivison, R.J., Kneib, J.-P.
1999, MNRAS, 302, 632.

\reference{benitez99} Benitez, N., Broadhurst, T., Bouwens, R.
Silk, J., Rosati, P. 1999, ApJL, 515, L65.

\reference{bershady98} Bershady, M.A., Lowenthal, J.D.,
 Koo, D.C. 1998, ApJ, 505, 50. 

\reference{Boyle00} 
Boyle, B.J., Shanks, T. Croom, S.M., Smith, R.J., Miller, L., Loaring, M., Heymans, C. 2000, MNRAS, 317, 1014.


\reference{calzetti95} Calzetti, D., Bohlin, R.C., Kinney, A.L.,
Stochi-Bergmann, T., Heckman, T.M. 1995, ApJ, 443, 136.

\reference{Clements99} Clements, D. L., 
Desert, F.-X., Franceschini, A., Reach, W. T.,
Baker, A. C., Davies, J. K., Cesarsky, C. 1999, A\&A, 346, 383.

\reference{Clements01} Clements, D. L., 
Desert, F.-X., Franceschini, A. 2001, 
MNRAS, 325, 665.

\reference{Chapman02} Chapman, S.C., Smail, I., Ivison, R.J.,
Helou, G., et al. 2002, ApJ, 573, 66.

\reference{Chary01} Chary, R.R., Elbaz, D. 2001, \apj, 556, 562.

\reference{Cohen02} Cohen, J.G. 2002, ApJ, 567, 672.

\reference{Cohen00} Cohen, J.G., Hogg, D.W.,
 Blandford, R., Cowie, L.L.,
 Hu, E., Songaila, A.,
 Shopbell, P., Richberg, K. 2000, \apj, 538, 29.

\reference{dol01} Dole, H., Gispert, R., Lagache, G., Puget, J-L., et al. 2001,
A\&A, 372, 364.

\reference{dwe98} Dwek, E., Arendt, R.G. 1998, ApJ, 508, L9.

\reference{dwe94} Dwek, E., Slavin, J. 1994, ApJ, 436, 696.

\reference{eales00} Eales, S.A., Lilly, S.J., Webb, T., Dunne, L., Gear, W.,
Clements, D., Yun, M. 2000, AJ, 120, 2244.

\reference{efstathiou88} 
Efstathiou, G., Ellis, R.S., Peterson, B.A. 1988, MNRAS, 232, 431.

\reference{efstathiou00} Efstathiou, A., Oliver, S., Rowan-Robinson,
M., Surace, C., et al. 2000, MNRAS, 319, 1169.

\reference{eggen62} Eggen, O.J., Lynden-Bell, D.,
Sandage, A.R. 1962, ApJ, 136, 748.


\reference{elb99} Elbaz, D., Cesarsky, C.J., Fadda, D. , et al. 1999,
\aap, 351, L37.

\reference{elb02} Elbaz, D., Cesarsky, C.J., Chanial, P., Franceschini, A.,
Fadda, D., Chary, R.R. 2002, \aap, in press (astro-ph/0201328).

\reference{ellis97} Ellis, R.S. 1997, ARAA, 35, 389.

\reference{Fang98} Fang, F., Shupe, D., Xu, C., Hacking, P.
1998, ApJ, 500, 693.

\reference{Fink00} Finkbeiner, D.P., Davis, M., Schlegel, D.J. 2000,
ApJ, 544, 81.

\reference{fix98} Fixsen, D.J., Dwek, E., Mather, J.C., Bennett,
C.L., Shafer, R.A. 1998, ApJ, 508, 123.

\reference{flores99a} Flores, H., Hammer, F., Des\'ert, F.X., C\'esarsky, C., 
 et al. 1999a, A\&A, 343, 389.

\reference{flores99b} Flores, H., Hammer, F., Thuan, T.X.,
Cesarsky, C., Desert, F.X., et al. 1999b, ApJ, 517, 148.

\reference{fox02} Fox, M.J., Efstathiou, A., Rowan-Robinson, M.,
Dunlop, J.S., et al. 2002, MNRAS, 331, 839.

\reference{fra97} Franceschini, A., Aussel,
H., Bressan, A., Cesarsky, C.J., Danese, L., de Zotti, G.,
Elbaz, D., Granato, G. L., Mazzei, P., Silva, L. 1997, in
{\it The Far Infrared and Submillimetre Universe}; ed. 
A. Wilson;  p.159 (Noordwijk, The Netherlands: ESA, 1997).

\reference{fra01} Franceschini, A., Aussel, H., Cesarsky, C.J., 
Elbaz, D.,  Fadda, D., 2001, \aap, 378, 1.

\reference{fra88} Franceschini, A., Danese, L., De Zotti, G.,, Xu, C. 1988,
MNRAS, 233, 175.

\reference{fra98} Franceschini, Silva, L., Fasano, G., 
Granato, G.L., Bressan, A., Arnouts, S., Danese, L. 1998, ApJ, 506, 600.

\reference{gardner96} Gardner, J. P.,
 Sharples, R. M., Carrasco, B. E.. Frenk, C. S.
 1996, MNRAS, 282, L1.

\reference{gallego} Gallego, J., Zamorano, J.,
 Aragon-Salamanca, A., Rego, M.  1995, ApJ, 455, L1.

\reference{gardner97} Gardner, J. P.,
 Sharples, R. M., Carrasco, B. E.. Frenk, C. S.
 1997, ApJ, 480, L99.

\reference{genzel2000} Genzel, R., Cesarsky, C.J. 2000, ARA\&A, 38, 761.

\reference {gill01} Gilli, R., Salvati, M. and Hasinger, G. 2001, A\&A, 
in press

\reference {gis00} Gispert, R., Lagache, G., Puget, J.L. 2000, A\&A, 360, 1.

\reference{Glazebrook99} Glazebrook, K.,
 Blake, C., Economou, F., Lilly, S., Colless, M. 1999, 
 MNRAS, 306, 843.


\reference{gorjian00} Gorjian, V., Wright, E.L., Chary, R.R.
ApJ, 536, 550.

\reference{granato00} Granato, G., Lacey, C.G., Silva, L.  Bressan, A.,
  Baugh, C.M., Cole, S., Frenk, C.S. 2000, \apj, 542, 710.

\reference{gre95} Gregorich, D.T., Neugebauer, G., Soifer, B.T., Gunn, J.E.,
Herter, T.L. 1995, AJ, 110,259.

\reference{Grup02}  
Gruppioni, C., Lari, C., Pozzi, C.F., Zamorani, G.,
Franceschini, A., Oliver, S., Rowan-Robinson, M., Serjeant, S.
2002, MNRAS, 335, 831.

\reference{hac87} Hacking, P. B., Condon, J. J., and Houck, J. R. 1987,
\apj, 316, L15.




\reference{Hogg00} Hogg, D.W., Pahre, M.A., Adelberger, K.L.,
 Blandford, R., Cohen, J.G., Gautier, T.N., Jarrett, T.,
Neugebauer, G., Steidel, C.C. 2000, APJS, 127, 1.

\reference{huang98} Huang, J.-S., Cowie, L., Luppino, G.
1998, ApJ, 396, 31.

\reference{Hughes90} Hughes, D.H., Gear, W.K., Robson, E.I. 
1990, MNRAS 244, 759.

\reference{Hughes98} Hughes, D.H., Serjeant, S., Dunlop, J., et al. 
1998, Nature, 394, 241.

\reference{im02} Im, M., Simard, L., Faber, S.M., Koo, D.C., Gebhardt,
et al. 2002, ApJ, 571, 1361.

\reference{ivison02} Ivison, R., Greve, T.R.,
Smail, I., Dunlop, J.S., et al. 2001, preprint (astro-ph/0206432).

\reference{ivison01} Ivison, R., Smail, I., Frayer, D.T., Kneib,
J.-P., Blain, A.W. 2001, ApJ, 561, L45.

\reference{Juvela00} Juvela, M., Mattila, K., Lemke, D. 2000,
A\&A, 360, 813.

\reference{kauffmann93}  Kauffmann, G., White, S., Guiderdoni, B.
1993, MNRAS, 308, 833.

\reference{kauffmann98}  Kauffmann, G., Charlot, S. 1998,
MNRAS, 297, L23.

\reference{kauffmann01}  Kauffmann, G., Charlot, S., Balogh, M.
2001, preprint (astro-ph/0103130).


\reference{Lan02} Lanzetta, K, Yahata, M., Pascarrela, S.,
Chen, X.W., Fernandez-Soto, A. 2002, ApJ, 570, 492.

\reference{lil96} Lilly, S.J., Le F\'evre, O., Hammer, F., Crampton, D. 
1996, ApJ, 460, L1.


\reference{kaw1998} Kawara, K., Sato, Y., Matsuhara, H., et al. 1998,
\aap, 336, L9.

\reference{kes96} Kessler, M.F., Steinz, J.A., Anderegg, M.E.
et al. 1996, \aap, {\bf 315}, L27.

\reference{koch} Kochanek, C.S., Pahre, M.A., Falco, E.E., Huchra, J.P., 
Mader, J., Jarrett, T.H., Chester, T., Cutri, R., Schneider, S. E.
2001, ApJ, 560, 566.

\reference{koo92} Koo, D., Kron, R. 1992, ARAA, 30, 613.


\reference{Kormendy92} Kormendy, J., Sanders, D.B. 1992, \apj 390, 53.

\reference{franca00} La Franca, F., Matute, I., Gruppioni, C., Alexander, D.
et al. 2000, to appear in the proceedings of the Fourth Italian
Conference on AGNs (MemSAIt) (astro-ph/0006177).




\reference{lag98} Lagache, G., Abergel, A., Boulanger, F., Puget, J.-L.
Puget 1998, \aap, 333, 709.

\reference{lil96} Lin, H., Yee, H.K., Carlberg, R.G.,
Morris, S.L., Sawicki, M., Patton, D.R., Wirth, G., Shepherd, C.
1999, ApJ, 518, 533.

\reference{linden-Voernle00} Linden-Voernle, M.J.D., Noergaard-Nielsen,
H.U., Joergensen, H.E., et al. 2000, A\&A, 395, 51.

\reference{lon99} Lonsdale, C.J. 1999, 
in {\it Astrophysics with Infrared Surveys: A Prelude to SIRTF},
ASPC Series {\bf 177}, eds. M.D. Bicay, C.A. Beichman, R.M. Cutri, and
B.F. Madore, p24.


\reference{lon90} Lonsdale, C.J., Hacking, P.B., Conrow, T.B.,
Rowan-Robinson, M. 1990, ApJ, 358, 20.

\reference{lon00} Lonsdale, C.J., et al.  2000,
http//sirtf.caltech.edu/SciUser/A\_GenInfo/SSC\_A1\_Legacy\_CL.html.

\reference{Mad96} Madau, P., Ferguson, H.C., Dickinson, M., Giavalisco, M.,
et al. 1996, MNRAS, 283, 1388.

\reference{Malkan01} Malkan, M.A., Stecker, F.W. 2001, \apj, 555, 641

\reference{Matsuhara00} Matsuhara, H., Kawara, K., Sato, Y.,
et al. 2000, A\&A, 361, 407.

\reference{Mazzei01} Mazzei, P., Aussel, H., Xu, C., Salvo, M., 
De Zotti, G., Franceschini, A. 2001, New A., 6, 265.


\reference{menanteau99}
Menanteau, F., Ellis, R.S., Abraham, R.G., Barger, A.J., Cowie, L.L.
1999, MNRAS, 309, 208.

\reference{metcalfe91}  Metcalfe, N., Shanks, T.,
 Fong, R., Jones, L.R. 1991, MNRAS, 249, 498.

\reference{metcalfe95}  Metcalfe, N., Shanks, T.,
 Fong, R., Roche, N. 1995, MNRAS, 273, 257.

\reference{milliard92} Milliard, B., Donas, J., Laget, M., Armand,
V., Vuillemin, A. 1992, \aap, 257, 24.

\reference{minezaki98} Minezaki, T., Kobayashi, Y.,
 Yoshii, Y., Peterson, B.A.  1998, ApJ, 494, 111.



\reference{pear01} Pearson, C. 2001, \mnras, 325, 1511.

\reference{pei95} Pei, Y., Fall, S.M. 1995, \apj, 454, 69.


\reference{poz96} Pozzetti, L., Bruzual, G., Zamorani, G.
1996, MNRAS, 281, 953.

\reference{press92} 
Press, W.H., Flannery, B.P., Teukolsky, S.A.,  Vetterling,
W.T. 1992, "Numerical Recipes in C: 2nd Edition", Cambridge
University Press, pp. 689-699.




\reference{pug99} Puget, J-L., Lagache, G., Clements, D.L, et al. 1999,
\aap, 345, 29.



\reference{roche99} Roche, N., Eales, S.A. 1999, MNRAS, 307, 111.

\reference{rodighiero01} Rodighiero, G., Franceschini, A.,
Fasano, G. 2001, MNRAS, 324, 491.

\reference{row01} Rowan-Robinson, M. 2001, ApJ, 549, 745.

\reference{row89} Rowan-Robinson, M., Crawford, J. 1989, MNRAS, 238, 523.

\reference{row90} Rowan-Robinson, M., Hughes, J., Vedi, K., Walker, D.W.
1990, MNRAS, 246, 473.

\reference{row97} Rowan-Robinson, M., Mann, R.G., Oliver, S.J., et
al. 1997, MNRAS, 289, 490.

\reference{rush93} Rush, B., Malkan, M.A., Spinoglio, L.
1993, ApJS. 89, 1.

\reference{sanm96} Sanders, D.B., Mirabel, I.F. 1996, ARA\&A, 34, 749.


\reference{searle72} Searle, L., Sargent, W.L. 1972, ApJ, 173, 25.

\reference{sau91} Saunders, W., Rowan-Robinson, M., Lawrence, A.,
et al. 1991, MNRAS, 242, 318.

\reference{scott00} 
Scott, D., Lagache, G., Borys, C., Chapman, S.C., 
Halpern, M., Sajina, A., Ciliegi, P., Clements, D.L., Dole, H.,
 Oliver, S., Puget, J.-L., Reach, W. T., Rowan-Robinson, M.
2000, A\&A, 357, L5.

\reference{scott02} 
Scott, S.E., Fox, M.J., Dunlop, J.S., Serjeant, S., et al. 2002,
MNRAS, 331, 817.

\reference{scoville00} 
Scoville, N.Z., Evans, A.S., Thompson, R., Rieke, M., et al. 2000,
AJ, 119, 991.

\reference{serjeant01} Serjeant, S., Efstathiou, A., Oliver, S.,
Surace, C., Héraudeau, P., Linden-Vørnle, M. J. D., Gruppioni, C., La
Franca, F., Rigopoulou, D., Morel, T., Crockett, H., Sumner, T.,
Rowan-Robinson, M., Graham, M. 2001, MNRAS, 322, 262.

\reference{serjeant00} 
 Serjeant, S., Oliver, S., Rowan-Robinson, M., Crockett, H., et al. 2000,
MNRAS, 316, 768.

\reference{schade98} 
Schade, D., Lilly, S.J., Crampton, D., Ellis, R.S., Le F\'evre, O., 
Hammer, F., Brinchmann, J.,
Abraham, R., Colless, M., Glazebrook, K., Tresse, L., Broadhurst, T.
1998, ApJ, 525, 31.

\reference{shp98} Shupe, D.L., Fan, F., Hacking, P.B., Huchra, J.P.
1998, ApJ, 501, 597.

\reference{silva99} Silva, L., Granato, G.L., Bressan, A., 
Danese, L. 1998, ApJ, 509, 103.


\reference{smail02} Smail, I., Ivison, R.J., Blain, A., Kneib, J.-P.
2002, MNRAS, 331, 495.


\reference{soif94} Soifer, B.T., Matthews, K., Djorgovski,
S., Larkin, J., et al. 1994, ApJL, 420, L1.


\reference{somerville01} Somerville, R.S., Primack, J.R., Faber, S.M.
2001, MNRAS, 320, 504.

\reference{Stanev98} Stanev, T., Franceschini, A. 1998, ApJ, 494, L159.

\reference{Steidel99} Steidel, C.C., Adelberger, K., Giavalisco, M., Dickinson,
M., Pettini, M., 1999, \apj, 519, 1

\reference{surace00} 
Surace, J.A., Sanders, D.B., Evans, A.S. 2000, ApJ, 529, 170.

\reference{taniguchi97} Taniguchi, Y., Cowie, L., Sato, Y., et al. 1997, A\&A,
328, L9.
 


\reference{t78} Toomre, A. 1978, in IAU Symp. 79, p109.

\reference{trentham99} Trentham, N., Kormendy, J., Sanders, D.B.
1999, AJ, 117, 2152.

\reference{Tresse} Tresse, L., Maddox, S. J. 1998, ApJ, 495, 691.

\reference{Yan} Yan, L., McCarthy, P. J., Freudling, W., Teplitz, Harry I., 
Malumuth, E. M., Weymann, . J., Malkan, M. A. 1999, ApJ, 519, L47.


\reference{will96} Williams, R.E., Blacker, B., Dickinson, M., 
Van Dyke, W., et al. 1996, \aj, 112, 1335.

\reference{worthy94} Worthy, G. 1994, ApJS, 95, 107.

\reference{Xu95} Xu, C., Buat, V. 1995, \aap, 293, L95.

\reference{Xu98} Xu, C., Hacking, P.B., Fan, F., Shupe, D.L., Lonsdale, C.J.,
Lu, N.Y., Helou, G.X. 1998, ApJ, 508, 576.

\reference{Xu00} Xu, C. 2000, \apj, 541, 134.

\reference{Xu01} Xu, C., Lonsdale, C.J., Shupe, D.L., 
O'linger, J., Masci, F. 2001, ApJ, 562, 179.

\end{references}
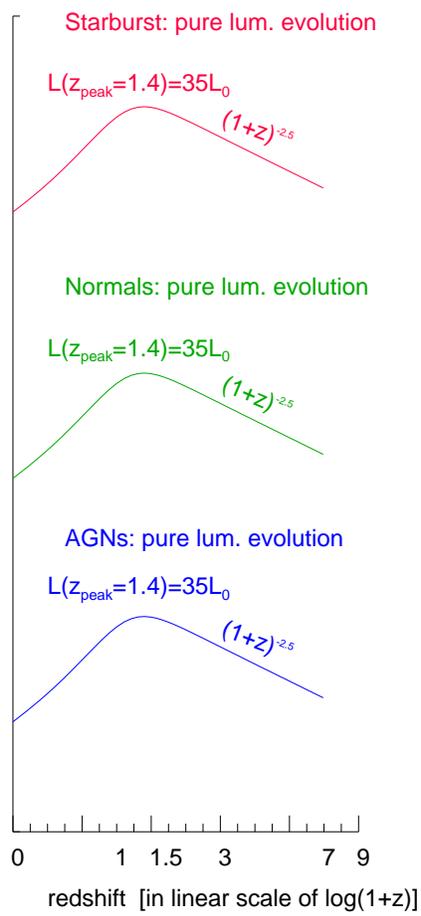
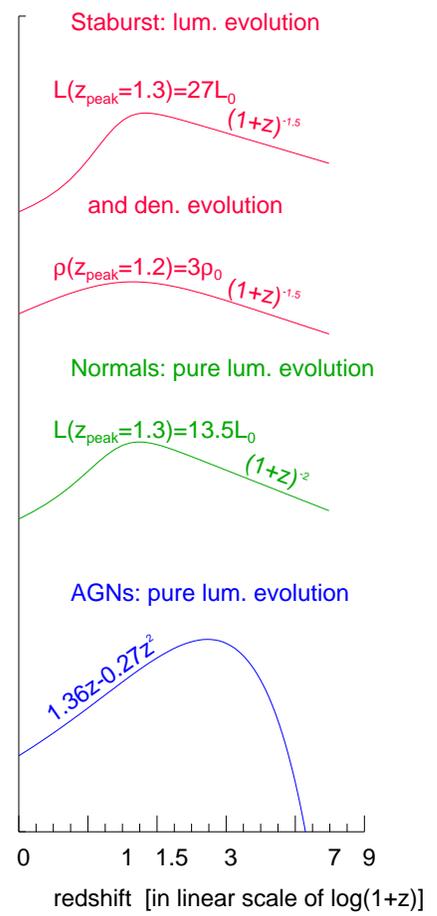
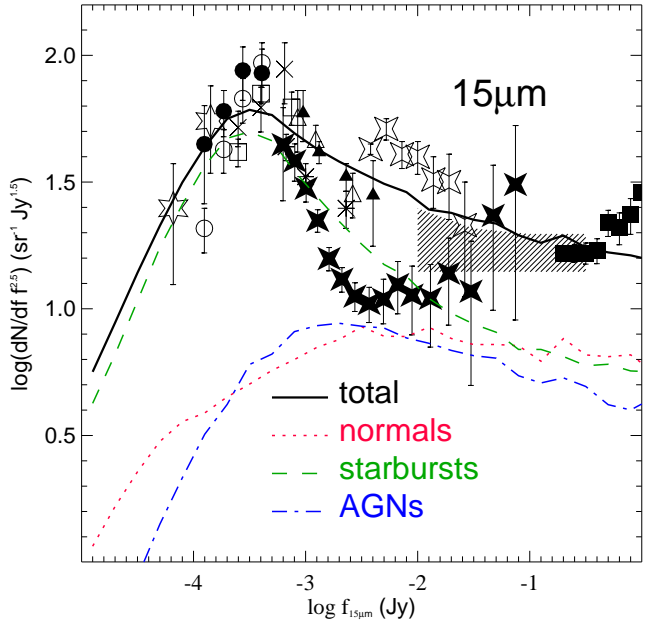
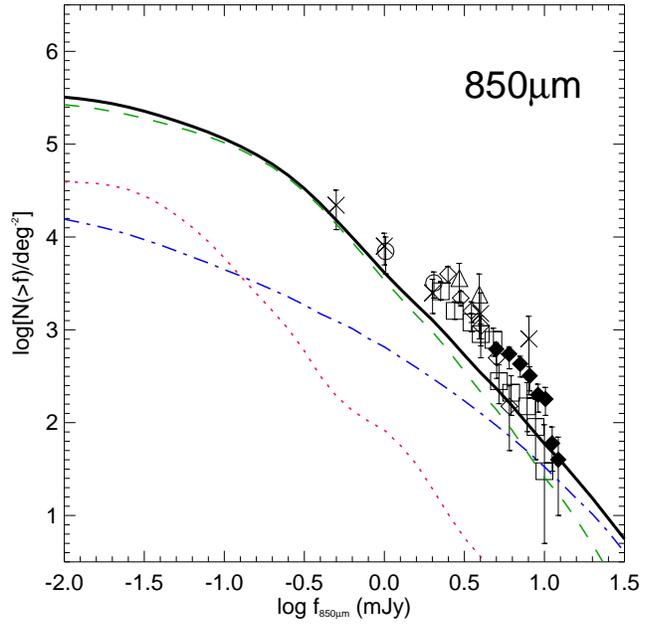
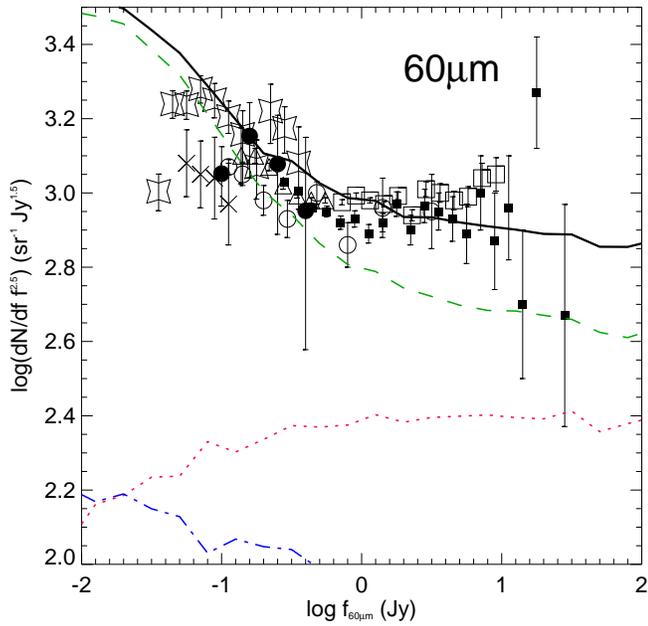
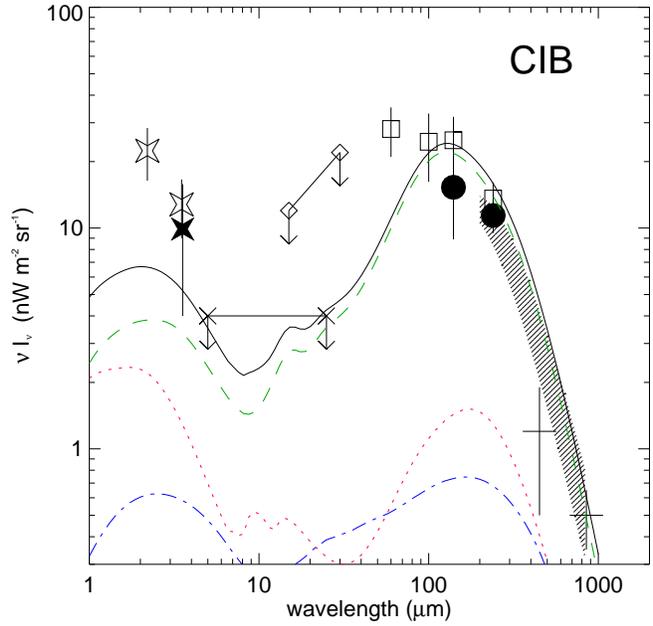
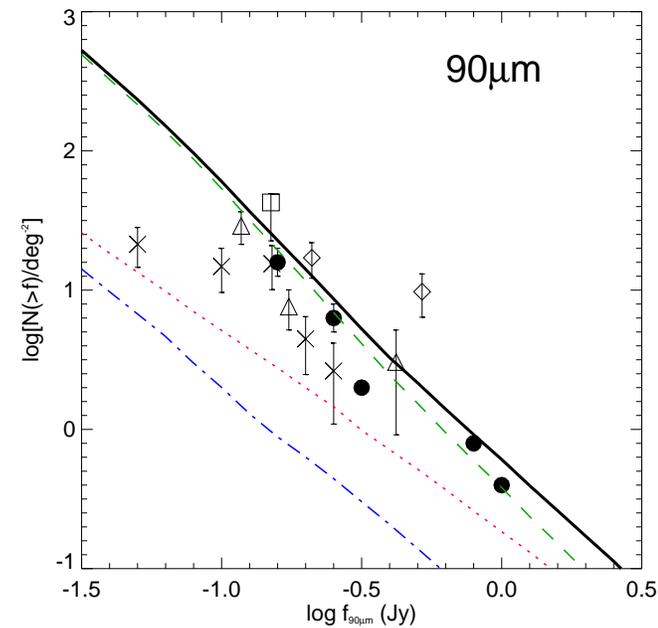
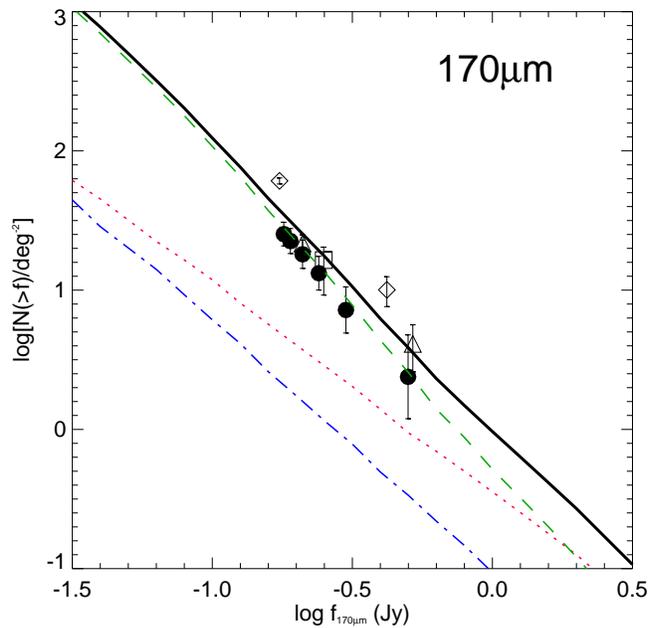
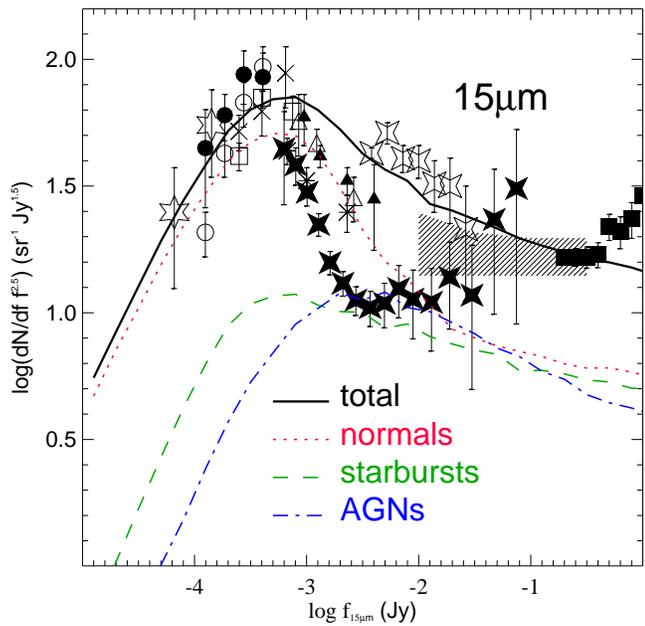
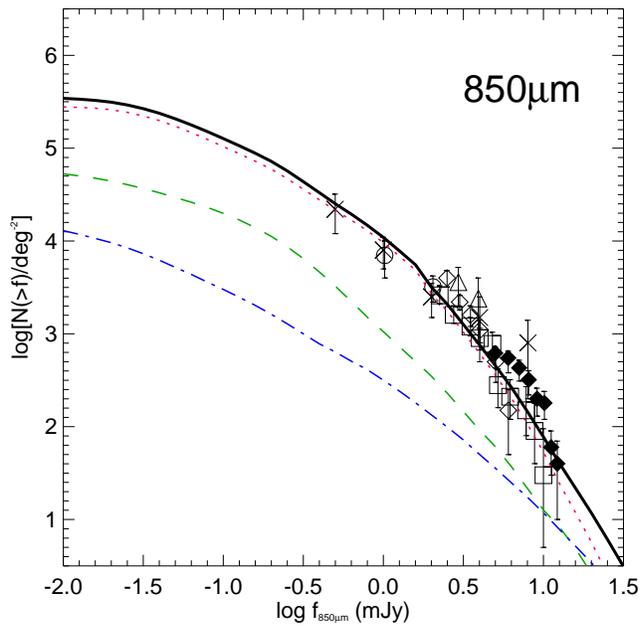
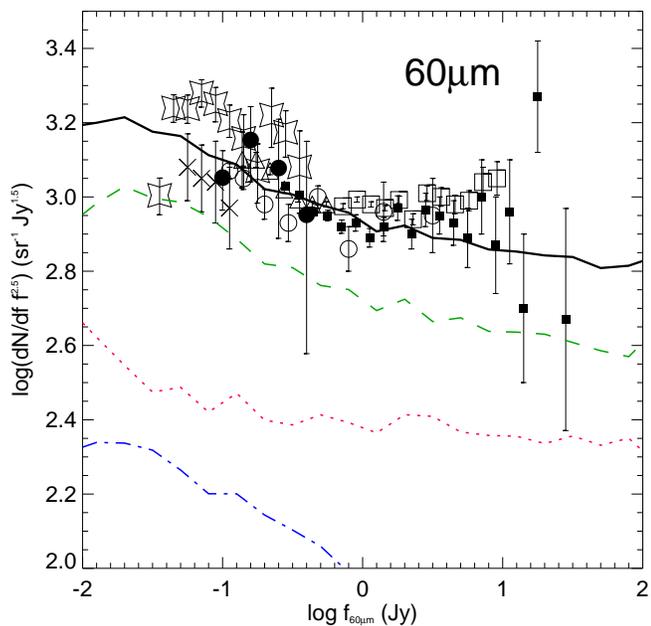
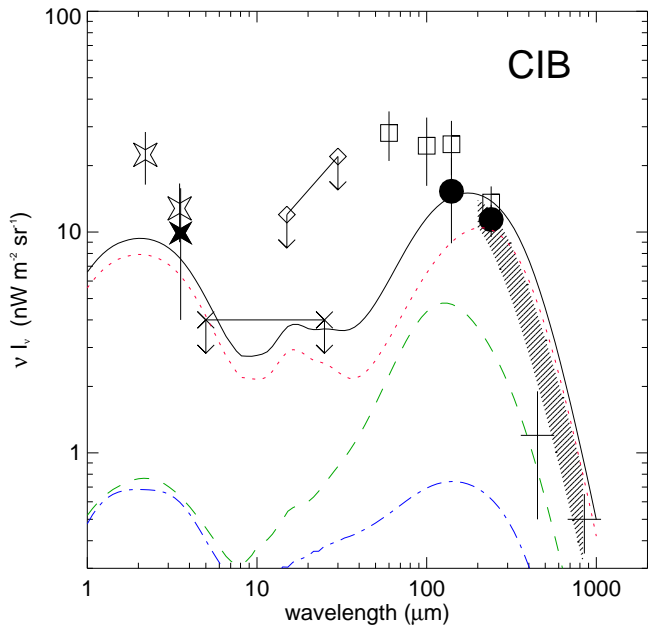
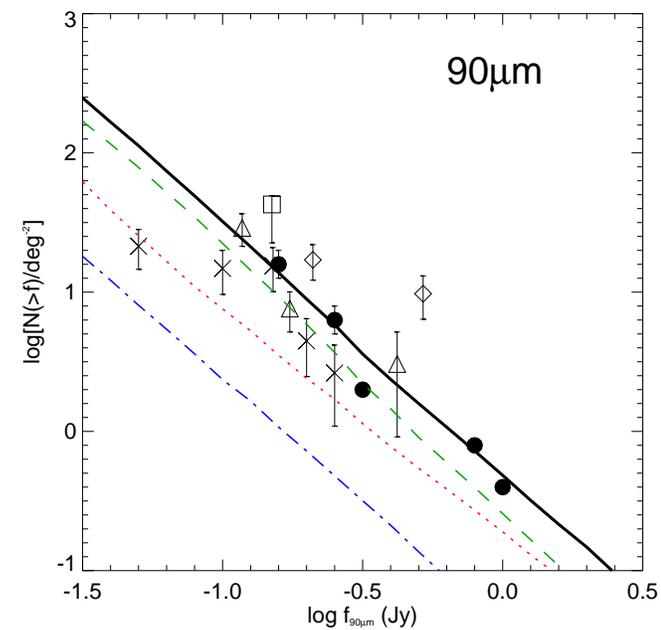
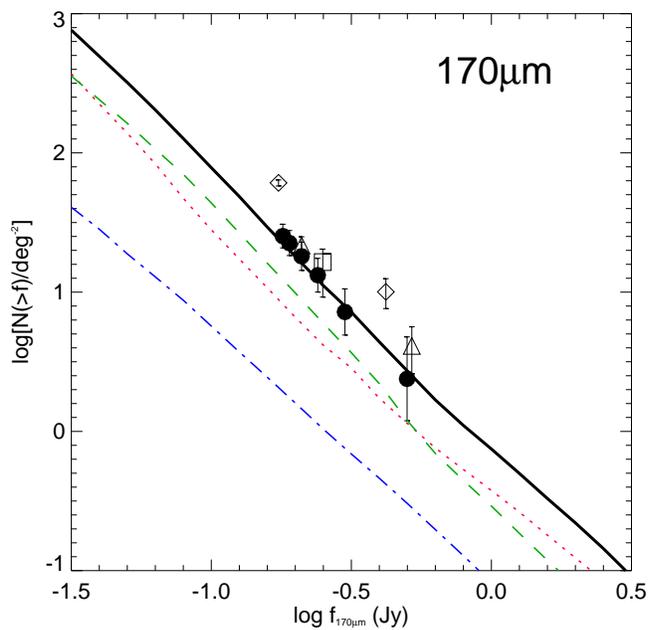
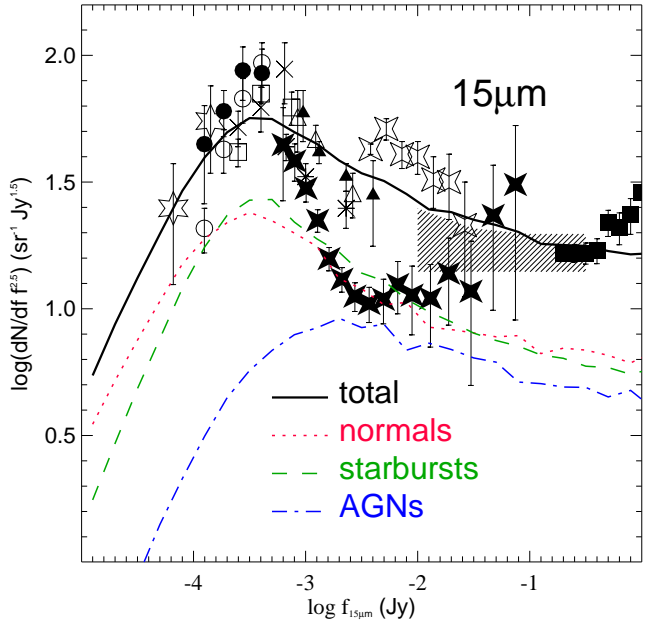
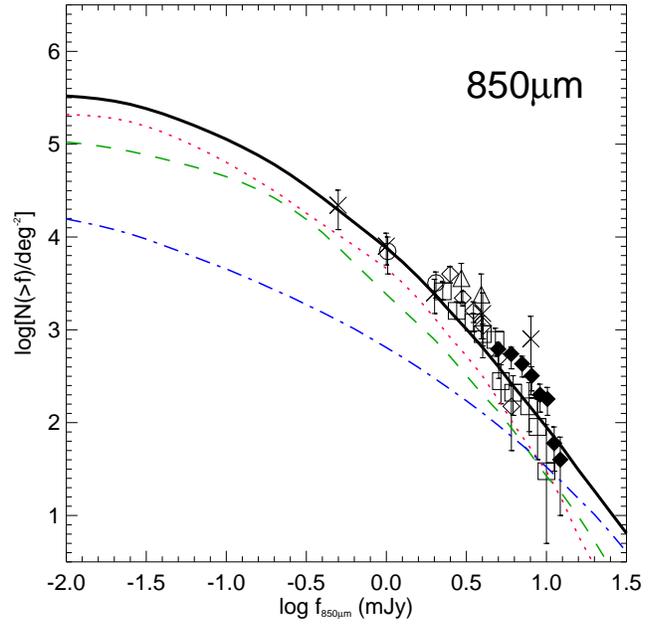
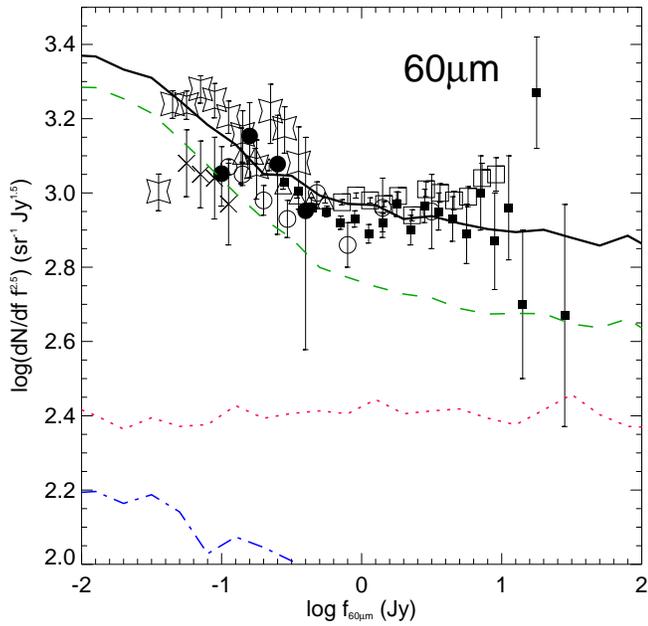
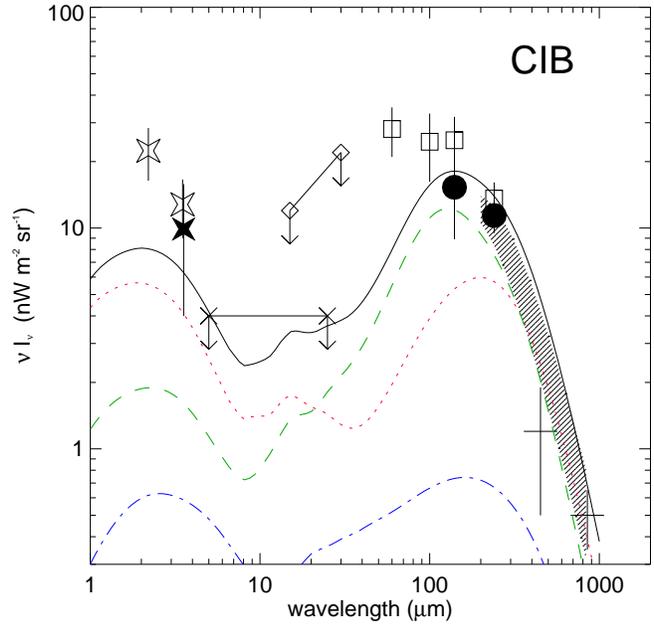
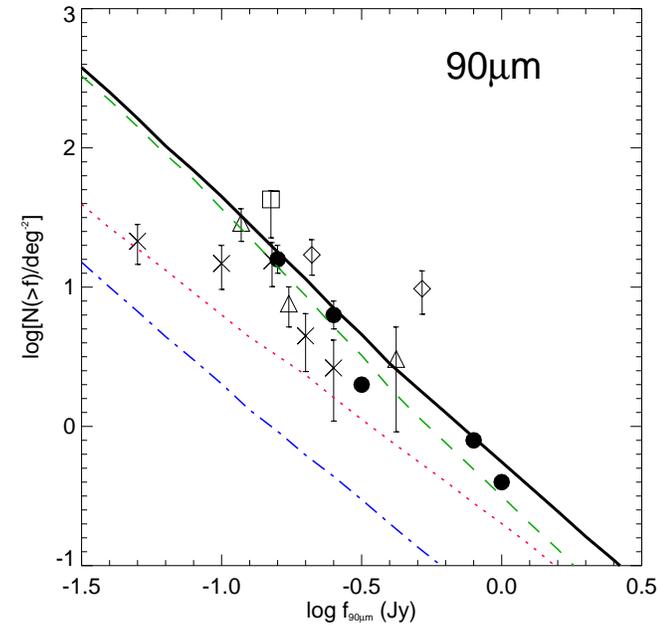
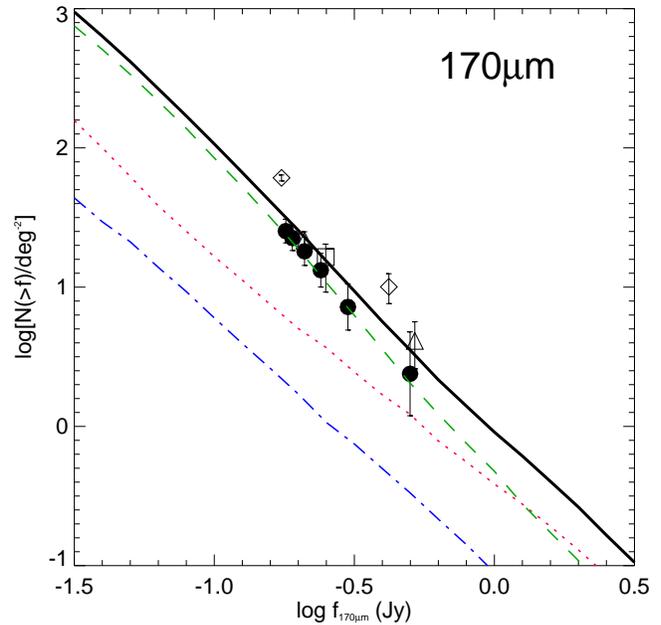
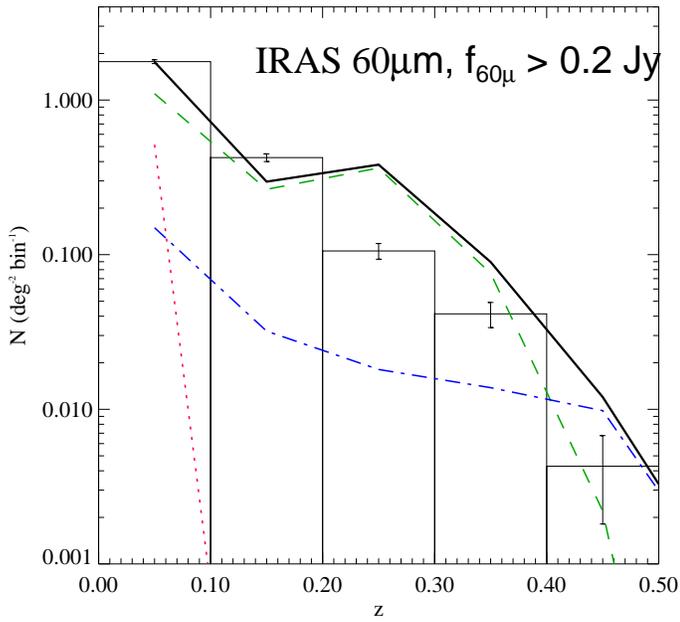
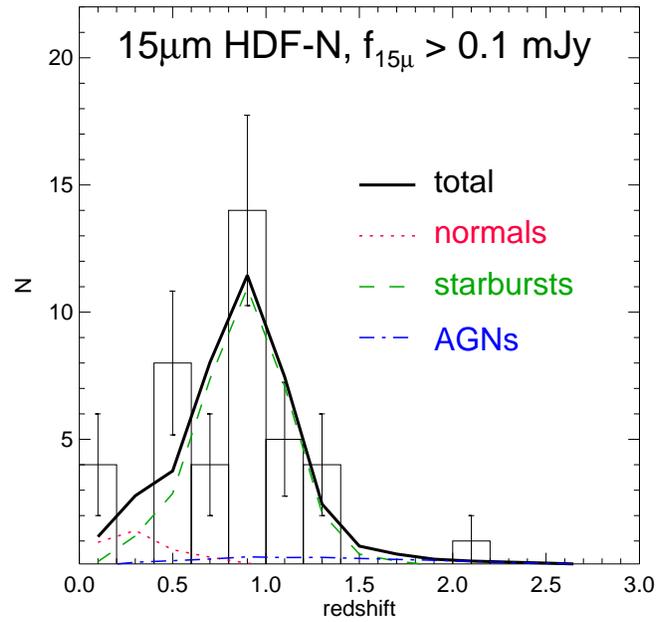
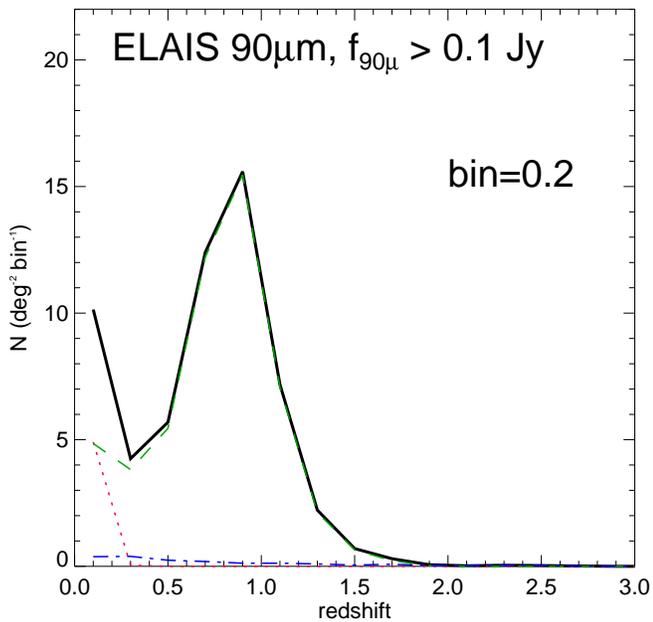
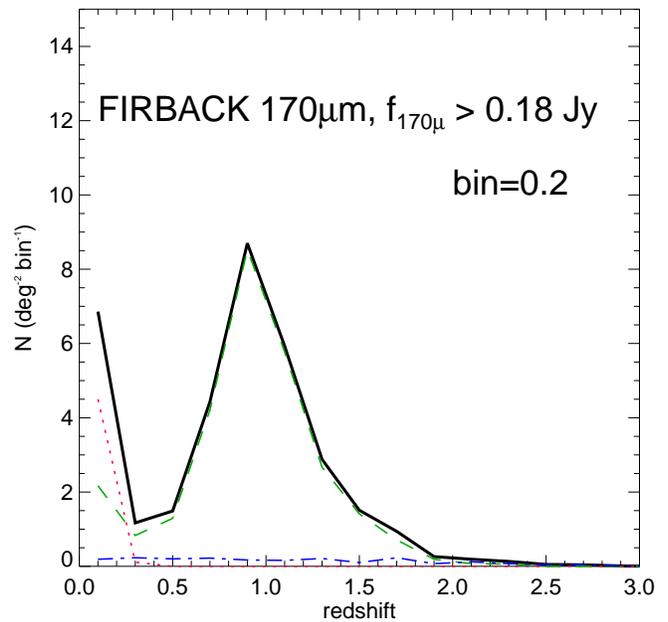
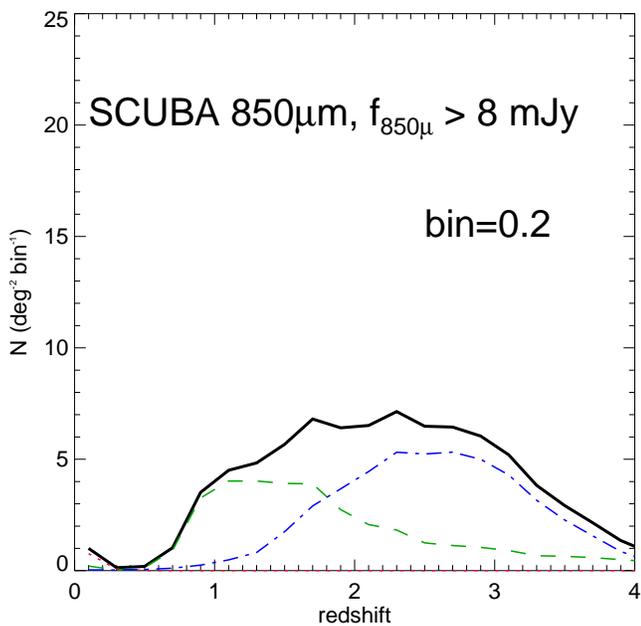
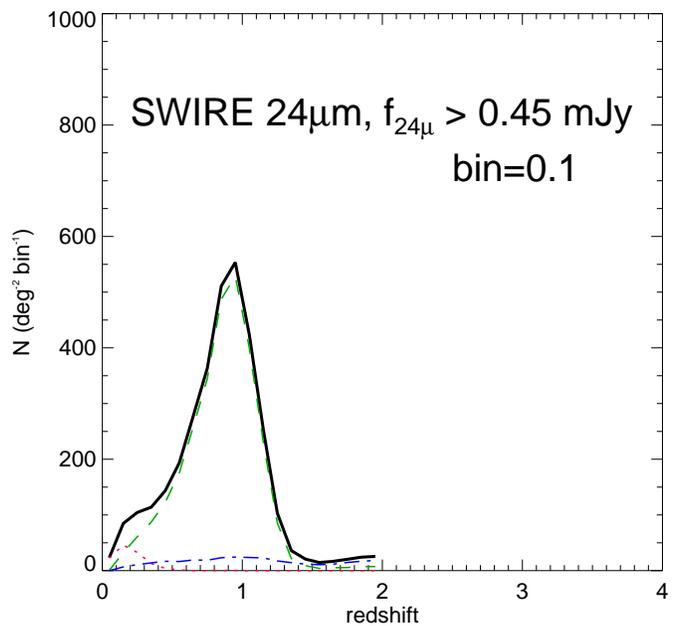
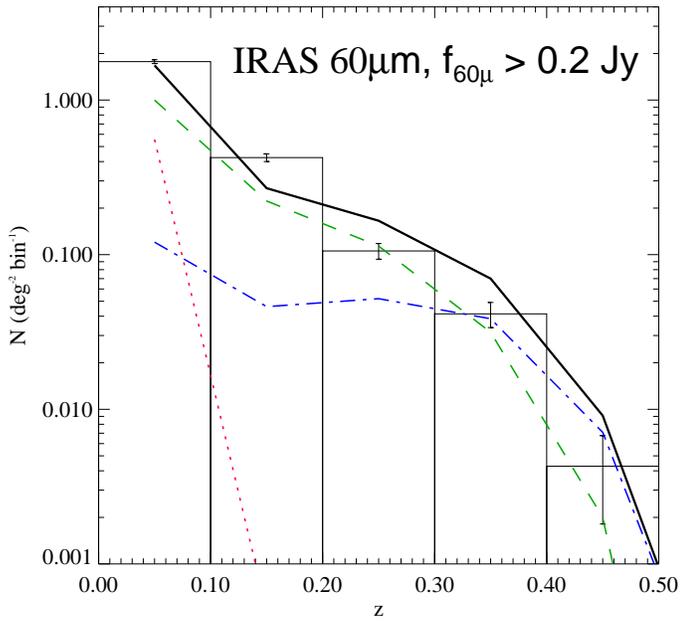
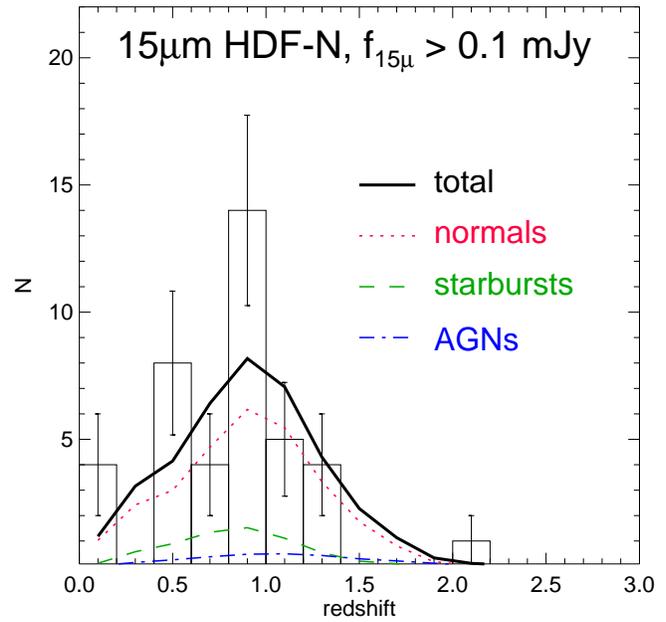
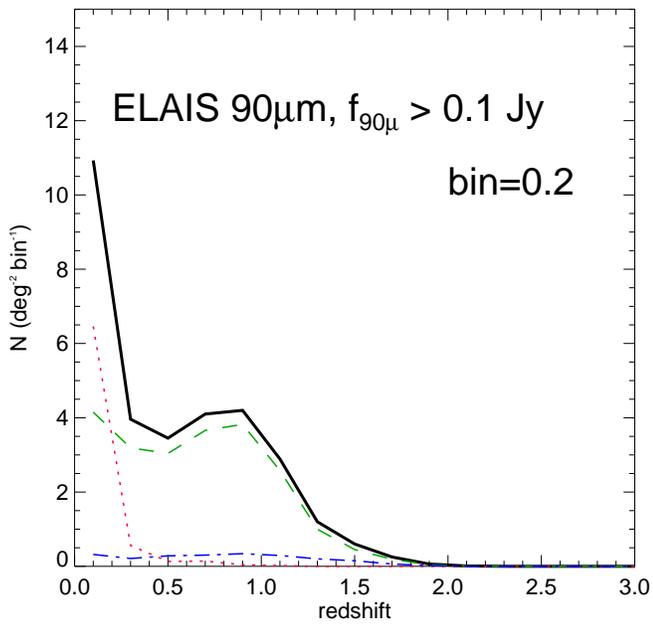
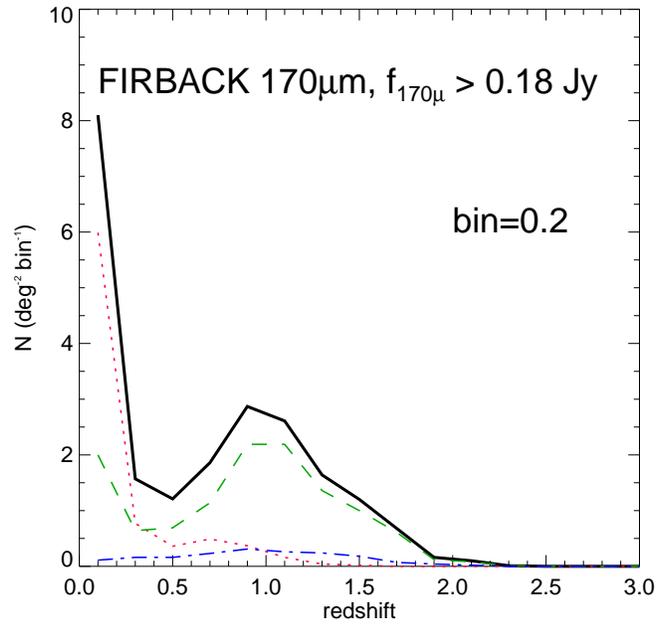
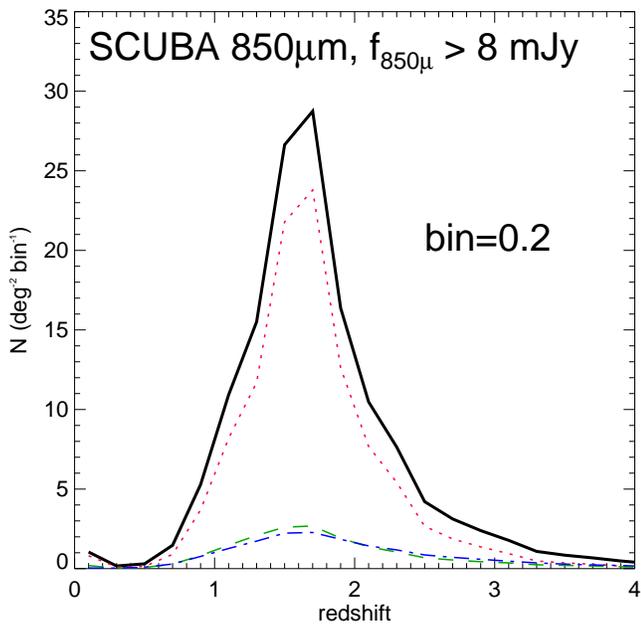
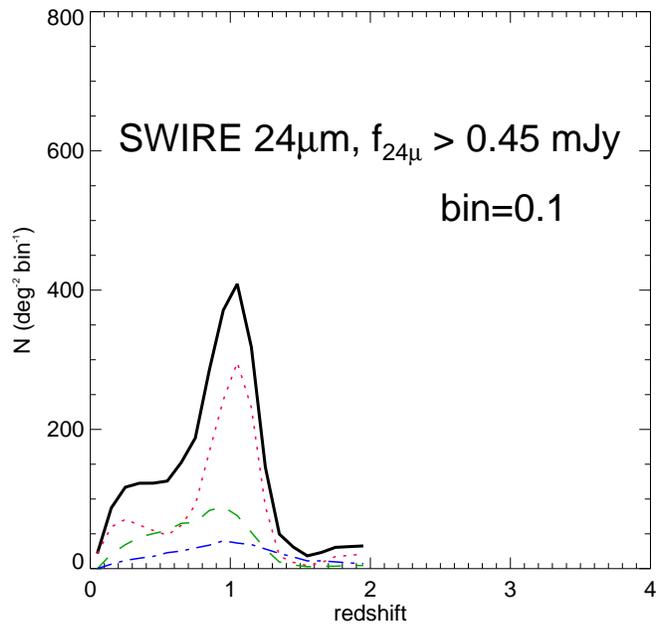
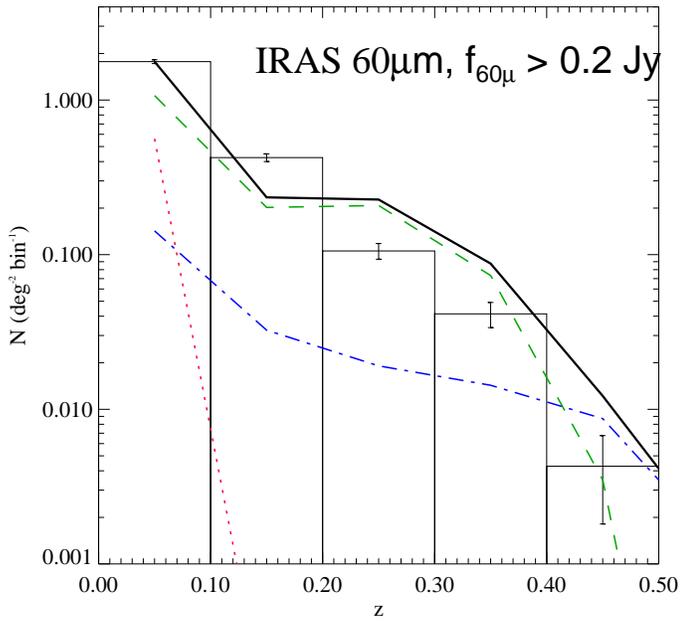
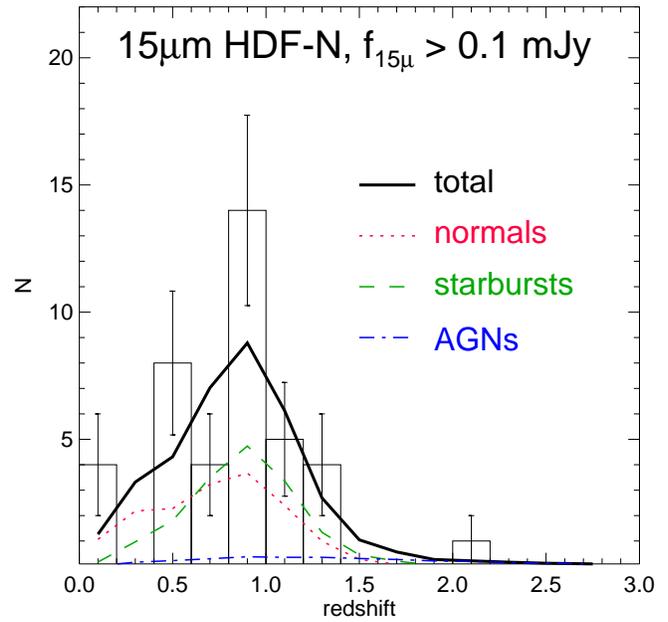
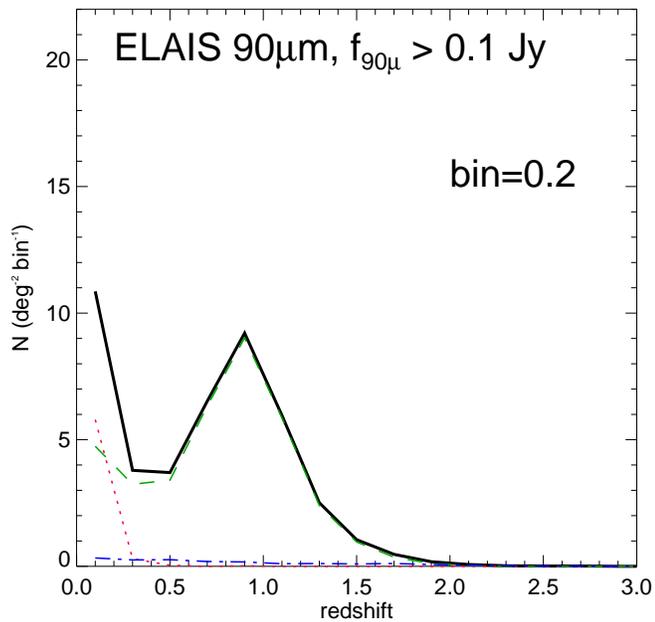
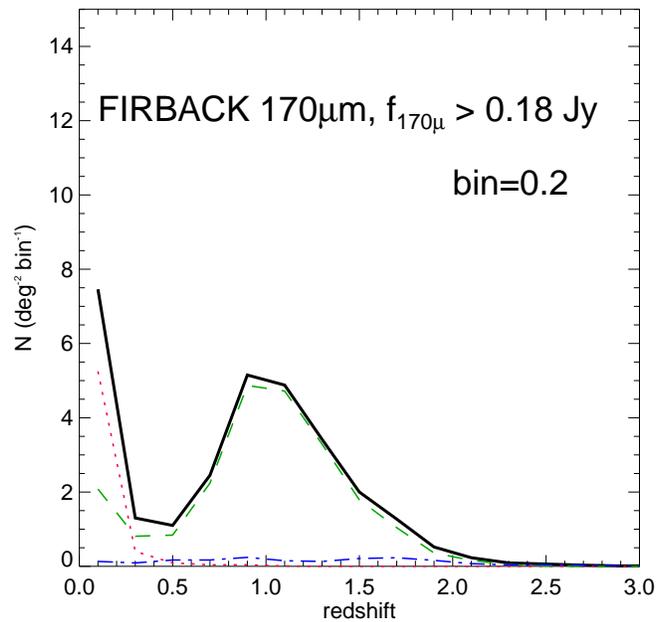
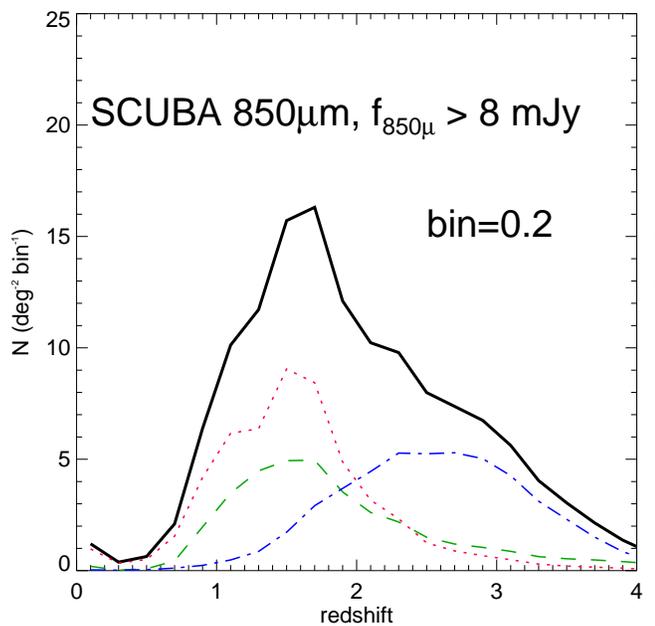
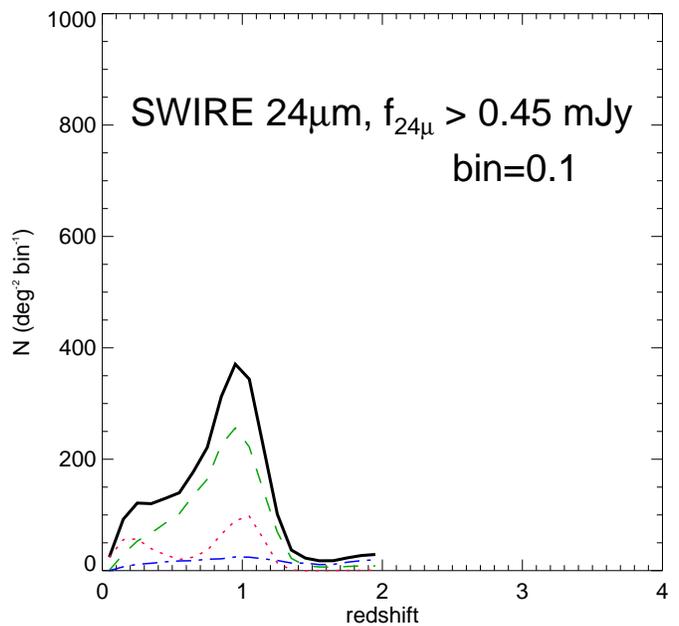
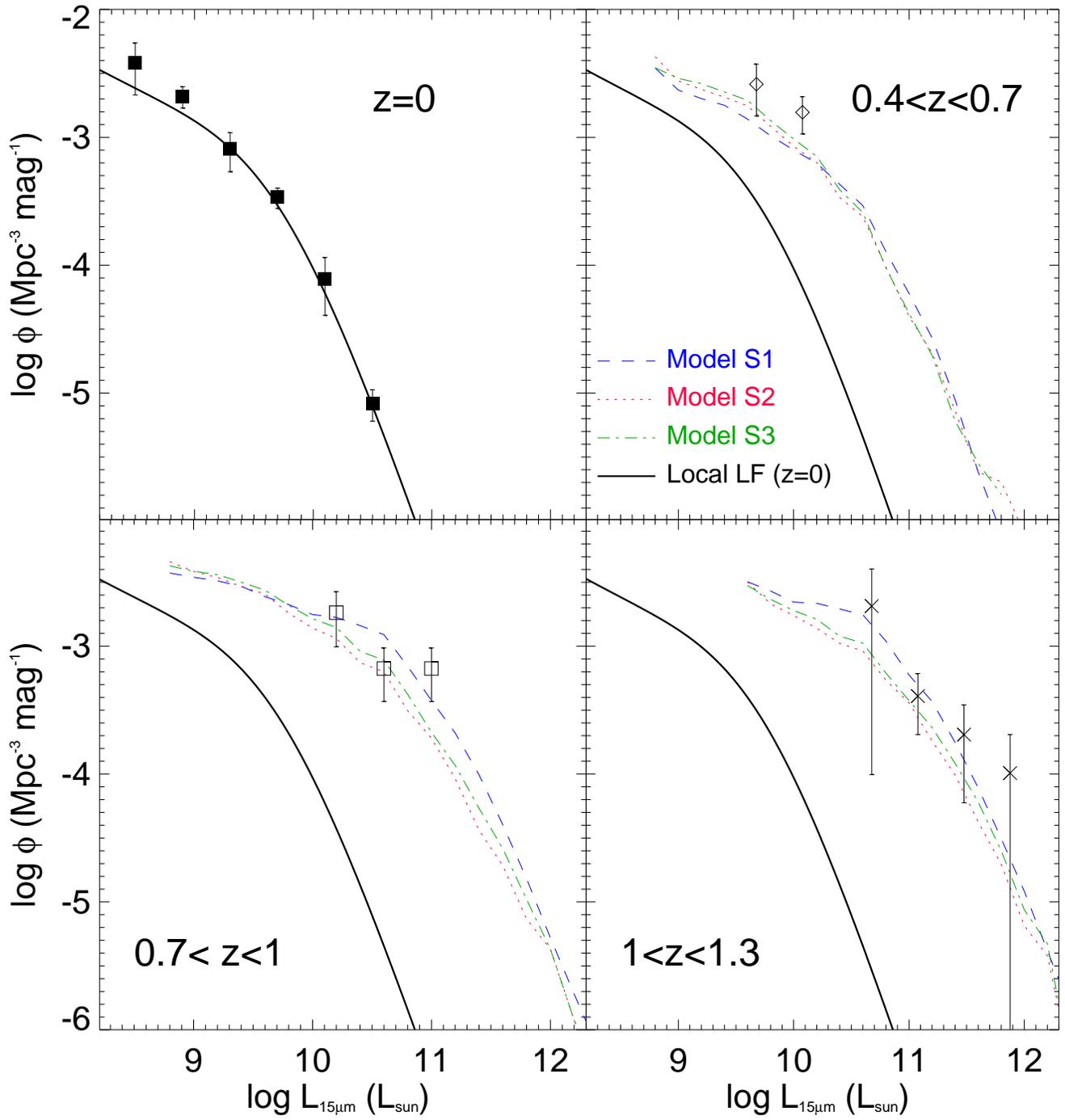
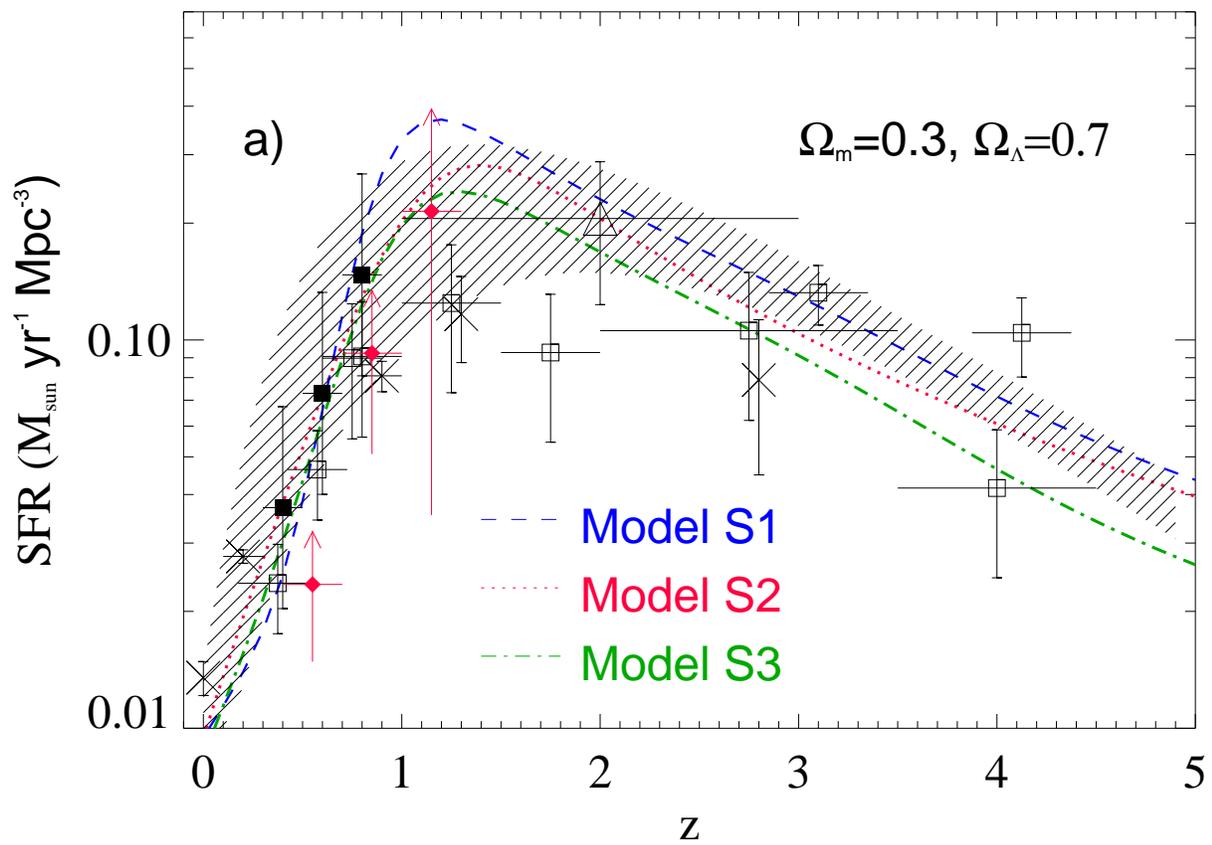
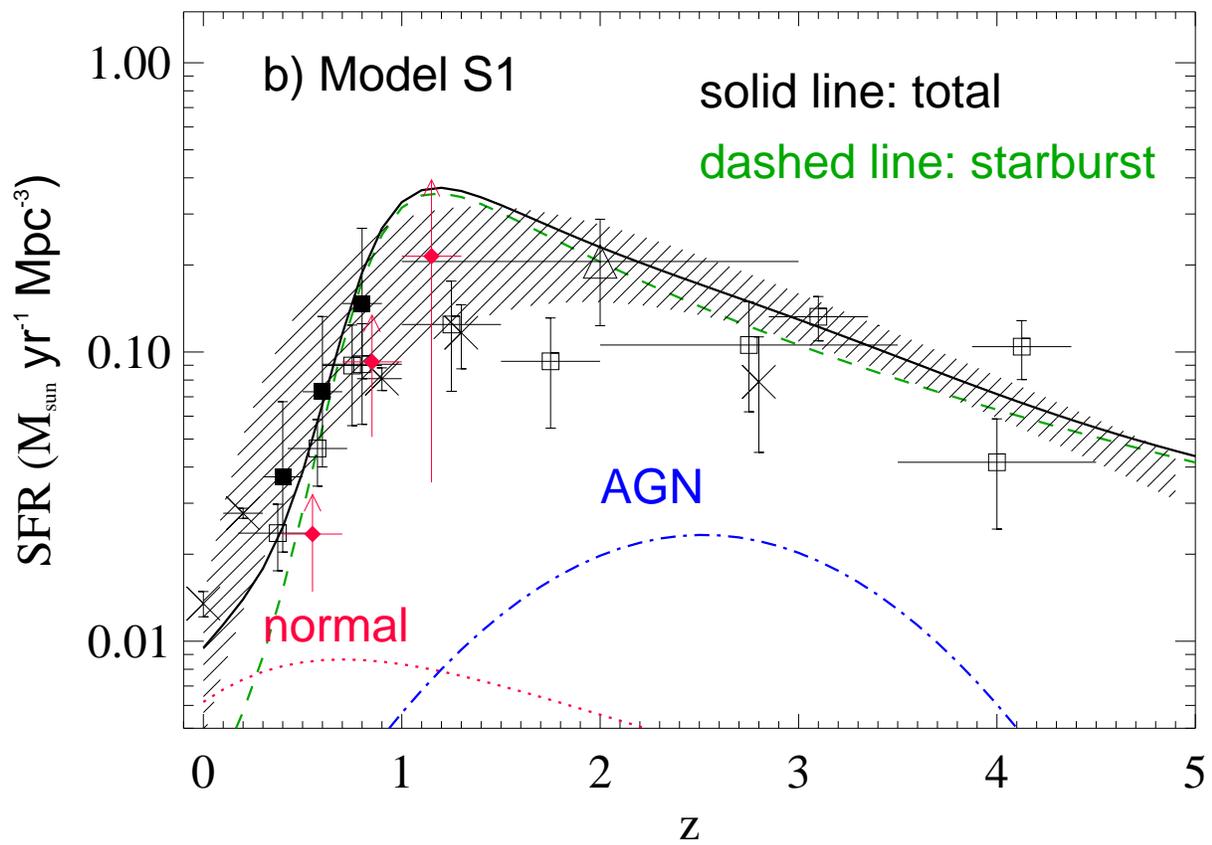
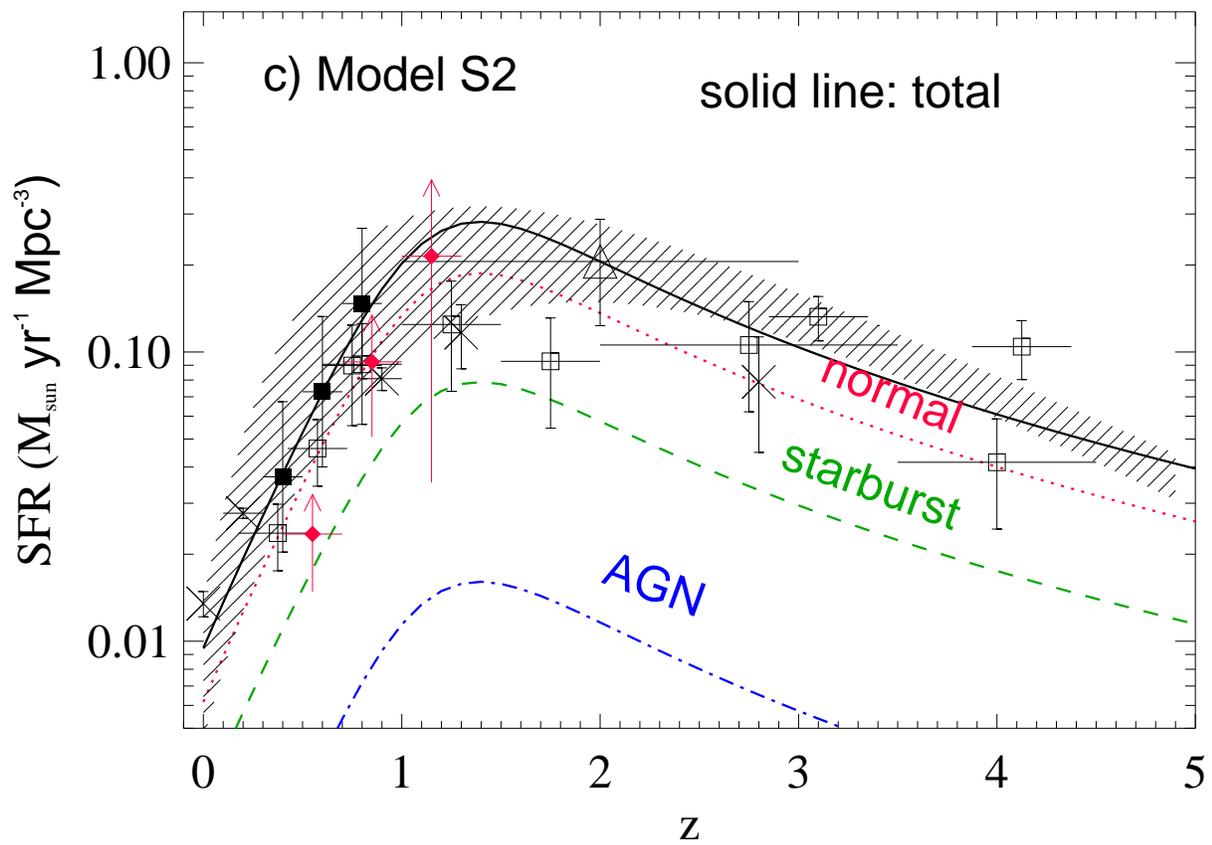
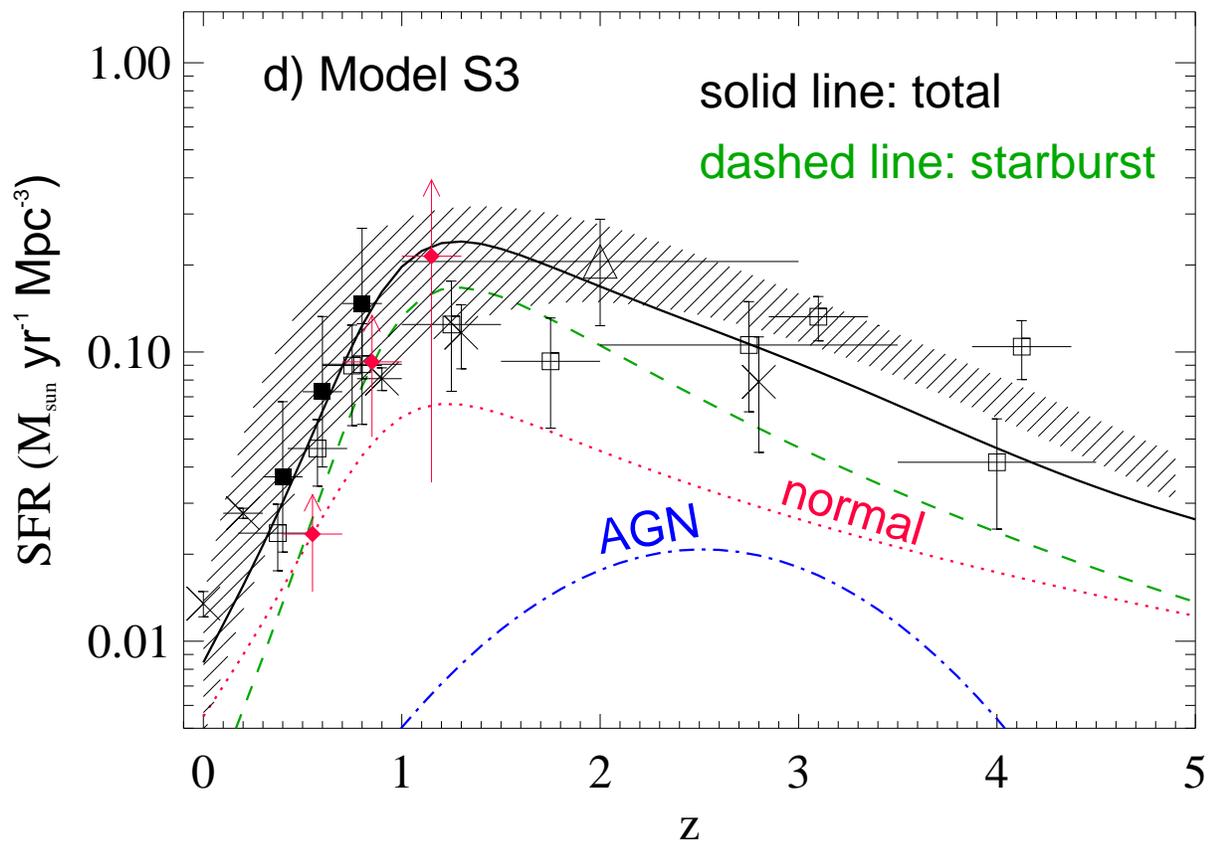
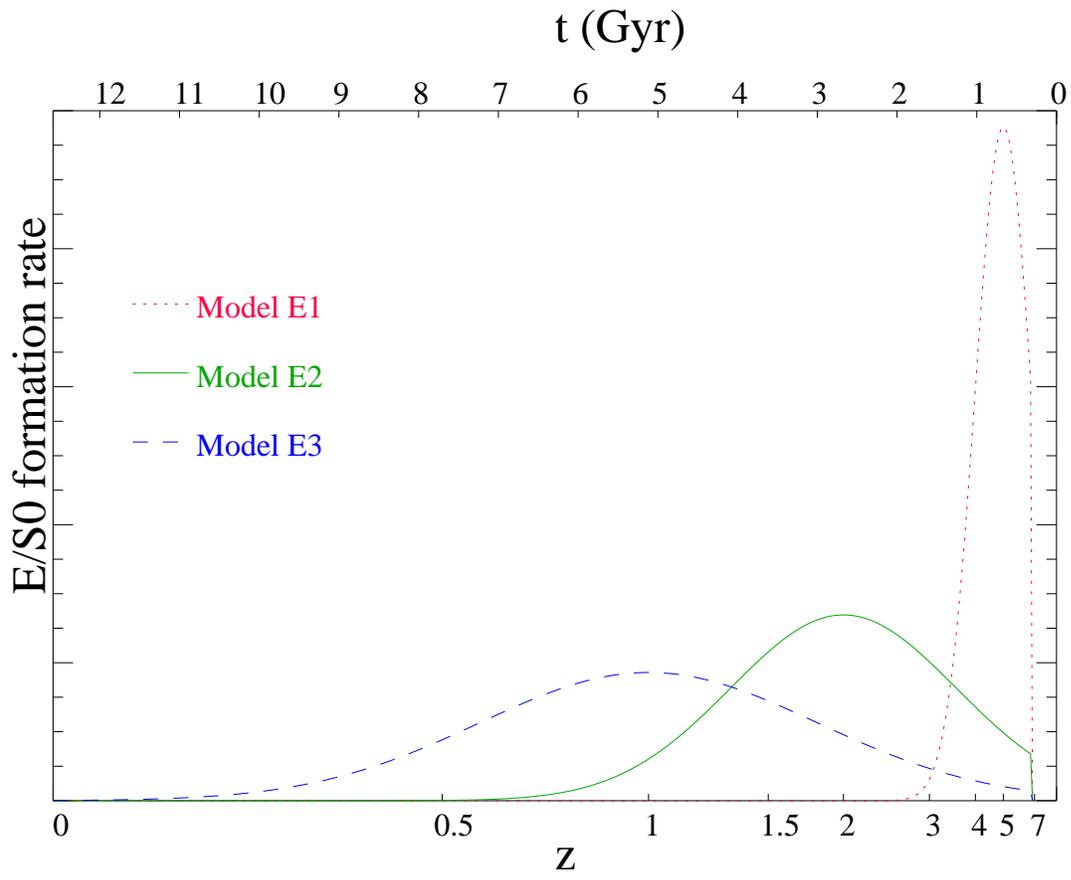
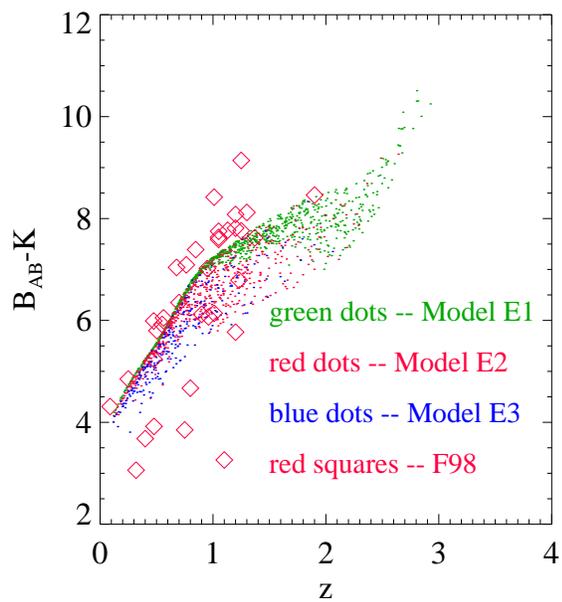
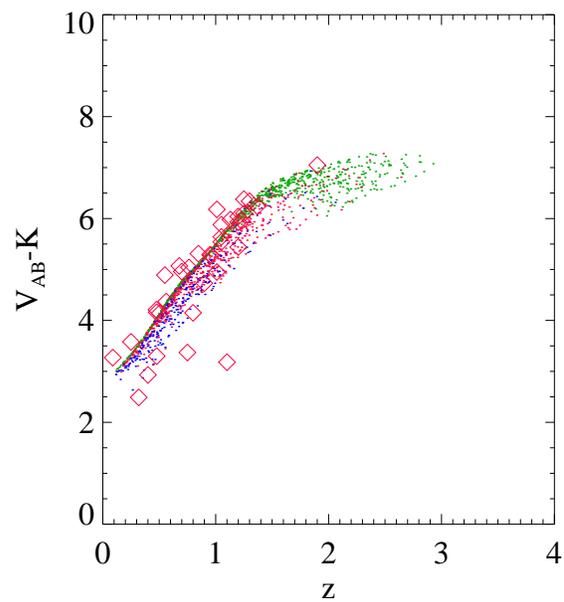
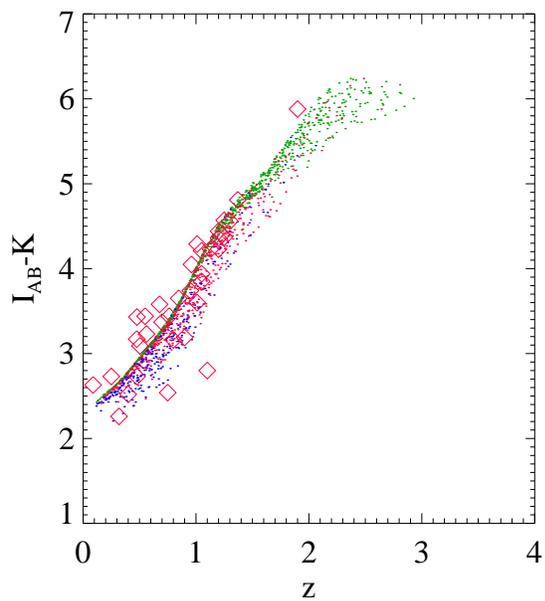
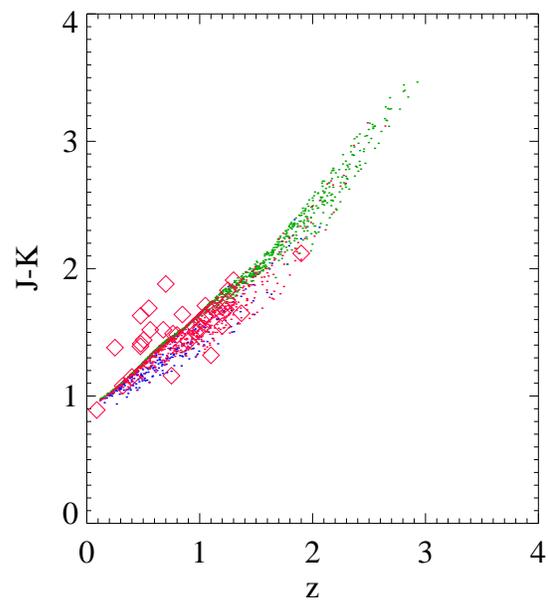
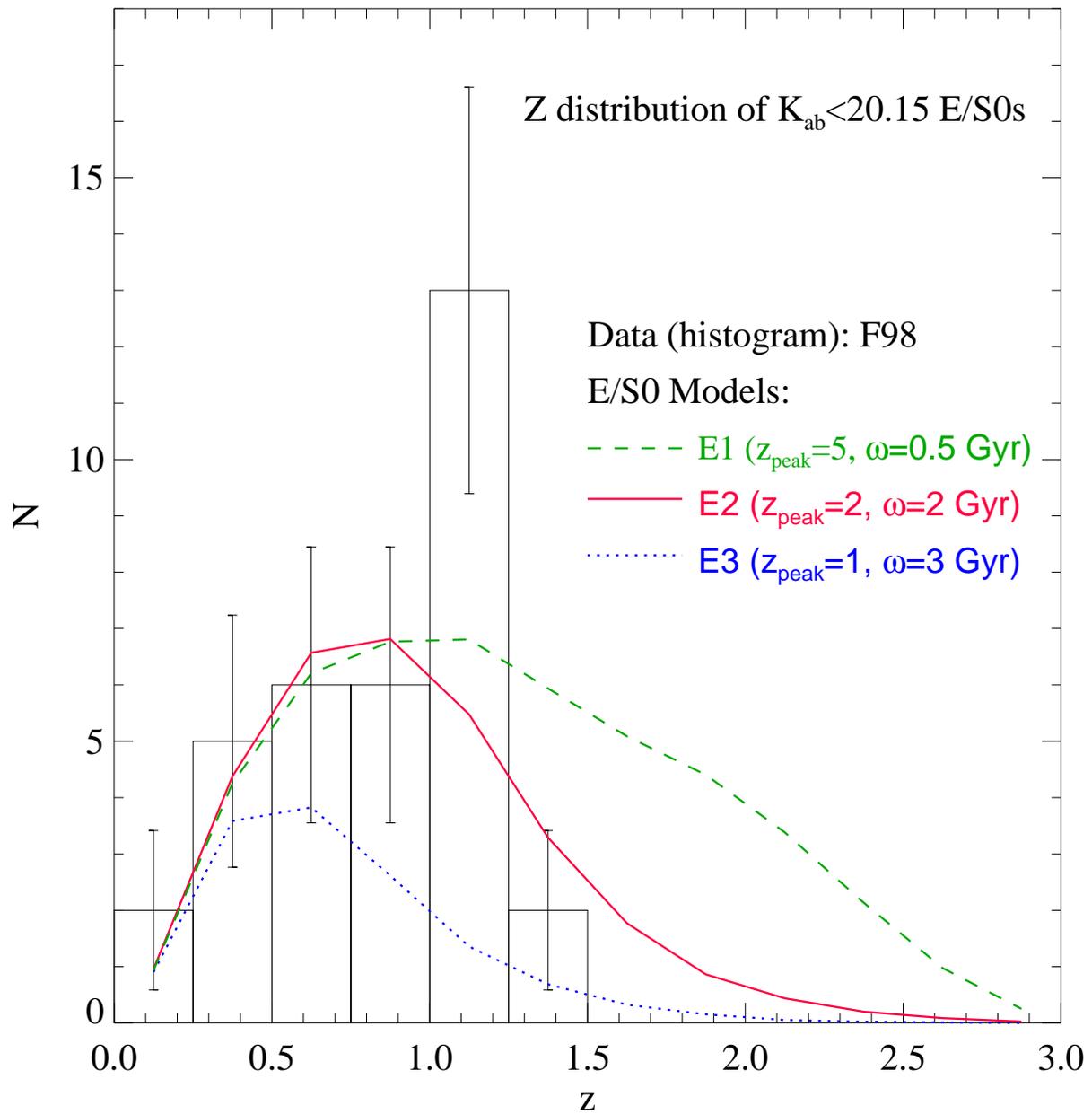
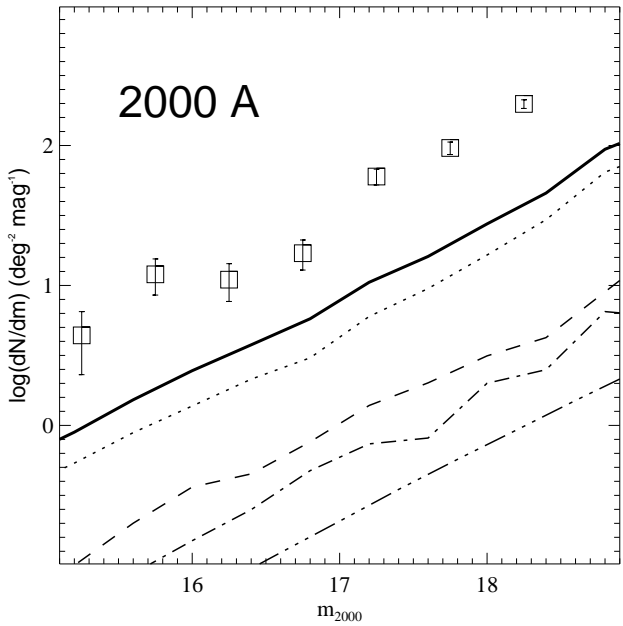
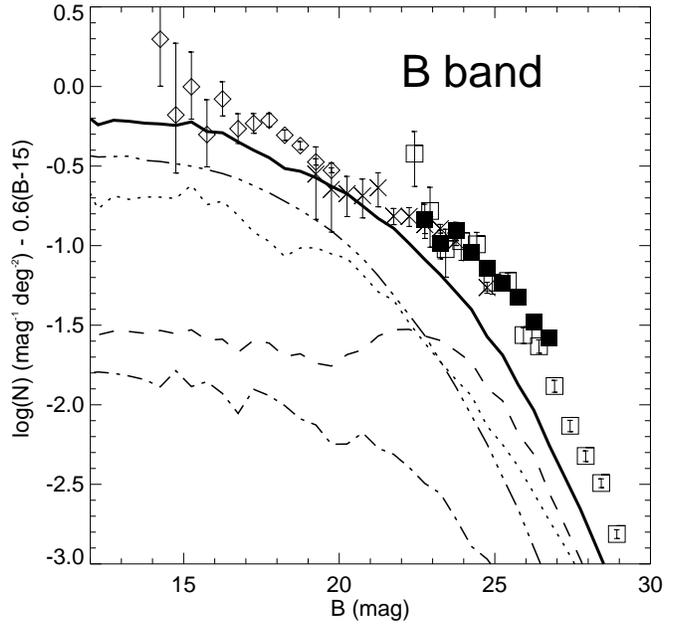
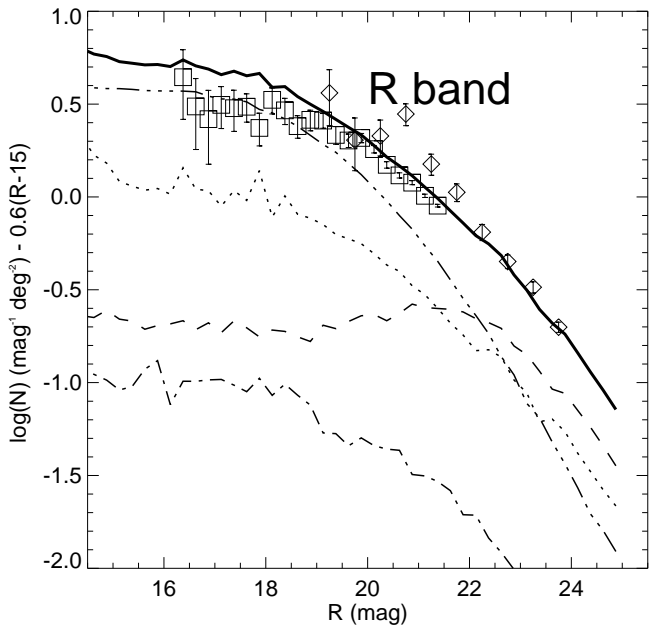
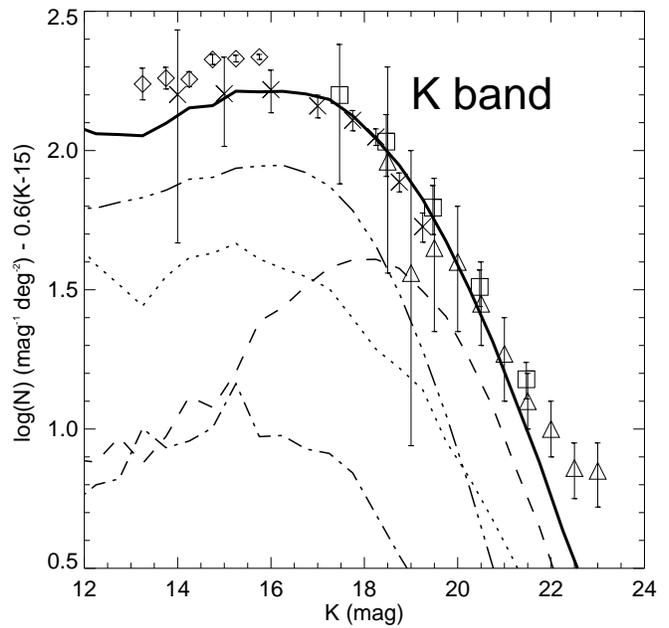
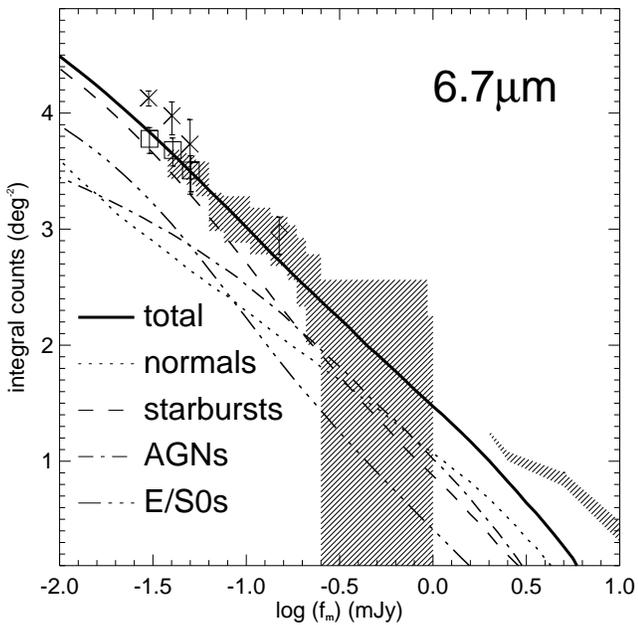
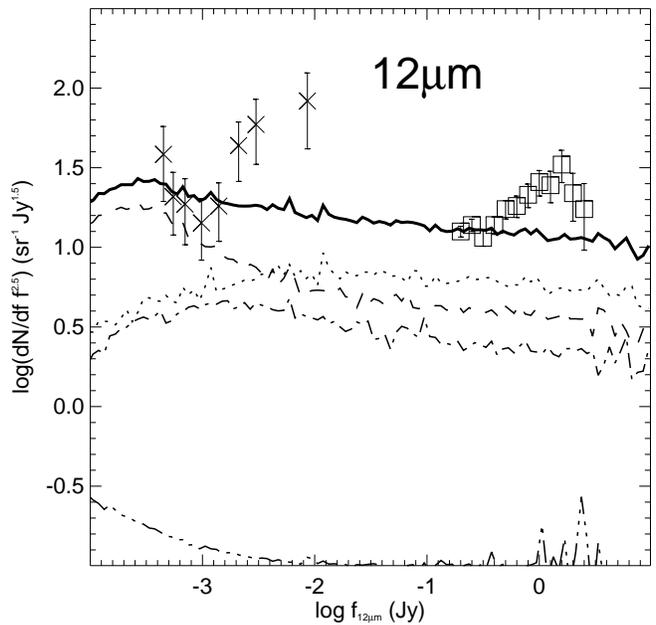
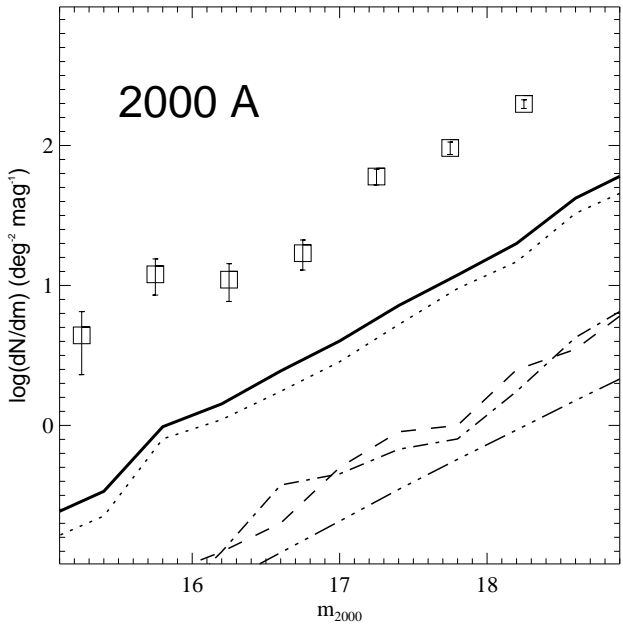
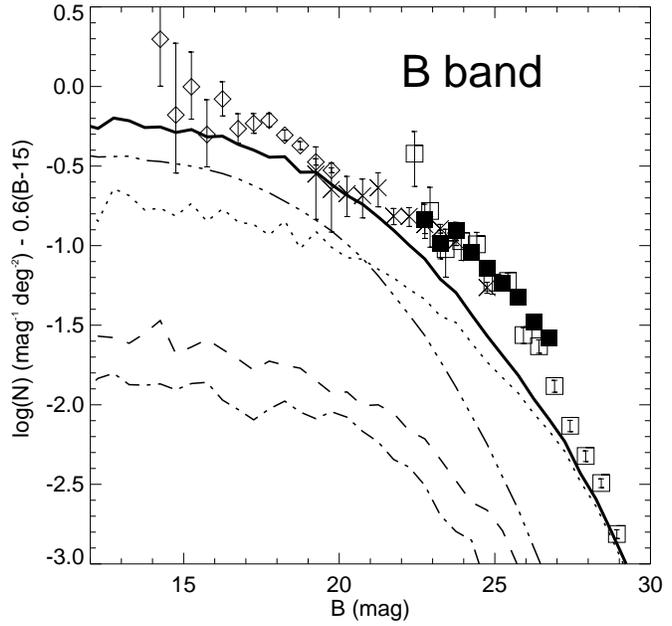
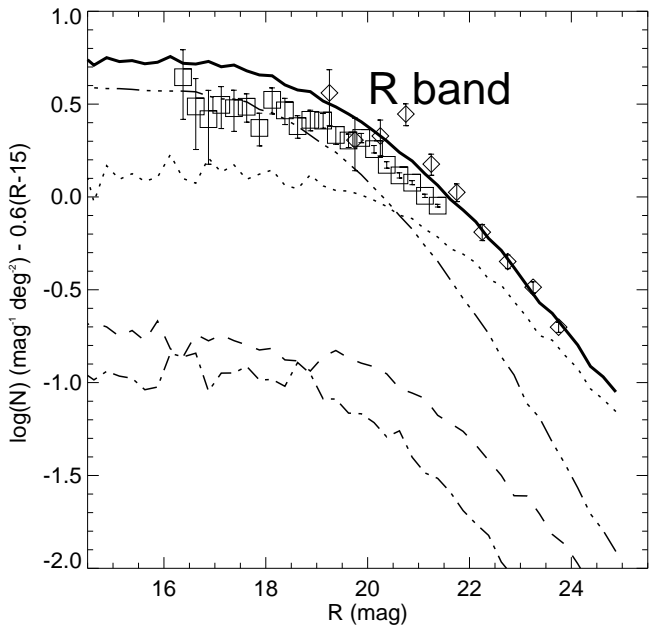
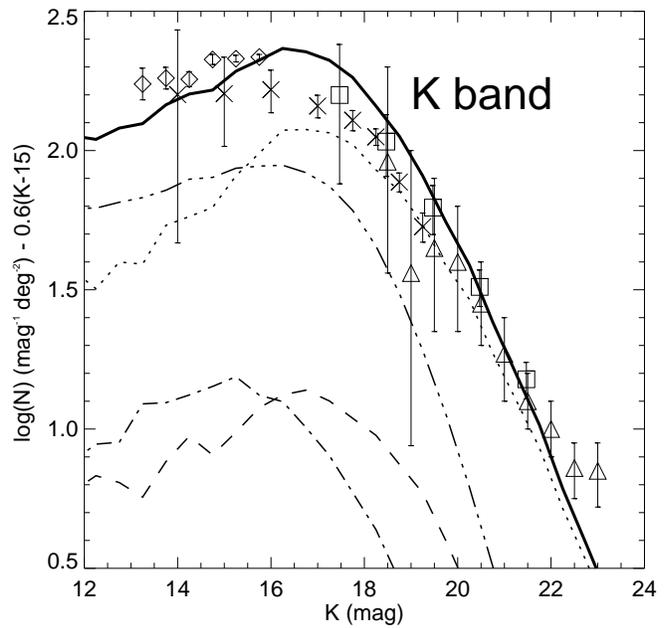
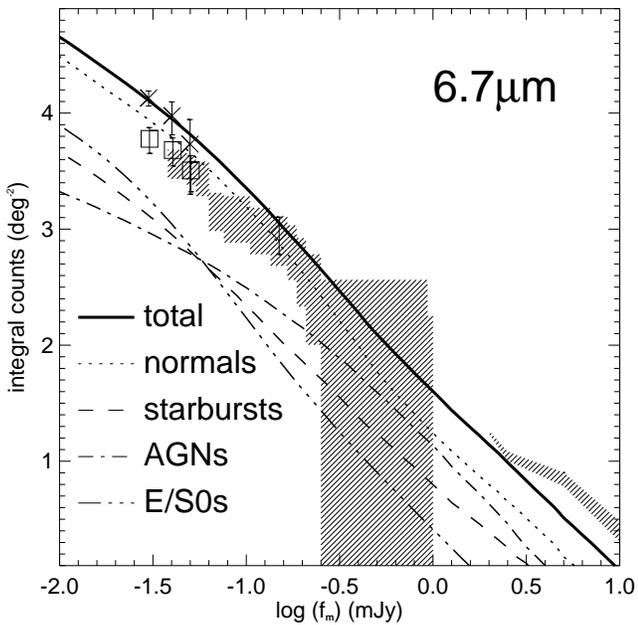
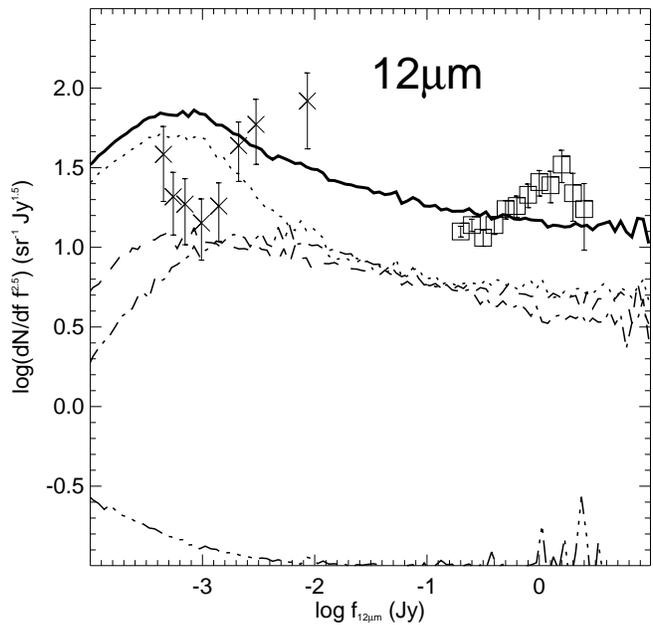
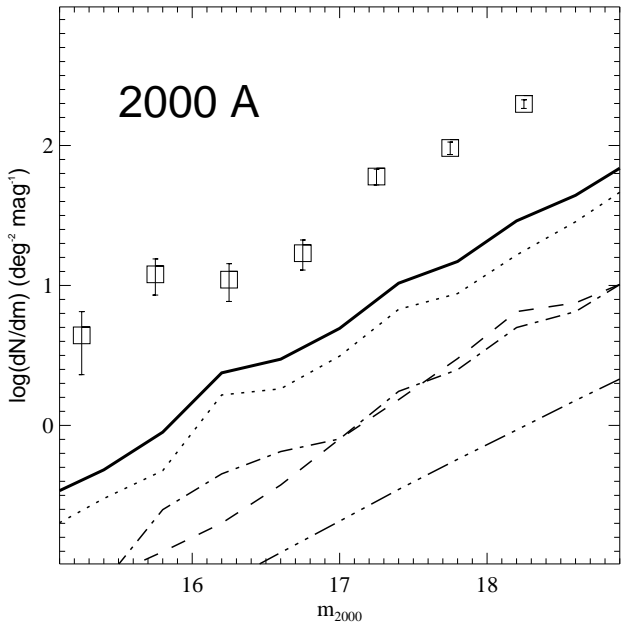
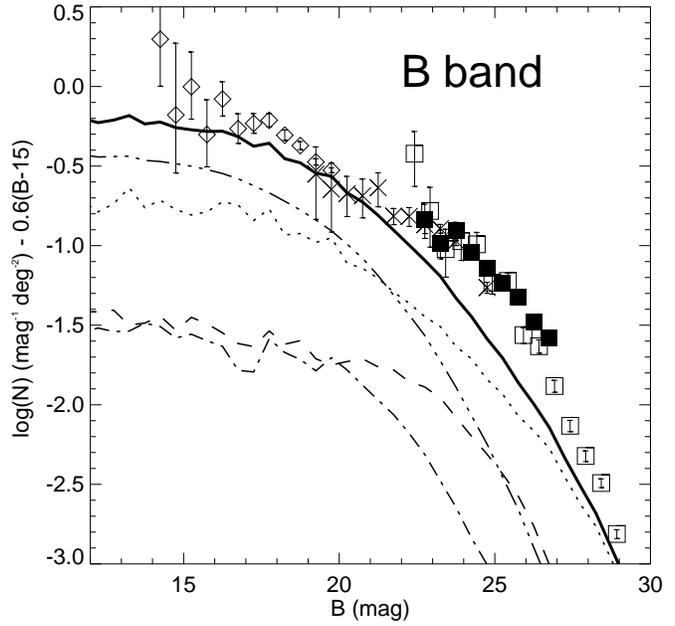
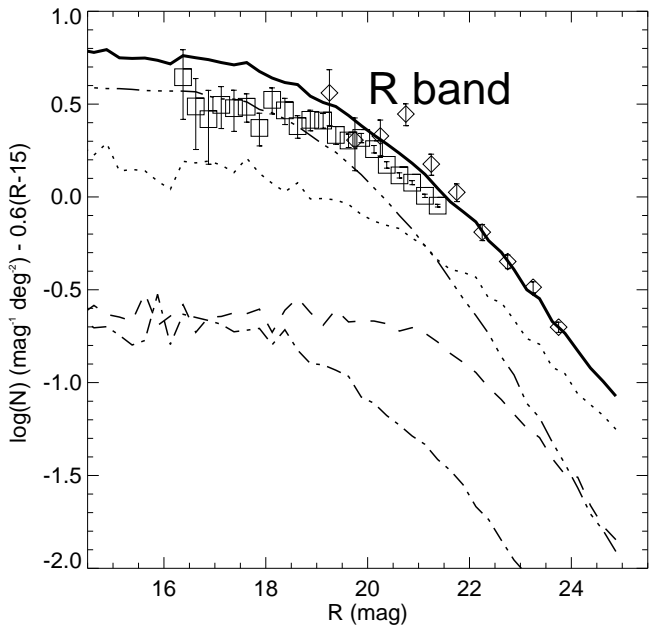
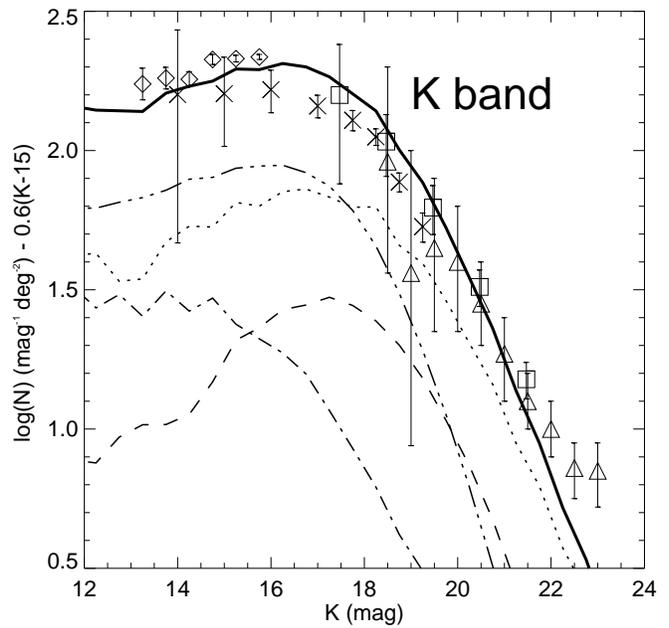
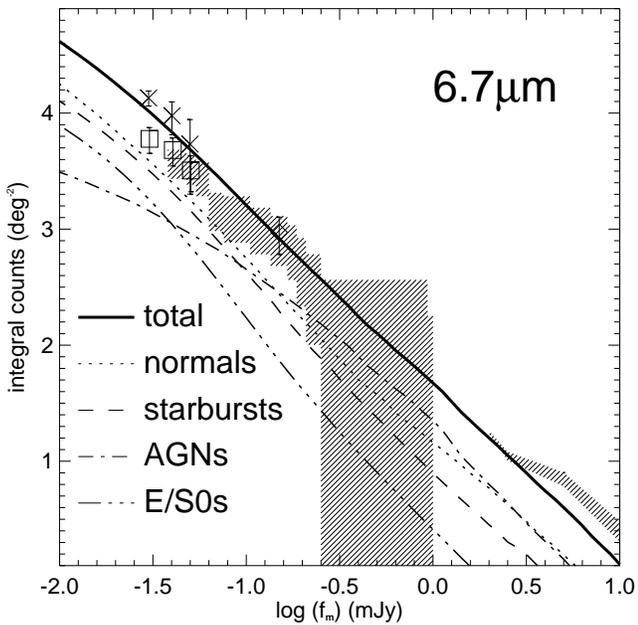
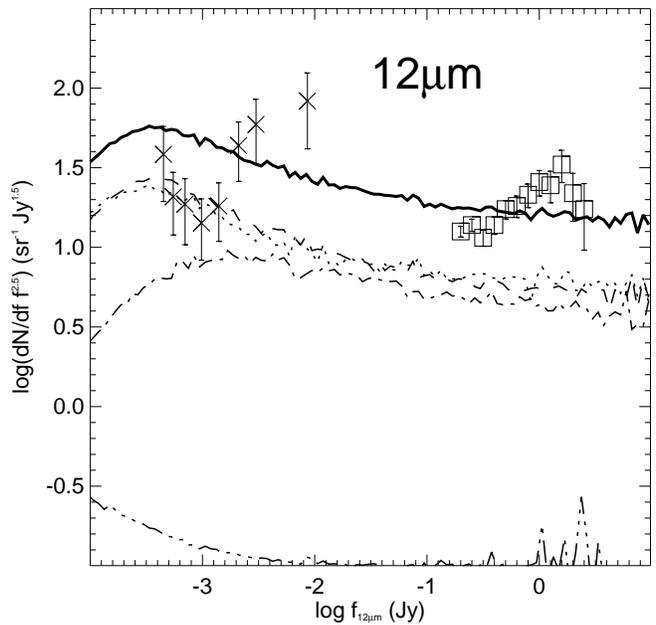
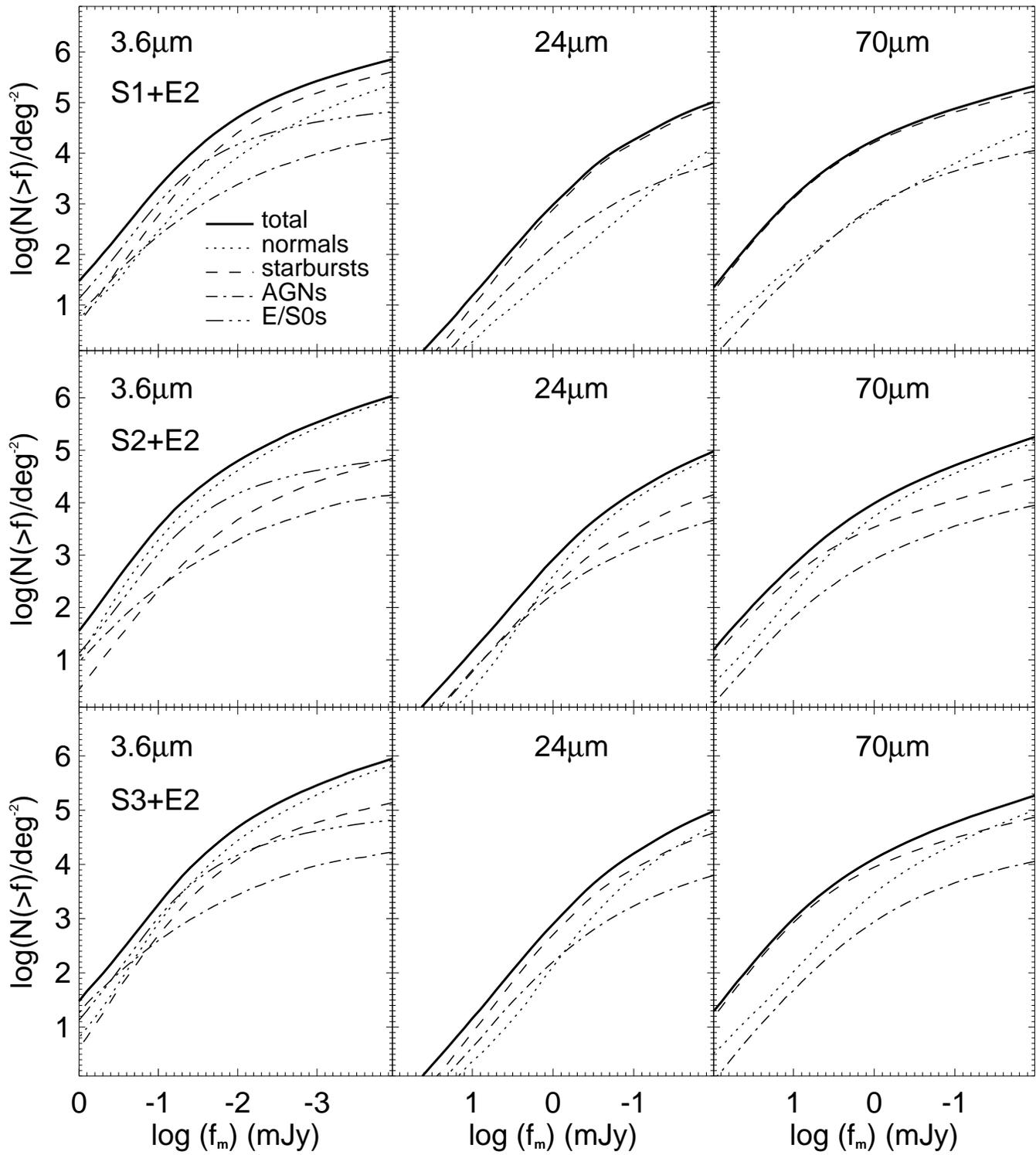
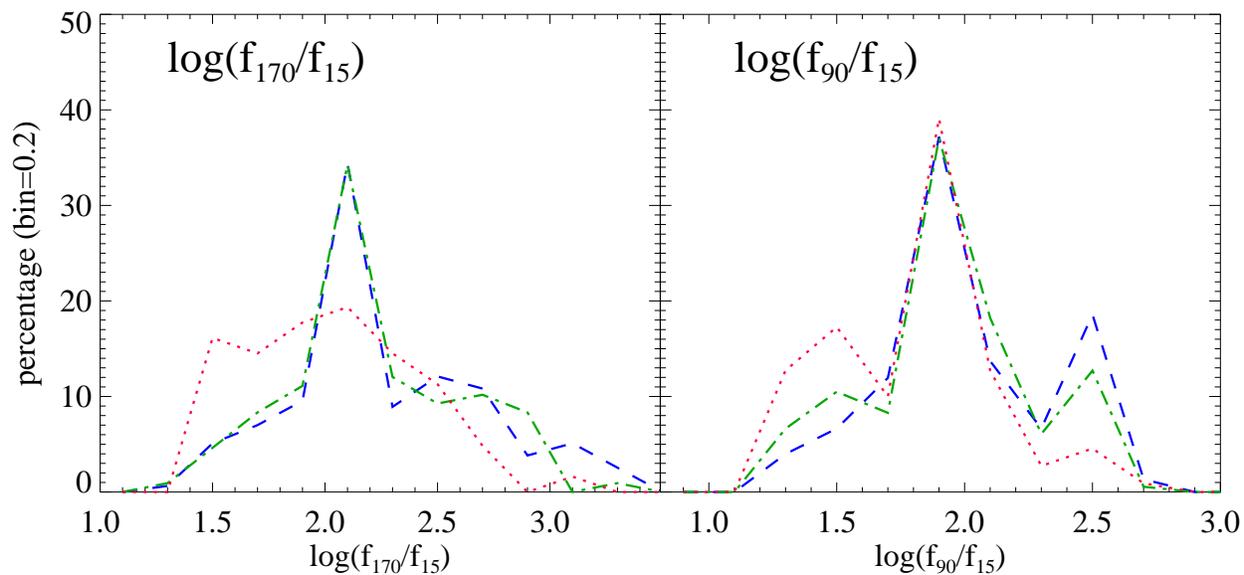
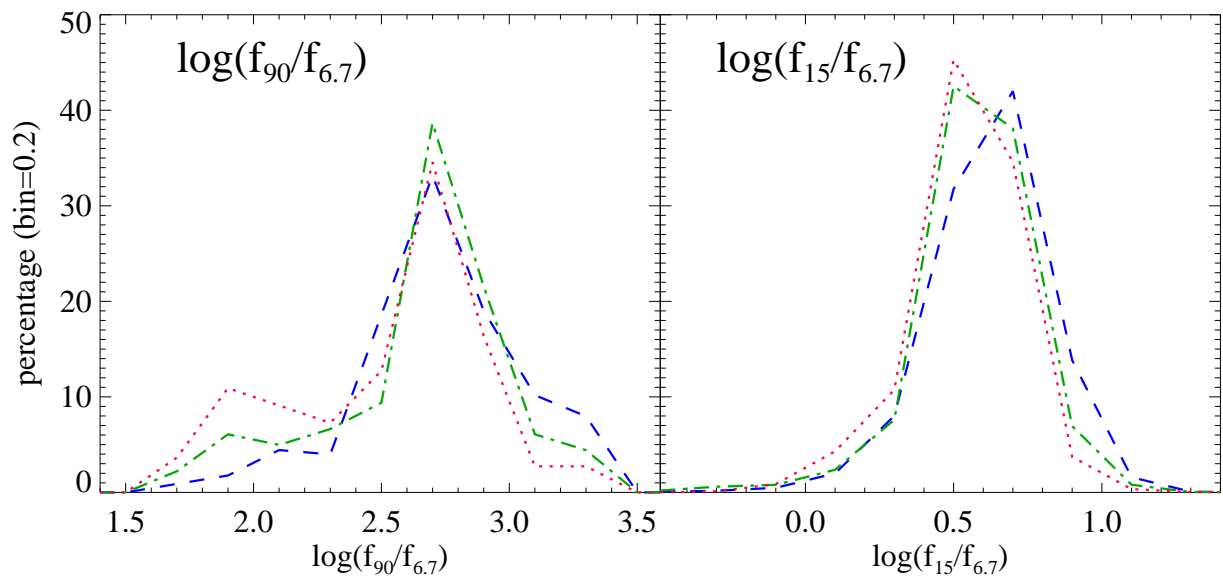
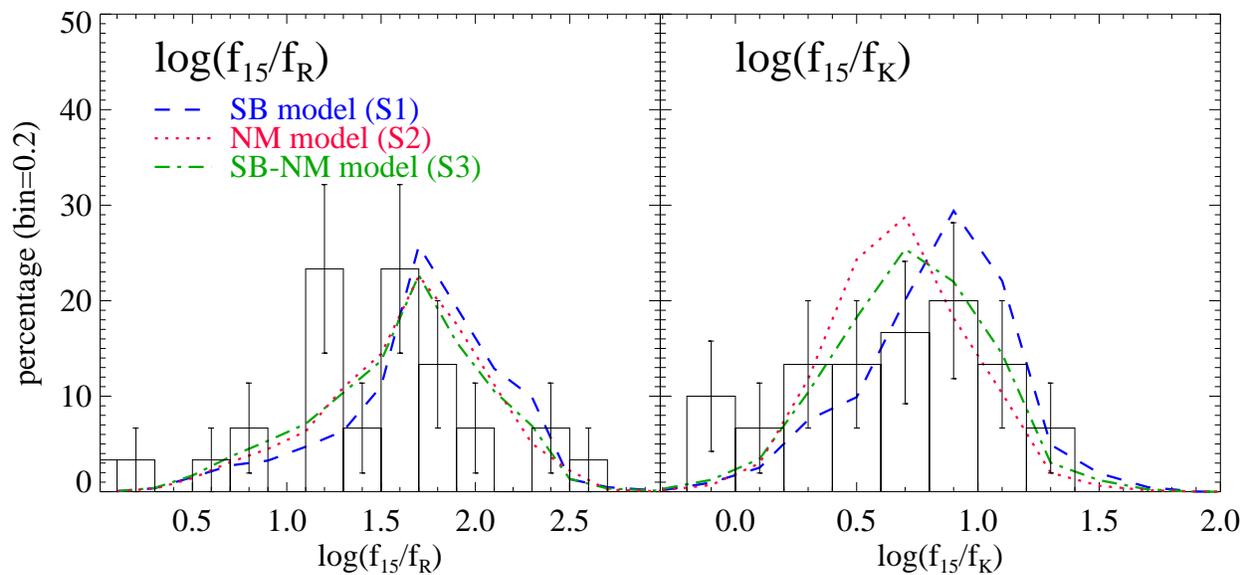
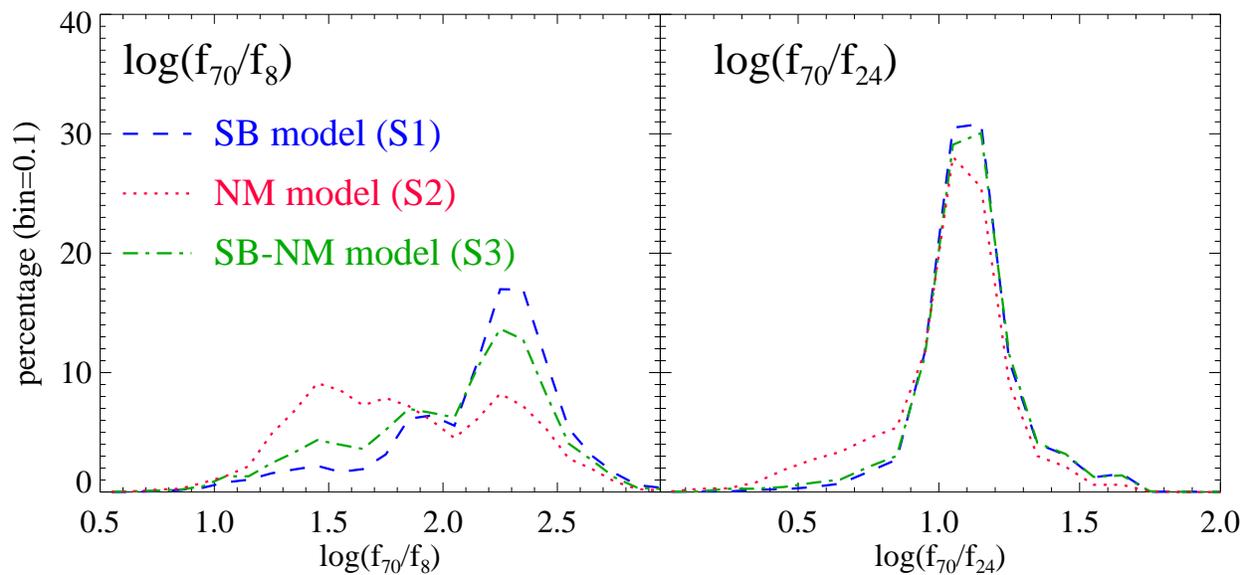
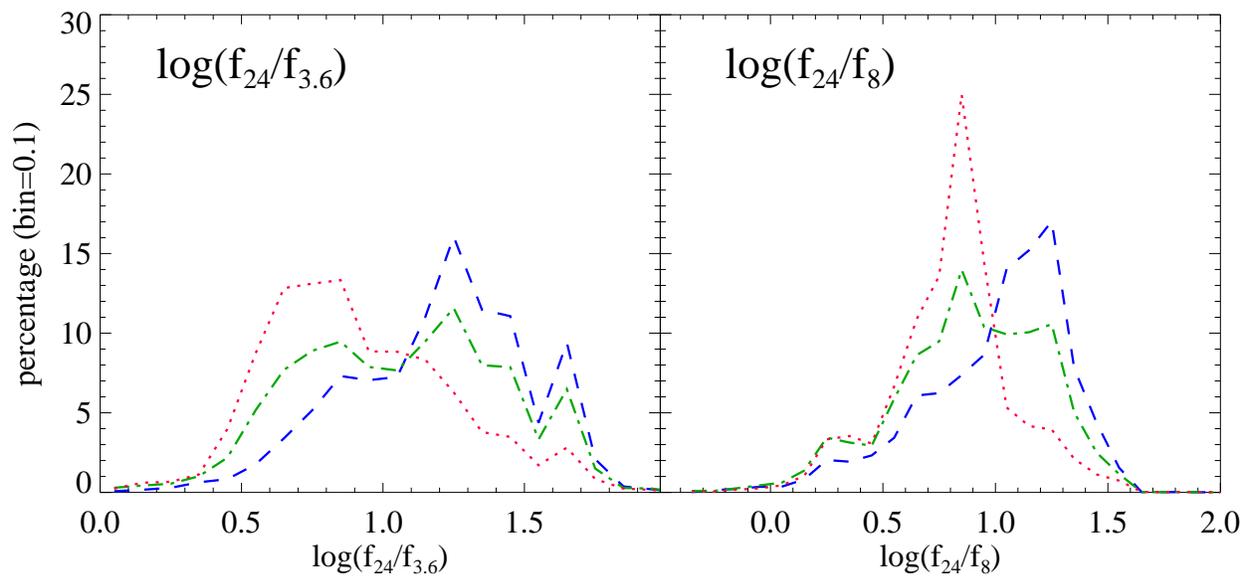
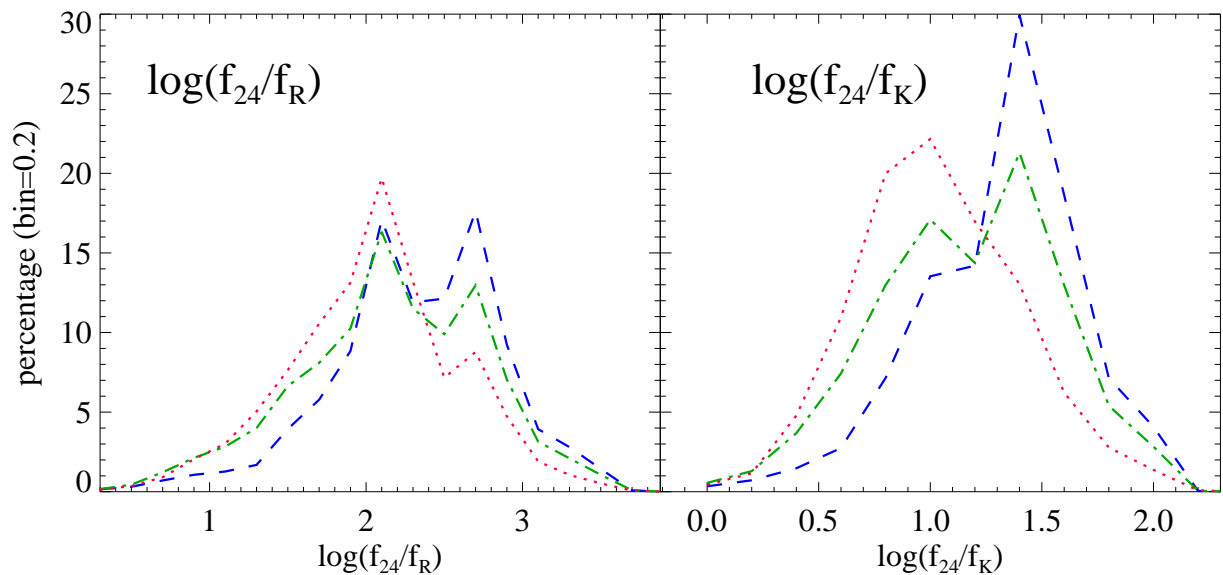
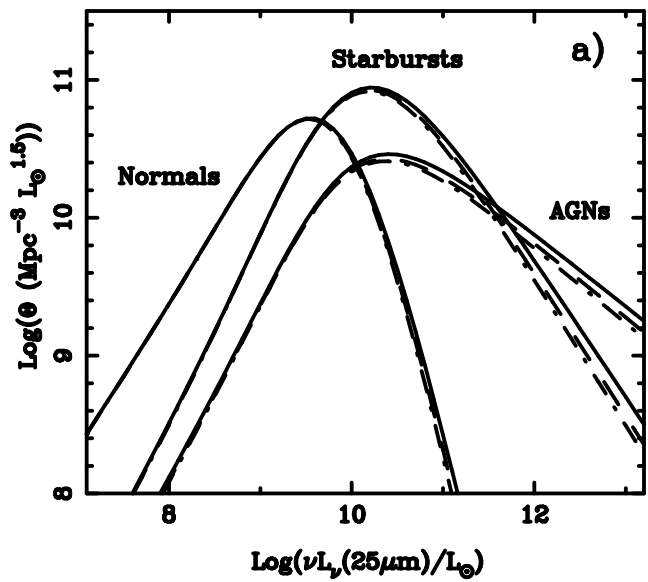
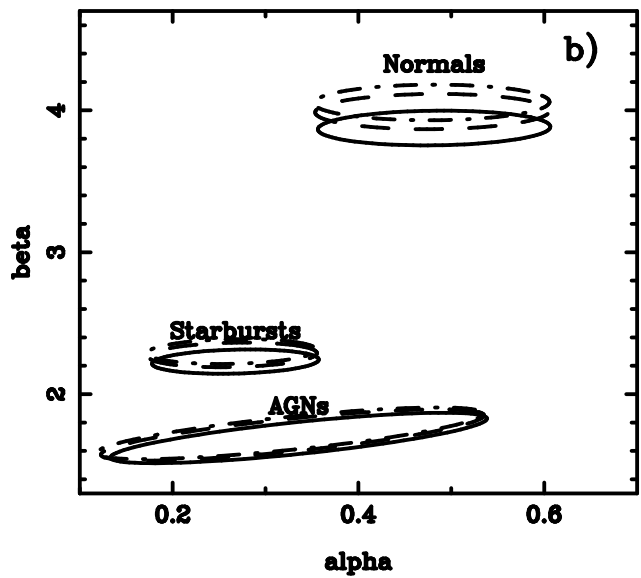
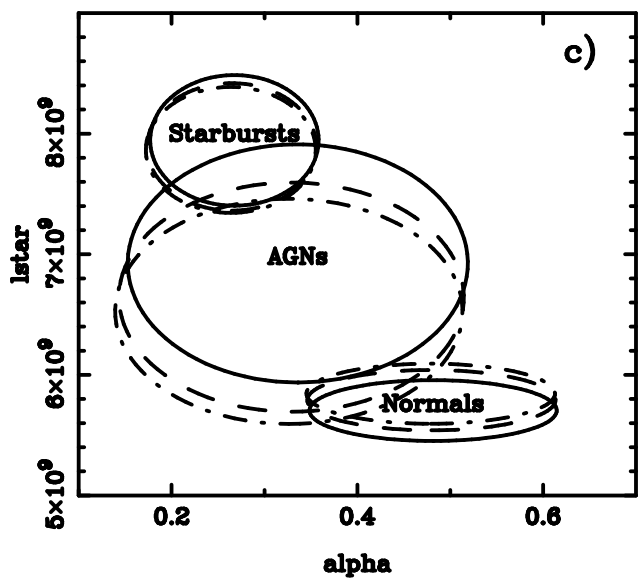
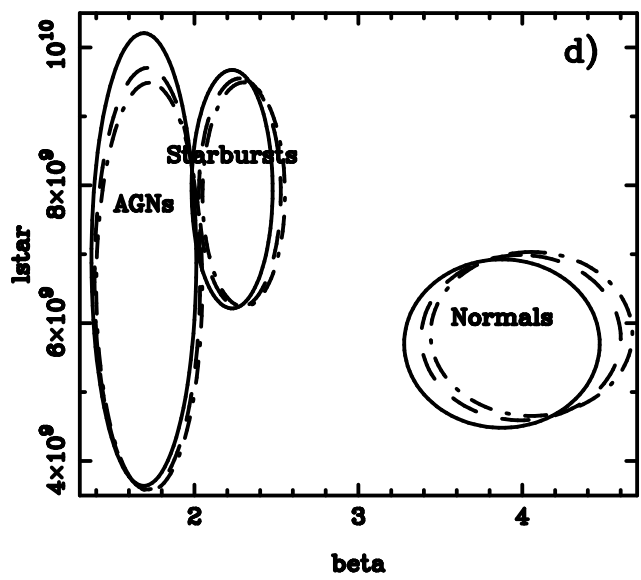
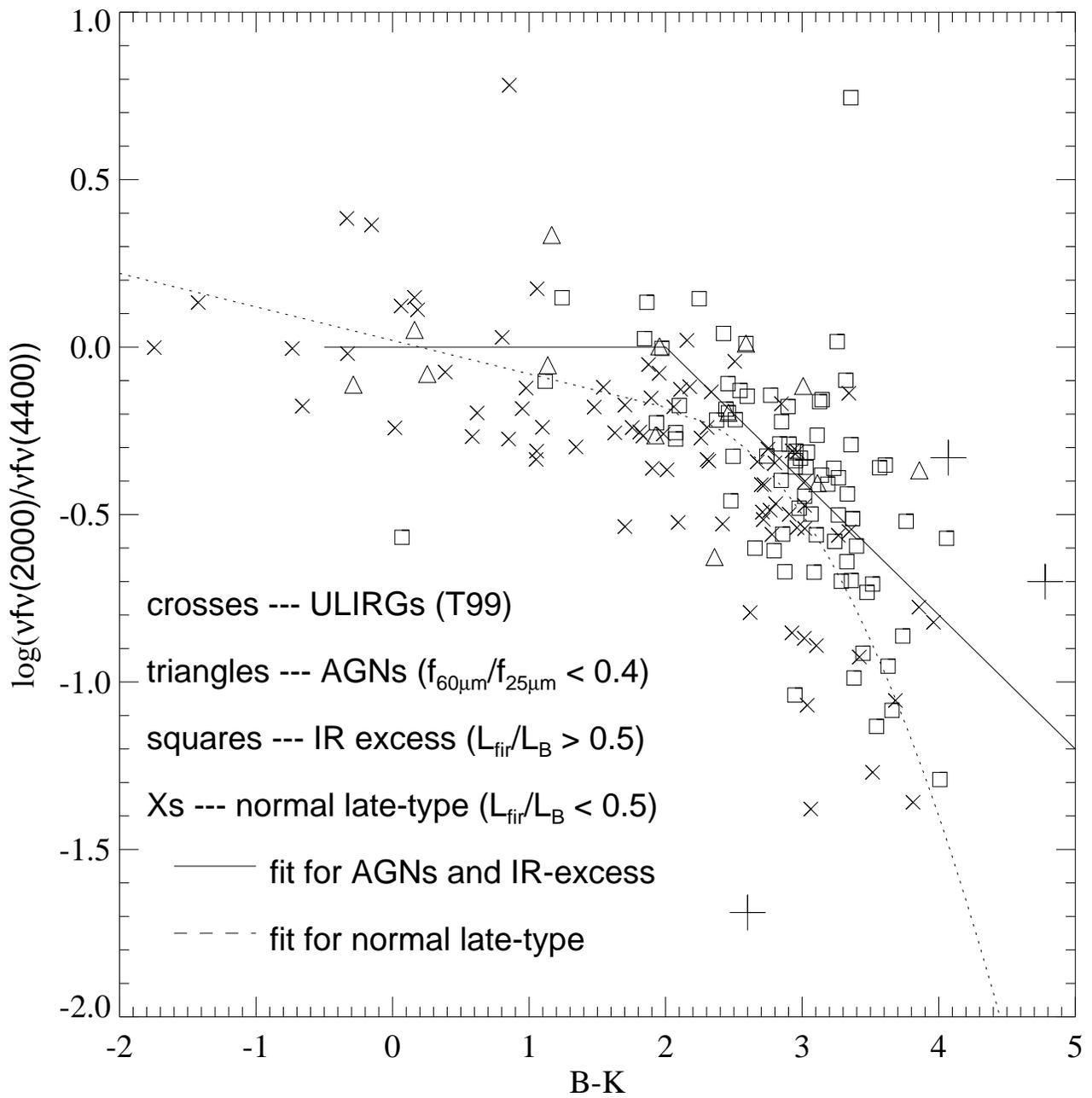
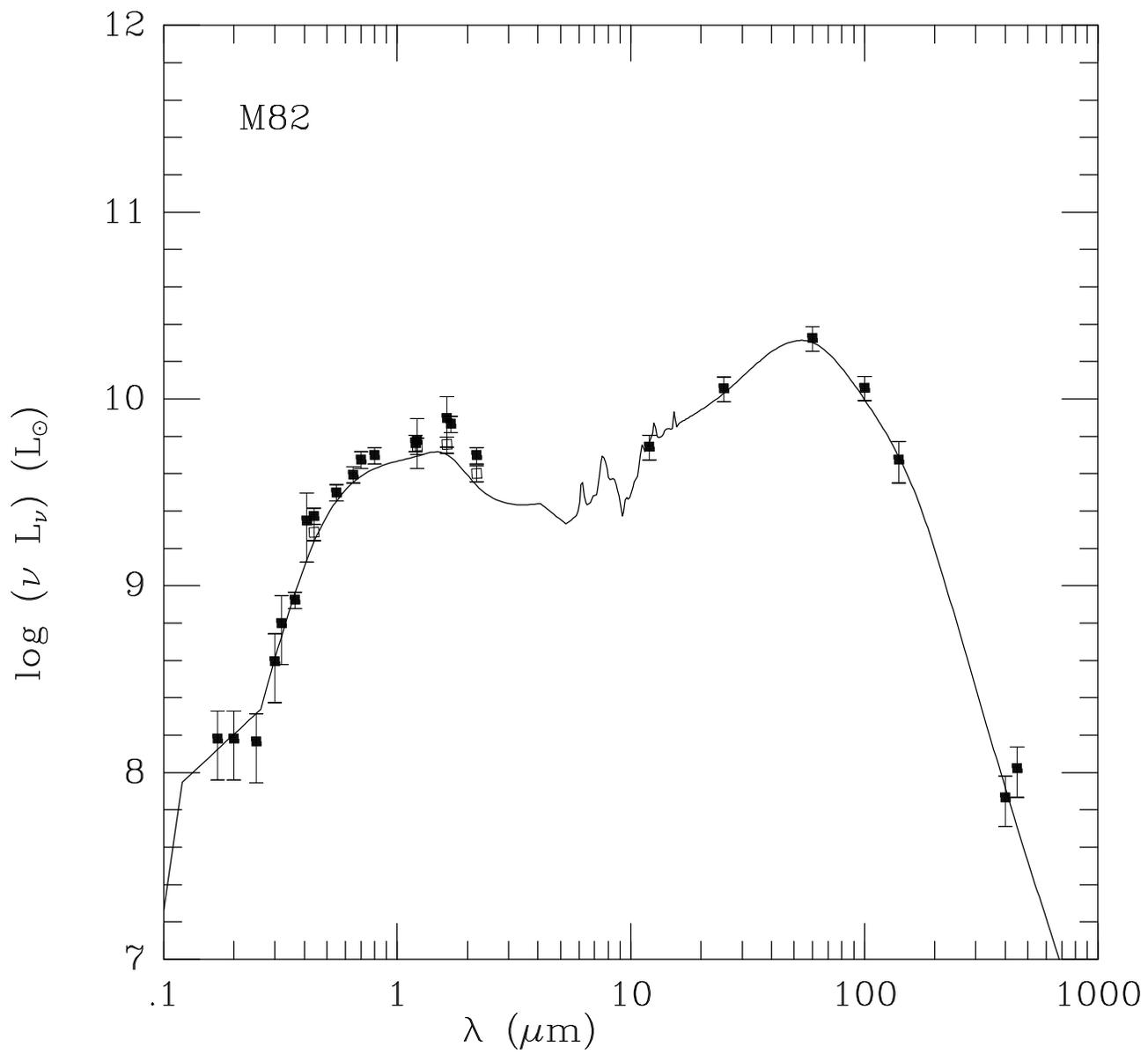
\end{document}